\documentclass[a4paper,11pt]{article}
\pdfoutput=1
\usepackage{jheppub}

\usepackage{dsfont}
\usepackage{graphicx}
\usepackage{subfigure}
\usepackage{caption}
\usepackage{tikz}
\usetikzlibrary{arrows,shapes}
\usetikzlibrary{trees}
\usetikzlibrary{patterns}
\usetikzlibrary{matrix,arrows} 				
\usetikzlibrary{positioning}				
\usetikzlibrary{calc,through}				
\usetikzlibrary{decorations.pathreplacing}  
\usepackage{pgffor}							

\usetikzlibrary{decorations.pathmorphing}	
\usetikzlibrary{decorations.markings}
\tikzset{
    vector/.style={decorate, decoration={snake}, draw},
    graviton/.style={decorate, double, decoration={snake}, draw},
	provector/.style={decorate, decoration={snake,amplitude=2.5pt}, draw},
	antivector/.style={decorate, decoration={snake,amplitude=-2.5pt}, draw},
        smallvector/.style={decorate, decoration={snake,amplitude=1.5pt,post length=0.5mm}, draw},
    fermion/.style={draw=black, postaction={decorate},
        decoration={markings,mark=at position .55 with {\arrow[draw=black]{>}}}},
    fermionbar/.style={draw=black, postaction={decorate},
        decoration={markings,mark=at position .55 with {\arrow[draw=black]{<}}}},
    fermionnoarrow/.style={draw=black},
    gluon/.style={decorate, draw=black,
        decoration={coil,amplitude=4pt, segment length=5pt}},
    scalar/.style={dashed,draw=black, postaction={decorate},
        decoration={markings,mark=at position .55 with {\arrow[draw=black]{>}}}},
    scalarbar/.style={dashed,draw=black, postaction={decorate},
        decoration={markings,mark=at position .55 with {\arrow[draw=black]{<}}}},
    scalarnoarrow/.style={dashed,draw=black},
    electron/.style={draw=black, postaction={decorate},
        decoration={markings,mark=at position .55 with {\arrow[draw=black]{>}}}},
    bigvector/.style={decorate, decoration={snake,amplitude=4pt}, draw},
    arrow/.style={draw=black, postaction={decorate},
        decoration={markings,mark=at position 1 with {\arrow[draw=black]{>}}}},
}

\tikzstyle{block} = [draw, rectangle, 
    minimum height=3em, minimum width=6earticlem]

\newcommand{\pvec}[1]{\vec{#1}\mkern2mu\vphantom{#1}}

\usepackage{xcolor}

\begin{document}

\title{Scattering Amplitudes and $N$-Body Post-Minkowskian Hamiltonians in General Relativity and Beyond}
\author[a]{Callum R. T. Jones,}
\emailAdd{cjones@physics.ucla.edu}
\affiliation[a]{
Mani L. Bhaumik Institute for Theoretical Physics, Department of Physics and Astronomy, University of California Los Angeles, Los Angeles, CA 90095, USA
}
\author[a]{Mikhail Solon,}
\emailAdd{solon@physics.ucla.edu }

\abstract{We present a general framework for calculating post-Minskowskian, classical, conservative Hamiltonians for $N$ non-spinning bodies in general relativity from relativistic scattering amplitudes. Novel features for $N>2$ are described including the subtraction of tree-like iteration contributions and the calculation of non-trivial many-body Fourier transform integrals needed to construct position space potentials. A new approach to calculating these integrals as an expansion in the hierarchical limit is described based on the method of regions. As an explicit example, we present the $\mathcal{O}\left(G^2\right)$ 3-body momentum space potential in general relativity as well as for charged bodies in Einstein-Maxwell. The result is shown to be in perfect agreement with previous post-Newtonian calculations in general relativity up to $\mathcal{O}\left(G^2 v^4\right)$. Furthermore, in appropriate limits the result is shown to agree perfectly with relativistic probe scattering in multi-center extremal black hole backgrounds and with the scattering of slowly-moving extremal black holes in the moduli space approximation.
}

\maketitle
\flushbottom


\section{Introduction}
\label{sec:intro}

The problem of predicting the motion of $N$ gravitating, compact bodies is one of the oldest and most basic problems in theoretical physics, and has important applications in astrophysics, cosmology, and gravitational wave science. In Newtonian gravity the equations of motion are simple, pairwise superpositions of 2-body Newtonian forces, but the corresponding solutions are famously chaotic~\cite{Poincare} and a complete understanding of periodic solutions for $N\geq 3$ remains an active area of research~\cite{PhysRevLett.110.114301}. In general relativity (GR) the corresponding problem is qualitatively different: even determining the equations of motion themselves is non-trivial since the underlying theory of gravitation is highly non-linear, and there are \textit{intrinsic} $N$-body forces that cannot be reduced to interactions among pairs of bodies. In the general relativistic $N$-body problem more is different.

Numerical solutions provide a wealth of information for $N$-body dynamics, including strong field effects like black hole mergers \cite{Campanelli:2007ea, Lousto:2007rj, Galaviz:2010mx}, but can be computationally expensive and impractical to implement even with only a few bodies. Since finding exact solutions to Einstein's equations for the motion of $N$ compact bodies appears intractable, it is important to consider a broad range of complementary approaches, including analytic methods. This strategy has been quite successful for the 2-body problem where numerical relativity~\cite{Pretorius:2005gq,Campanelli:2005dd,Baker:2005vv}, effective one-body~\cite{Buonanno:1998gg,Buonanno:2000ef}, and gravitational self-force~\cite{Mino:1996nk,Quinn:1996am,Poisson:2011nh,Barack:2018yvs} have been applied in concert with perturbative analyses in post-Newtonian (PN)~\cite{Einstein:1938yz,Einstein:1940mt,Ohta:1973je,Jaranowski:1997ky,Damour:1999cr,Blanchet:2000nv,Damour:2001bu,Jaranowski:2015lha}, post-Minkowskian (PM)~\cite{Bertotti:1956pxu,Kerr:1959zlt,Bertotti:1960wuq,Portilla:1979xx,Westpfahl:1979gu,Portilla:1980uz,Bel:1981be,Westpfahl:1985tsl,Damour:2016gwp}, and non-relativistic general relativity~\cite{Goldberger:2004jt,Gilmore:2008gq,Foffa:2011ub,Foffa:2016rgu,Porto:2017dgs,Foffa:2019hrb,Blumlein:2019zku,Foffa:2019yfl,Blumlein:2020pog,Blumlein:2020pyo}.

Recently, new techniques based on powerful tools from theoretical high energy physics, such as scattering amplitudes and effective field theory (EFT), have also been applied to the 2-body problem~\cite{Bjerrum-Bohr:2018xdl,Cheung:2018wkq,Kosower:2018adc}. These approaches build on the field theoretic description of gravitons~\cite{Iwasaki:1971vb,Iwasaki:1971iy,Okamura:1973my,Amati:1990xe,Donoghue:1993eb,Donoghue:1994dn,Bjerrum-Bohr:2002gqz,Bjerrum-Bohr:2013bxa} by bringing in cutting-edge tools such as the double copy~\cite{Kawai:1985xq,Bern:1998sv,Bern:2008qj,Bern:2010ue,Bern:2019prr}, on-shell methods~\cite{Bern:1994zx,Bern:1994cg,Bern:1997sc,Britto:2004nc,Bern:2007ct}, EFT~\cite{Goldberger:2004jt,Neill:2013wsa}, and advanced multiloop integration~\cite{Chetyrkin:1981qh,Laporta:2000dsw,Smirnov:2008iw,Kotikov:1990kg,Bern:1993kr,Remiddi:1997ny,Gehrmann:1999as,Henn:2013pwa,Henn:2013nsa,Parra-Martinez:2020dzs}. The program is motivated in large part by advancing the development of highly accurate waveform models for future gravitational wave detectors~\cite{Antonelli:2019ytb,Khalil:2022ylj}; see e.g.~\cite{Buonanno:2022pgc,Adamo:2022dcm,Bjerrum-Bohr:2022blt,Kosower:2022yvp} for recent reviews. Scattering amplitudes have now been applied for deriving a number of state-of-the-art predictions in the PM expansion (fixed order in $G$ and all-orders in velocity), such as the 2-body Hamiltonian at 3PM~\cite{3PM,3PMLong} and 4PM~\cite{Bern:2021dqo,Bern:2021yeh}, as well as for modeling spin~\cite{Arkani-Hamed:2017jhn,Bern:2020buy,Chiodaroli:2021eug,Aoude:2020onz,Jakobsen:2021zvh,Maybee:2019jus,Guevara:2019fsj,Guevara:2018wpp,Kosmopoulos:2021zoq,Chung:2019duq,Chung:2020rrz,Chen:2021qkk,Jakobsen:2022fcj,Bern:2022kto,Aoude:2022trd,Aoude:2022thd,FebresCordero:2022jts}, tidal corrections~\cite{Cheung:2020sdj,Haddad:2020que,Aoude:2020ygw,Bern:2020uwk,Cheung:2020gbf,AccettulliHuber:2020dal}, and radiative effects~\cite{DiVecchia:2020ymx,DiVecchia:2021bdo,Bjerrum-Bohr:2021din,Damgaard:2021ipf,Brandhuber:2021eyq,Herrmann:2021lqe,Herrmann:2021tct,DiVecchia:2021ndb,Heissenberg:2021tzo,Alessio:2022kwv}.

In this paper we extend these amplitudes-based methods to the $N$-body problem, focusing on the calculation of PM Hamiltonians from high-multiplicity relativistic scattering amplitudes. We emphasize novel features that arise for the $N>2$ case, such as gauge (coordinate) ambiguities, the subtraction of iteration contributions, and non-trivial Fourier transform integrals. As in the 2-body case, deriving classical $N$-body dynamics from scattering amplitudes leads to vast simplifications in the structure of perturbation theory, allowing for both efficient calculation and insight into underlying theoretical structures. These developments offer a path for advancing the state-of-the-art in classical $N$-body dynamics, which can be applied, e.g., for modeling sequential and hierarchical mergers~\cite{Naoz:2012bx,Lim:2020cvm,Martinez:2020lzt,Fragione:2020gly}, and may also enhance our understanding of binaries, and, more broadly, of the theoretical structures that connect scattering amplitudes and classical dynamics.

Previous PN results (an expansion in $v \sim \sqrt{Gm/r} \ll 1$) for the $N$-body problem in GR have been obtained using a variety of purely classical approaches. Important results include the calculation of the leading 1PN $N$-body potential in \cite{Einstein:1938yz}. The calculation of the 2PN contribution to the 3-body potential was undertaken in \cite{Ohta:1974pq} and completed in \cite{SCHAFER1987336}. More recently, by making use of the EFT formalism of \cite{Goldberger:2004jt}, these results were extended to include 4-body interactions at 2PN order where the result was given in terms of certain un-evaluated Feynman integrals. The 1PM Hamiltonian was calculated in \cite{Ledvinka:2008tk}, although at this order there are no genuine 3-body interactions. Using worldline based methods, a formal 2PM effective Lagrangian was calculated in \cite{Loebbert:2020aos} and explicit velocity expanded expressions obtained up to $\mathcal{O}\left(G^2 v^4\right)$. As will be discussed further in Section \ref{sec:MSA}, the $\mathcal{O}\left(v^2\right)$, and all orders in $G$, $N$-body potential between extremal (charged) black holes has also been calculated using the moduli space approximation \cite{Ferrell:1987gf}. The main explicit result obtained in this paper, the 3-body, 2PM, momentum space Hamiltonian for charged bodies in Einstein-Maxwell theory (\ref{3bdypotential}), is shown to be in perfect agreement with these previous results.

This paper is organized as follows. In Section \ref{sec:framework} we introduce the general framework for calculating PM (and PN) conservative $N$-body potentials from $N$-body scattering amplitudes. The problem of gauge ambiguity is discussed in Section \ref{subsec:gauge}, we introduce a generalization of isotropic gauge to arbitrary reference frames for 2-body dynamics. The systematics of calculating ``tree" iteration subtraction contributions is described in Section \ref{sec:matterpole}. The family of non-trivial Feynman integrals relevant for the 3-body potential are described in detail in Section \ref{sec:NbdyFT}, a new approach to calculating these integrals perturbatively in the hierarchical limit is described based on the method of regions. In Section \ref{sec:3bdycalc} the explicit 3-body, 2PM, momentum space Hamiltonian for charged bodies is given. Further details of the calculation of the subtraction contribution are given in Appendix \ref{app:3bdSubtraction}. Verification of this result is described by comparison with previous PN results in Section \ref{sec:PN}, with relativistic probe dynamics on multi-center (extremal) black hole backgrounds in Section \ref{sec:probe} and with the moduli space approximation to extremal black hole interactions in Section \ref{sec:MSA}. In Section \ref{sec:discussion} important open problems and future directions are described. Finally, the explicit Feynman rules used to calculate the 3-body amplitudes used in Section \ref{sec:3bdycalc} are given in Appendix \ref{app:FeynRules}.

\section{General Framework}
\label{sec:framework}

In this section we will describe a generalization of the framework for calculating 2-body conservative PM potentials from scattering amplitudes described in \cite{Cheung:2018wkq} to $N$-bodies. 

\subsection{Classical Dynamics from Scattering Amplitudes}
We model an $N$-body system of gravitationally and electromagnetically interacting compact, spinless bodies (e.g. black holes or neutron stars) with masses $m_i$ and electric charges $Q_i$ by $N$ distinguishable scalar fields $\phi_i$:
\begin{align}\label{eq:scalars}
   S = \int {\rm d}^{d+1} x \sqrt{-g} \left[-{R \over 16 \pi G} -\frac{1}{4}F_{\mu\nu}F^{\mu\nu}+ \sum_{i=1}^{N} \left( |\nabla_\mu \phi_i|^2 -m_i^2 |\phi_i|^2 \right)  \right] + S_{\rm GF} + S_{\text{HD}} \,. 
\end{align}
In this action and throughout this paper we will use the $(+,-,...,-)$ metric convention. Here $S_{\rm GF}$ denotes the gauge fixing terms and $S_{\text{HD}}$ denotes higher-derivative interactions encoding (sub-leading) finite-size effects and higher (gravitational and electromagnetic) multipole moments (see e.g. \cite{Cheung:2020sdj,Bern:2020uwk}). This effective description is valid in the regime where we have a hierarchy of scales $R\ll r$, where $R$ the typical size of the bodies (for a black hole $R\sim 2Gm$ the Schwarzschild radius) and $r$ is the typical inter-body separation. In the 2-body problem this EFT is appropriate for describing the inspiral phase of a binary black hole merger \cite{Goldberger:2004jt,Cheung:2018wkq}. 

We consider the elastic $N$-to-$N$ process where the incoming scalars have four-momenta $p_i^\mu = (E_i, \vec{p}_i)$, while the outgoing scalars have momenta $p_i^\mu - q_i^\mu$. Conservation of total four-momentum implies $\sum_i q_i^\mu = 0$, while the on-shell condition for external legs implies $p_i \cdot q_i = q_i^2 / 2$. In the two-body case, it is convenient to work in the center-of-mass (COM) frame where the momentum transfer is purely spatial. This yields results for the Hamiltonian in isotropic gauge. With $N$ bodies, however, there is no frame choice where all momentum transfers are purely spatial. We will thus work in a generic frame and use the on-shell conditions to fix the energy component of the momentum transfers, $q_i^0$.

In general, scattering amplitudes are manifestly relativistic objects, and so naturally encode dynamics at all orders in velocity. For the $N$-body problem in particular, the non-trivial dynamical input comes from high-multiplicity scattering amplitudes, which can be efficiently calculated using powerful on-shell methods including the double-copy \cite{Kawai:1985xq,Bern:1998sv,Bern:2008qj,Bern:2010ue,Bern:2019prr}, on-shell recursion, and unitarity methods \cite{Bern:1994zx,Bern:1994cg,Bern:1997sc,Britto:2004nc,Bern:2007ct}. 

The scattering amplitude is calculated as a PM expansion in powers of $G$
\begin{equation}
\label{amplitudePM}
    \mathcal{M}_N = \underbrace{\mathcal{M}_N^{\text{1PM}}}_{G} + \underbrace{\mathcal{M}_N^{\text{2PM}}}_{G^2}  + \underbrace{\mathcal{M}_N^{\text{3PM}}}_{G^3} + \cdots \, .
\end{equation}
The classical limit of the full quantum scattering amplitudes $\mathcal{M}_2$, $\mathcal{M}_3$, ... , $\mathcal{M}_N$ encode conservative $N$-body scattering dynamics. The classical limit follows from having large gravitational charges, $m_i \gg M_{\rm Planck}$, and large angular momenta in the scattering process, $J \gg \hbar$. From here on we set $\hbar = 1$. The classical limit is implemented by rescaling all graviton momenta $\ell \to \lambda \ell$ and then expanding in small $\lambda$. For example, $\ell$ could be a loop momentum or one of the momentum transfers $q_i$. These soft graviton momenta can be further separated into potential and radiation subregions, defined by the respective energy-momentum scalings $\ell^\mu \sim (v \lambda, \lambda)$ and $\ell^\mu \sim (v \lambda, v \lambda) $, where $v$ is the typical velocity of the $N$ bodies that defines the PN expansion. We focus here on the conservative dynamics described by potential gravitons.

Order-by-order in PM we subsequently expand in the classical limit
\begin{equation}
\label{classicalexpansion}
    \mathcal{M}_N^{n\text{PM}} = \sum_{k\in \mathds{Z}}  \mathcal{M}_N^{n\text{PM}(k)} , \hspace{5mm} \text{where} \hspace{5mm} \mathcal{M}_N^{n\text{PM}(k)}(\{\vec{p},\lambda \vec{q}\}) =\lambda ^k \mathcal{M}_N^{n\text{PM}(k)} (\{\vec{p},\vec{q}\}).
\end{equation}
Terms in this series are separated according to the power $k$\footnote{This scaling can be deduced by dimensional analysis. The $N$-body (quantum) position space potential is a function of the available dimensionful quantities $\{G,\hbar,c,m,p,x\}$ with $[V_N]=M L^2 T^{-2}$. The classical part of the potential at $n$PM, is necessarily homogeneous in re-scaling of the position variables $V_N\vert_{G^n \hbar^0}(\lambda x) = \lambda^{-n(d-2)} V_N\vert_{G^n \hbar^0}(x)$. Fourier transforming to momentum space, the intrinsic $N$-body potential is multiplied by an overall factor $\delta^{(3)}(\vec{q}_1+...\vec{q}_N)$, and so $V^{(N)}\vert_{G^n \hbar^0}(\lambda q) = \lambda^{n(d-2)-d(N-1)} V^{(N)}\vert_{G^n \hbar^0}(q)$. The classical part of the amplitude $M_N$ is then defined as the piece generated by the classical momentum space potential in the first Born approximation $M_N \approx -V^{(N)}$.}:
\begin{itemize}
    \item \textit{Super-classical}: \hspace{10mm}$k<n(d-2)-d(N-1)$.
    \item \textit{Classical}:\hspace{21.5mm}$k=n(d-2)-d(N-1)$.
    \item \textit{Quantum}:\hspace{21mm}$k>n(d-2)-d(N-1)$.
\end{itemize}
Super-classical terms do not contribute to the potential, and must cancel with contributions arising from the iteration of lower-PM-order (and lower-multiplicity) potentials. Quantum terms are discarded as soon as possible since they do not contribute to the classical potential. Note that in (\ref{classicalexpansion}) we are ignoring the overall energy conserving delta function which is \textit{not} homogeneous in the classical scaling $\vec{q}\rightarrow \lambda \vec{q}$. Until a resolution of this constraint is imposed the precise separation between super-classical and classical contributions is ill-defined.

\subsection{$N$-Body Conservative Potentials}

After integrating out potential mode gravitons, the resulting conservative dynamics can be described by an effective $N$-body Hamiltonian of the form
\begin{equation}\label{eq:hamiltonian}
    H(\{\vec{p},\vec{x}\}) = \sum_{i=1}^N \sqrt{|\vec{p}_i|^2+m_i^2} + V\left(\{\vec{p},\vec{x}\}\right),
\end{equation}
where $\vec{p}_i$ and $\vec{x}_i$ denote the 3-momenta and positions of the $i$-th particle. Obtaining the potential from matching to a scattering amplitude naturally produces a potential in \textit{momentum space}, this is related to the position space potential by a Fourier transform which in our conventions will be\footnote{Note that here we are making a common abuse of notation using the symbol $\vec{q}_i$ to denote two strictly different quantities. In (\ref{amplitudePM}), $\vec{q}_i$ denotes the momentum transfer in a scattering amplitude, while in (\ref{fourierpotential}), $\vec{q}_i$ denotes the Fourier conjugate variable to the position vector $\vec{x}_i$. These two quantities only coincide when the potential is used to calculate a scattering amplitude using the Born series (\ref{Born}).} 
\begin{equation}
\label{fourierpotential}
    V\left(\{\vec{p},\vec{x}\}\right) = \prod_{i=1}^N\int \frac{\text{d}^d \vec{q}_i}{(2\pi)^d} e^{i\vec{q}_i \cdot \vec{x}_i} V\left(\{\vec{p},\vec{q}\}\right) \,.
\end{equation}
To regularize divergences at intermediate stages of calculation, the Fourier transform integrals in this paper will be calculated using dimensional regularization as an expansion around $d=3-2\epsilon$ dimensions. Overall translation invariance of the position space potential will always produce $d$ delta functions after Fourier transforming. However, the $N$-body potential will also contain terms that depend only on a subset of the bodies, and the Fourier transform of these will produce more singular terms that encode \textit{intrinsically} $n$-body interactions for $n<N$. For example for $N=3$ 
\begin{align}
    \label{fourierpotentialexpanded}
    V\left(\{\vec{p},\vec{q}\}\right) &= \sum_{(i,j,k)\in S_3}\Biggr[(2\pi)^{2d}\delta^{(d)}\left(\vec{q}_i+\vec{q}_j\right)\delta^{(d)}\left(\vec{q}_k\right)\times \frac{1}{2} \times V_{ij}^{(2)}\left(\vec{p}_i,\vec{p}_j,\vec{q}_i,\vec{q}_j\right) \nonumber\\
    &\hspace{20mm}+(2\pi)^d\delta^{(d)}\left(\vec{q}_i+\vec{q}_j+\vec{q}_k\right)V_{ijk}^{(3)}\left(\vec{p}_i,\vec{p}_j,\vec{p}_k,\vec{q}_i,\vec{q}_j,\vec{q}_k\right)\Biggr].
\end{align}
In this formula and subsequently in this paper, the notation $(i,j,k)\in S_3$ means that we sum over the $3!$ distinct permutations of $\{1,2,3\}$. The factor of $1/2$ multiplying the intrinsic 2-body interaction  $V^{(2)}_{ij}$ in this expression is purely conventional.

Classical scaling for the momentum space potential is defined analogously to (\ref{classicalexpansion}). For $N$-bodies in $d$ spatial dimensions at $n$PM, the classical potential (including the delta functions) is homogeneous
\begin{equation}
    V\vert_{G^n}\left(\{\vec{p},\lambda \vec{q}\}\right) = \lambda^{n(d-2)-dN}V\vert_{G^n}\left(\{\vec{p},\vec{q}\}\right).
\end{equation}
Unlike the scattering amplitude (\ref{classicalexpansion}), there are no super-classical pieces.

\subsection{Matching Calculation}
\label{sec:matching}

The $N$-body effective potential (\ref{eq:hamiltonian}) is determined by integrating out potential gravitons through an EFT matching calculation. The same physical observable, the classical part of an $N$-to-$N$ scattering amplitude, is calculated from (\ref{eq:scalars}) using manifestly relativistic Feynman-Dyson perturbation theory and from (\ref{eq:hamiltonian}) by (formally) solving the Lippmann-Schwinger equation. Requiring that these two calculations agree defines the effective potential.

In spite of the manifestly relativistic nature of the scattering amplitude $\mathcal{M}_N$ in \eqref{amplitudePM}, the corresponding scattering amplitudes computed from the effective Hamiltonian \eqref{eq:hamiltonian} are not manifestly Lorentz invariant since this object depends on a choice of time coordinate. We therefore parametrize the scattering kinematics using 3-momenta. In a general $N$-body scattering process we will denote the incoming 3-momenta by $\vec{p}_i$ and the outgoing 3-momenta by $\pvec{p}'_i \equiv \vec{p}_i - \vec{q}_i$, where $\vec{q}_i$ is the 3-momentum transfer. Conservation of 3-momentum corresponds to the constraint $\sum_i \vec{q}_i = 0$. All external particles are assumed to be on-shell, and for a given particle of mass $m_i$ and 3-momentum $\vec{k}_i$ we denote the relativistic energy as $E_i \big( \vec{k}_i \big) = \sqrt{|\vec{k}_i|^2+m_i^2}$. Energy is also conserved in scattering which is given as a non-linear constraint
\begin{equation}
\label{consen}
    \sum_i E_i\left(\vec{p}_i\right) =  \sum_i E_i\left(\vec{p}_i-\vec{q}_i\right), 
\end{equation}
which, to leading classical order, takes the form
\begin{equation}
\label{classicalenergycons}
    \sum_i \frac{\vec{p}_i \cdot \vec{q}_i}{E_i(\vec{p}_i)} = \mathcal{O}\left(|\vec{q}|^2\right).
\end{equation}
Using the momentum space potential we can calculate a scattering amplitude by solving the (relativistic) Lippmann-Schwinger equation \cite{Lippmann:1950zz,Cristofoli:2019neg}
\begin{align}
    \label{LippmannSchwinger}
    T\left(\{\vec{p},\vec{q}\}\right) \overset{!}{=} -V\left(\{\vec{p},\vec{q}\}\right) + \int_{\{\vec{k}\}} \frac{V\left(\{\vec{k},\vec{k}-\vec{p}+\vec{q}\}\right)T\left(\{\vec{p},\vec{p}-\vec{k}\}\right)}{\sum_{j=1}^N\left[E_j(\vec{p})-E_j(\vec{k})\right] + i\epsilon},
\end{align}
where
\begin{equation}
    \int_{\{\vec{k}\}} \equiv \prod_{i=1}^N \int \frac{\text{d}^d \vec{k}_i}{(2\pi)^d},
\end{equation}
and
\begin{equation}
    \label{Tmatrix}
   \langle \{\vec{p}-\vec{q}\}|T|\{\vec{p}\}\rangle \equiv 2\pi\delta\left(\sum_{j=1}^N\left[E_j(\vec{p})-E_j(\vec{p}-\vec{q})\right]\right)T\left(\{\vec{p},\vec{q}\}\right), 
\end{equation}
is the non-trivial part of the S-matrix defined in the convention $S=1+iT$. In (\ref{LippmannSchwinger}) we have introduced the notation $\overset{!}{=}$ to denote equality on the constraint surface associated with conservation of energy (\ref{consen}). Note that for $N>2$, the $T$-matrix element is not exactly the same as the scattering amplitude since it contains both fully- and partially-connected contributions; for example for $N=3$
\begin{align}
    T\left(\{\vec{p},\vec{q}\}\right) &\overset{!}{=} (2\pi)^{2d}\delta^{(d)}\left(\vec{q}_3\right)\delta^{(d)}\left(\vec{q}_1+\vec{q}_2\right)M_{2}^{(12)}\left(\vec{p}_1,\vec{p}_2,\vec{q}_1,\vec{q}_2\right)\nonumber\\
    &\hspace{5mm}+(2\pi)^{2d}\delta^{(d)}\left(\vec{q}_2\right)\delta^{(d)}\left(\vec{q}_1+\vec{q}_3\right)M_{2}^{(13)}\left(\vec{p}_1,\vec{p}_3,\vec{q}_1,\vec{q}_3\right) \nonumber\\
    &\hspace{5mm}+(2\pi)^{2d}\delta^{(d)}\left(\vec{q}_1\right)\delta^{(d)}\left(\vec{q}_2+\vec{q}_3\right)M_{2}^{(23)}\left(\vec{p}_2,\vec{p}_3,\vec{q}_2,\vec{q}_3\right) \nonumber\\
    &\hspace{5mm}+(2\pi)^d\delta^{(d)}\left(\vec{q}_1+\vec{q}_2+\vec{q}_3\right)M_{3}\left(\vec{p}_1,\vec{p}_2,\vec{p}_3,\vec{q}_1,\vec{q}_2,\vec{q}_3\right).
\end{align}
The amplitude $M_N$ differs from the amplitude $\mathcal{M}_N$ in (\ref{amplitudePM}) by a (purely conventional) non-relativistic normalization factor
\begin{equation}
    M_N\left(\{\vec{p},\vec{q}\}\right) = \prod_{i=1}^N \frac{1}{2\sqrt{E_i\left(\vec{p}_i\right) E_i\left(\vec{p}_i-\vec{q}_i\right)}}\mathcal{M}_N\left(\{\vec{p},\vec{q}\}\right).
\end{equation}
The Lippmann-Schwinger equation (\ref{LippmannSchwinger}) is formally solved by the Born series
\begin{align}
\label{Born}
    T\left(\{\vec{p},\vec{q}\}\right) &\overset{!}{=} -V\left(\{\vec{p},\vec{q}\}\right) - \int_{\{\vec{k}\}} \frac{V\left(\{\vec{k},\vec{k}-\vec{p}+\vec{q}\}\right)V\left(\{\vec{p},\vec{p}-\vec{k}\}\right)}{\sum_{j=1}^N\left[E_j(\vec{p})-E_j(\vec{k})\right] + i\epsilon} \\
    &\hspace{5mm}-\int_{\{\vec{k}\}}\int_{\{\vec{k}'\}} \frac{V\left(\{\vec{k},\vec{k}-\vec{p}+\vec{q}\}\right)V\left(\{\vec{k}',\vec{k}'-\vec{k}\}\right)V\left(\{\vec{p},\vec{p}-\vec{k}'\}\right)}{\left[\sum_{j=1}^N\left[E_j(\vec{p})-E_j(\vec{k})\right] + i\epsilon\right]\left[\sum_{j=1}^N\left[E_j(\vec{p})-E_j(\vec{k}')\right] + i\epsilon\right]}+...\nonumber
\end{align}
The connected components of the $T$-matrix can be calculated in a Feynman diagrammatic expansion by treating the expanded potential components as Wilson coefficients in a non-relativistic EFT \cite{Neill:2013wsa,Cheung:2018wkq}
\begin{align}\label{eq:EFT}
     L_{\text{EFT}} = &\sum_{i} \int_{\vec{k}_i}\phi_i^\dagger\left(-\vec{k}_i\right)\left[i\partial_t -\sqrt{|\vec{k}_i|^2+m_i^2}\right] \phi_i\left(\vec{k}_i\right)\nonumber\\
    &-\sum_{ij} \int \frac{\text{d}^d\vec{k}_i}{(2\pi)^d}\frac{\text{d}^d\vec{k}_j}{(2\pi)^d}(2\pi)^d \delta^{(d)}\left(\vec{k}_i+\vec{k}_j-\vec{k}_i'-\vec{k}_j'\right)V^{(2)}_{ij}\left(\{\vec{k},\vec{k}-\vec{k}'\}\right)\nonumber\\
    &\hspace{45mm}\times\phi_i^\dagger\left(\vec{k}_i'\right)\phi_j^\dagger\left(\vec{k}_j'\right)\phi_i\left(\vec{k}_i\right)\phi_j\left(\vec{k}_j\right)\nonumber\\
     &-\sum_{ijk} \int \frac{\text{d}^d\vec{k}_i}{(2\pi)^d}\frac{\text{d}^d\vec{k}_j}{(2\pi)^d}\frac{\text{d}^d\vec{k}_k}{(2\pi)^d} \delta^{(d)}\left(\vec{k}_i+\vec{k}_j+\vec{k}_k-\vec{k}_i'-\vec{k}_j'-\vec{k}_k'\right)V^{(3)}_{ijk}\left(\{\vec{k},\vec{k}-\vec{k}'\}\right)\nonumber\\
     &\hspace{45mm}\times\phi_i^\dagger\left(\vec{k}_i'\right)\phi_j^\dagger\left(\vec{k}_j'\right)\phi_k^\dagger\left(\vec{k}_k'\right)\phi_i\left(\vec{k}_i\right)\phi_j\left(\vec{k}_j\right)\phi_k\left(\vec{k}_k\right)\nonumber\\
     &+\;\;\;(\text{4-body interactions}) \;\;\;+ \;\;\; ...
\end{align}
One of the advantages of the EFT is that the tree-like contributions to (\ref{Born}), arising for $N>2$ from the lower-multiplicity contributions to the potential (\ref{fourierpotentialexpanded}), become literal tree Feynman diagrams.

The \textit{matching} procedure for calculating the potential from a scattering amplitudes is a simple rearrangement of the Born series (\ref{Born}). For example at 2PM
\begin{align}  
\label{matching}
    V\vert_{G^2}\left(\{\vec{p},\vec{q}\}\right) &\overset{!}{=} -T\vert_{G^2}\left(\{\vec{p},\vec{q}\}\right) - \int_{\{\vec{k}\}} \frac{V\vert_{G}\left(\{\vec{k},\vec{k}-\vec{p}+\vec{q}\}\right)V\vert_{G}\left(\{\vec{p},\vec{p}-\vec{k}\}\right)}{\sum_{j=1}^N\left[E_j(\vec{p})-E_j(\vec{k})\right] + i\epsilon}.
\end{align}
The $V$-dependent terms on the right-hand-side are called \textit{iteration} contributions, and are shown in Figure~\ref{fig:EFTamps}. They are discussed in greater detail in Section \ref{sec:matterpole}.

\subsection{Gauge Ambiguity and Conservation of Energy}
\label{subsec:gauge}

The effective potential between $N$-bodies (\ref{eq:hamiltonian}) depends on the choice of coordinate system and is therefore not a gauge invariant physical quantity. By contrast, scattering amplitudes \textit{are} gauge invariant. It may seem puzzling, or even inconsistent, that the matching procedure (\ref{matching}) seems to relate a gauge invariant quantity to a gauge non-invariant quantity. The resolution is that equations (\ref{Tmatrix}) and (\ref{matching}) hold only on the constraint surface of energy conservation, and therefore do not admit a unique solution for the potential. The correct statement about the gauge invariance of the scattering amplitude is that the $T$-matrix element, including the energy conserving delta function, is gauge invariant and unambiguous. The function $T\left(\{\vec{p},\vec{q}\}\right)$ in (\ref{Tmatrix}) is defined up to the freedom to add an arbitrary function that vanishes on the support of the energy conserving delta function, and this ambiguity encompasses the coordinate dependence of the potential. 

To be more explicit, consider the calculation of the 1PM 2-body potential from the physical scattering amplitude $M_2$. In this case, in $d=3$, the matching equation reduces to
\begin{equation}
    V\left(\vec{p}_1,\vec{p}_2,\vec{q}_1,\vec{q}_2\right) \overset{!}{=} -(2\pi)^3\delta^{(3)}\left(\vec{q}_1+\vec{q}_2\right)M_2\left(\vec{p}_1,\vec{p}_2,\vec{q}_1,\vec{q}_2\right),
\end{equation}
where on both sides we are truncating at $\mathcal{O}\left(G |\vec{q}|^{-2}\right)$. It is straightforward to see that this equation does not uniquely determine the potential, suppose we take any solution and modify it as
\begin{equation}
    V\left(\vec{p}_1,\vec{p}_2,\vec{q}_1,\vec{q}_2\right) \rightarrow V\left(\vec{p}_1,\vec{p}_2,\vec{q}_1,\vec{q}_2\right) + (2\pi)^3 \delta^{(3)}\left(\vec{q}_1+\vec{q}_2\right) \times \frac{cG}{|\vec{q}_1|^4}\left(\frac{\vec{p}_1\cdot \vec{q}_1}{E_1(\vec{p}_1)}+\frac{\vec{p}_2\cdot \vec{q}_2}{E_2(\vec{p}_2)}\right)^2,
\end{equation}
where $c$ is an arbitrary real number. From (\ref{classicalenergycons}) we see that the second term on the right-hand-side is zero on the energy conserving constraint surface, up to terms that vanish in the classical limit. This expression defines a distinct, but perfectly physical, momentum space potential that is related to the original potential by some change of coordinates in position space. All physical observables calculated using either potential must agree.

For $N=2$ there is a choice of resolution of this ambiguity that dramatically simplifies calculations. Using (\ref{classicalenergycons}) and 3-momentum conservation $\vec{q}_1+\vec{q}_2=0$
\begin{equation}
    \vec{p}_i\cdot \vec{q}_j \overset{!}{=} \frac{E_i(\vec{p}_i)\left((\vec{p}_1+\vec{p}_2)\cdot \vec{q}_j\right)}{E_1(\vec{p}_1)+E_2(\vec{p}_2)} + \mathcal{O}\left(|\vec{q}|^2\right).
\end{equation}
Order-by-order in the classical expansion we can use this identity to rewrite all $\vec{p}\cdot \vec{q}$ dot products in terms of $(\vec{p}_1+\vec{p}_2)\cdot \vec{q}_1$. Making this choice in the definition of $T\left(\{\vec{p},\vec{q}\}\right)$ uniquely fixes the ambiguity in a way that reduces the total number of independent kinematic invariants that appear in the momentum space potential. This choice is especially useful if we further restrict to the COM frame $\vec{p}_1+\vec{p}_2=0$, where this choice is usually called \textit{isotropic gauge}, in which case all $\vec{p}\cdot \vec{q}$ dot products are absent. As will be discussed further below, the calculation of the iteration contributions to $N$-body potentials with $N>2$ requires the $2$-body potential in a general frame $\vec{p}_1+\vec{p}_2\neq 0$. In this case we will refer to this resolution of the ambiguity as \textit{generalized isotropic gauge}, and will be used for the 3-body calculation of the 2PM potential in Section \ref{sec:3bdycalc}.

There is an analogous ambiguity present in the definition of $T(\{\vec{p},\vec{q}\})$ for the scattering of $N>2$ particles. For $N=3$, classical energy conservation gives the constraint
\begin{equation}
    \frac{\vec{p}_1\cdot \vec{q}_1}{E_1(\vec{p}_1)} + \frac{\vec{p}_2\cdot \vec{q}_2}{E_2(\vec{p}_2)} + \frac{\vec{p}_3\cdot \vec{q}_3}{E_3(\vec{p}_3)} = \mathcal{O}\left(|\vec{q}|^2\right).
\end{equation}
Together with $3$-momentum conservation $\vec{q}_1+\vec{q}_2+\vec{q}_3=0$, this can be used to reduce the total number of $\vec{p}_i\cdot \vec{q}_j$ invariants appearing from 9 to 5. Unfortunately, any complete fixing of the ambiguity of this kind will produce a 3-body potential that does not have the expected $S_3$ relabelling symmetry. This situation is not improved by restricting to the 3-body center of mass frame where $\vec{p}_1+\vec{p}_2+\vec{p}_3 =0$, although the number of invariants is further reduced to 3.

This inherent ambiguity in the potential makes it difficult to compare the result of our calculation to existing results in the literature. One approach following \cite{Bern:2019crd} would be to construct an explicit coordinate transformation relating different potentials, we will not attempt this. Another approach is to compare different potentials by calculating gauge invariant quantities. In Section \ref{sec:comparison} we will verify our calculation by comparing gauge invariant classical scattering amplitudes. By generalizing the formalism of KMOC \cite{Kosower:2018adc} to $N$-bodies, the classical scattering amplitude should determine a range of asymptotic classical observables including the linear $\Delta p_i^\mu$ and angular $\Delta J_i^{\mu\nu}$ impulses. This will be described in a separate paper.

\subsection{Matter Poles and Tree Iteration}
\label{sec:matterpole}

\begin{figure}
\centering
\begin{subfigure}{}
\centering
    \begin{tikzpicture}
     \draw[-] (1,-2)--(5,0);
     \draw[-] (1,0)--(5,-2);
     \draw[-] (1,-1)--(5,-1);
    \filldraw[color=black!60, fill=black!5, very thick](3,-1) circle (0.6);
    \node at (3,-1) {$V^{(3)}$};
    \end{tikzpicture}
\end{subfigure}
\hspace{15mm}
\begin{subfigure}{}
\centering
    \begin{tikzpicture}
     \draw[-] (3,0)--(5,0);
     \draw[-] (1,-2)--(3,-2);
     \draw[-] (1,0)--(5,-2);
     \draw[-] (1,-1)--(3,0);
     \draw[-] (3,-2)--(5,-1);
    \filldraw[color=black!60, fill=black!5, very thick](2,-0.5) circle (0.6);
    \filldraw[color=black!60, fill=black!5, very thick](4,-1.5) circle (0.6);
    \node at (2,-0.5) {$V^{(2)}$};
    \node at (4,-1.5) {$V^{(2)}$};
    \end{tikzpicture}
\end{subfigure}
 \caption{\textit{(Left)} contact contribution and \textit{(right)} tree-like iteration contribution of 2-body interactions to the 3-body potential. }
 \label{fig:EFTamps}
\end{figure}
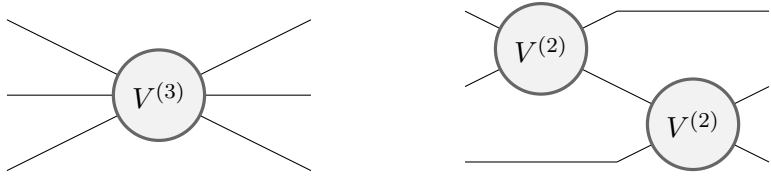

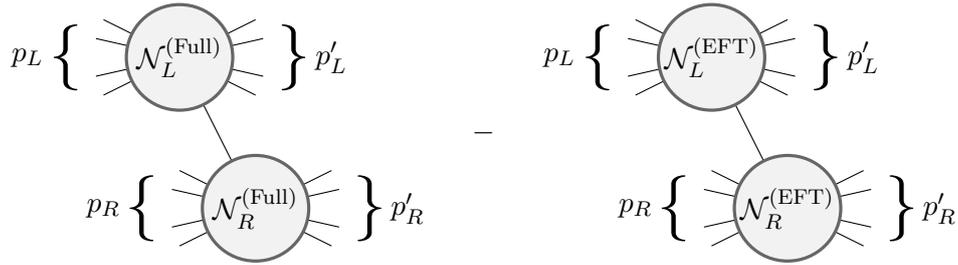
\begin{figure}
\centering
    \begin{tikzpicture}
    \begin{scope}[xshift=0cm, yshift=0cm]
     \draw[-] (1,0)--(3,-1);
     \draw[-] (0.9,-0.25)--(3.1,-0.75);
     \draw[-] (0.9,-0.75)--(3.1,-0.25);
     \draw[-] (1,-1)--(3,0);
     \draw[-] (2,-3)--(4,-2);
     \draw[-] (1.9,-2.25)--(4.1,-2.75);
     \draw[-] (1.9,-2.75)--(4.1,-2.25);
     \draw[-] (2,-2)--(4,-3);
     \draw[-] (2,-0.5)--(3,-2.5);
    \filldraw[color=black!60, fill=black!5, very thick](2,-0.5) circle (0.7);
    \filldraw[color=black!60, fill=black!5, very thick](3,-2.5) circle (0.7);
    \node at (2,-0.5) {$\mathcal{N}^{(\text{Full})}_L$};
    \node at (0.5,-0.5) {\Huge $\{$};
    \node at (0,-0.5) {$p_L$};
    \node at (3.5,-0.5) {\Huge $\}$};
    \node at (4,-0.5) {$p_L'$};
    \node at (3,-2.5) {$\mathcal{N}^{(\text{Full})}_R$};
    \node at (6,-1.5) {$-$};
    \node at (1.5,-2.5) {\Huge $\{$};
    \node at (4.5,-2.5) {\Huge $\}$};
    \node at (1,-2.5) {$p_R$};
    \node at (5,-2.5) {$p_R'$};
    \end{scope}
    \begin{scope}[xshift=7cm, yshift=0cm]
     \draw[-] (1,0)--(3,-1);
     \draw[-] (0.9,-0.25)--(3.1,-0.75);
     \draw[-] (0.9,-0.75)--(3.1,-0.25);
     \draw[-] (1,-1)--(3,0);
     \draw[-] (2,-3)--(4,-2);
     \draw[-] (1.9,-2.25)--(4.1,-2.75);
     \draw[-] (1.9,-2.75)--(4.1,-2.25);
     \draw[-] (2,-2)--(4,-3);
     \draw[-] (2,-0.5)--(3,-2.5);
    \filldraw[color=black!60, fill=black!5, very thick](2,-0.5) circle (0.7);
    \filldraw[color=black!60, fill=black!5, very thick](3,-2.5) circle (0.7);
    \node at (2,-0.5) {$\mathcal{N}^{(\text{EFT})}_L$};
    \node at (0.5,-0.5) {\Huge $\{$};
    \node at (0,-0.5) {$p_L$};
    \node at (3.5,-0.5) {\Huge $\}$};
    \node at (4,-0.5) {$p_L'$};
    \node at (3,-2.5) {$\mathcal{N}^{(\text{EFT})}_R$};
    \node at (1.5,-2.5) {\Huge $\{$};
    \node at (4.5,-2.5) {\Huge $\}$};
    \node at (1,-2.5) {$p_R$};
    \node at (5,-2.5) {$p_R'$};
     \end{scope}
\end{tikzpicture}
 \caption{Schematic representation of a ``tree" iteration subtraction contribution to the effective potential. In a physically acceptable representation of the potential, the spurious matter pole must cancel between full-theory and EFT.}
 \label{fig:matterpole}
\end{figure}

A qualitatively new feature that appears in the matching calculation for $N>2$ body interactions is the appearance of iteration diagrams that can be separated into two disconnected pieces by cutting a single matter propagator; examples are shown in Figures~\ref{fig:EFTamps} and~\ref{fig:matterpole}. Non-local terms in the potential arise from integrating out degrees-of-freedom, such as potential mode gravitons. The matter fields in the EFT are not integrated out, and so we should not expect a singularity in the potential to arise from matter propagators. Whether such singularities are present in a given explicit expression for the potential is sensitive to the chosen resolution of the gauge ambiguity described in Section \ref{subsec:gauge}. We will impose as an additional condition that in a \textit{physically acceptable} $N$-body momentum space potential, the only singularities should correspond to on-shell (soft) graviton exchange. As we will describe below, it is always possible to construct a potential such that the spurious matter singularities cancel between full-theory and EFT iteration. 

It is useful to write the amplitude (\ref{amplitudePM}) with the graviton propagator in standard relativistic form but with the matter propagators separated into positive energy (matter) and negative energy (anti-matter) contributions
\begin{equation}
    \frac{1}{p_i^2-m_i^2} = \frac{1}{2E_i\left(\vec{p}_i\right)}\left[\frac{1}{p_i^0-E_i(\vec{p}_i)}-\frac{1}{p_i^0+E_i(\vec{p}_i)}\right].
\end{equation}
The negative energy pole appears in time-ordered diagrams describing pair production, a purely quantum mechanical process, and so the classical part of the \textit{anti-matter piece} does not contain a kinematic singularity. This piece will however, still give a non-singular contribution to the classical potential (excepting graviton poles) and cannot be discarded.

For the matter piece of the full theory amplitude, there are genuine singularities at classical order. These must cancel with corresponding poles arising from denominator factors of the Born series (\ref{Born}). The difference of the contributions defines a \textit{subtraction} contribution to the potential. Diagrammatically, there is a correspondence between full theory and iteration graph topologies with a common matter pole. The general form of the associated subtraction contribution is given in Figure \ref{fig:matterpole}. It is tempting to imagine that the corresponding diagrams will cancel entirely and so can be simply excluded from the Feynman diagram expansion of the full theory amplitude. Unfortunately this is not the case. The general form of the difference depicted in Figure \ref{fig:matterpole} is
\begin{equation}
\label{wouldbesubtraction}
    V_{\text{subtraction}} \overset{!}{\supset} \frac{\mathcal{N}^{(\text{Full})}_L\mathcal{N}^{(\text{Full})}_R}{(p_L-p_L')^0-E(\vec{p}_L-\vec{p}_L')} - \frac{\mathcal{N}^{(\text{EFT})}_L\mathcal{N}^{(\text{EFT})}_R}{(p_L-p_L')^0-E(\vec{p}_L-\vec{p}_L')},
\end{equation}
where the numerator factors $\mathcal{N}$ collectively denote the rest of the diagram on one side of the matter pole. Since both full theory and EFT calculate the same scattering amplitude, on the support of energy conservation the residues of the matter singularities must match, and therefore it must be possible to write
\begin{equation}
	\label{deltaN}
    \mathcal{N}^{(\text{Full})} \overset{!}{=} \mathcal{N}^{(\text{EFT})} + \left((p_L-p_L')^0-E(\vec{p}_L-\vec{p}_L')\right)\delta \mathcal{N},
\end{equation}
where $\delta \mathcal{N}$ is non-singular on the matter pole (though it may be singular elsewhere). We then iterate this procedure on a different spurious matter pole (this is non-trivial only if there are diagrams with overlapping matter poles), including the subtraction contribution (\ref{wouldbesubtraction}) defined with (\ref{deltaN}) as part of the EFT amplitude. When all spurious poles have been eliminated we continue the result away from the support of the energy conserving delta function (dropping the $!$ above the equals sign) to define a potential. 

While a potential constructed in this way has only physical singularities, there are further conditions we should impose to define a completely physical result: cancellation of super-classical terms and manifest $S_N$ symmetry among the bodies. As an explicit example, in Appendix \ref{app:3bdSubtraction} we show how to define the subtraction contribution to the 3-body potential at 2PM such that super-classical contributions cancel between diagrams related by symmetry. A completely systematic approach to engineering such cancellations is not known. 

The explicit calculation in Appendix \ref{app:3bdSubtraction} also demonstrates clearly why the naive subtraction (\ref{wouldbesubtraction}) is not trivially zero. In the PM amplitudes the singularities corresponding to potential graviton exchange are not at the same location, and so cannot cancel. Explicitly for a 3-body 2PM subtraction contribution with $p_L = p_i + p_j$, $p_L' = p_i-q_i$ and $p_R = p_k$, and with the 2-body potential calculated in generalized isotropic gauge, the graviton pole factors in the full theory and EFT are
\begin{equation}
 \mathcal{N}_L^{(\text{Full})} \sim \frac{1}{\left(\frac{\vec{p}_k\cdot \vec{q}_k}{E_k}\right)^2-|\vec{q}_k|^2} \,,  \hspace{10mm}     \mathcal{N}_L^{(\text{EFT})} \sim \frac{1}{\left(\frac{(\vec{p}_j+\vec{p}_k)\cdot \vec{q}_k}{E_j +E_k}\right)^2-|\vec{q}_k|^2} \, .
\end{equation}
The misalignment of poles arises from the different \textit{energy transfer} prescriptions of the 4-momentum transfer in full theory and EFT. This appears to be a generic feature of the subtraction and no choice of 2-body gauge will cause the poles to align. The resulting PM subtraction contribution, after the cancellation of the spurious matter pole, will necessarily have ``doubled" graviton poles with factors containing both prescriptions. In the PN expansion the energy transfer component are always subleading to the 3-momentum transfer, so the subtraction contribution will only have a graviton pole at $\vec{q}_k=0$ as expected.

For 3-bodies at 3PM and beyond (as well as the more familiar 2-bodies at 2PM), spurious branch cuts are present in both full theory and iteration amplitudes arising from loop integrals with internal matter propagators. The logic in this case is identical to the above: these must match between full theory and EFT iteration and so it must always be possible to define a subtraction contribution without spurious non-analyticities. It is left to future work to carry out such a calculation in detail.

\subsection{$N$-Body Fourier Transform Integrals}
\label{sec:NbdyFT}

The scattering amplitude method for calculating the effective potential described in the previous sections produces a potential in \textit{momentum space}. For real-time $N$-body simulations however we require a potential in \textit{position space}, and therefore need to calculate a Fourier transform. For the case $N=2$ this step of the calculation is essentially trivial, in dimensional regularization
\begin{equation}    
    \label{2bdyFourier}
    \int \frac{\text{d}^d \vec{q}}{(2\pi)^d}\frac{e^{-i\vec{q}\cdot \vec{x}}}{|\vec{q}|^\alpha} = \frac{1}{2^\alpha \pi^{d/2}} \frac{\Gamma\left(\frac{d-\alpha}{2}\right)}{\Gamma\left(\frac{\alpha}{2}\right)} \frac{1}{|\vec{x}|^{d-\alpha}}.
\end{equation}
For 2-body potentials calculated in COM frame (in isotropic gauge) this is all that is needed for the complete PM result. For 2-body potentials in a general frame (in generalized isotropic gauge) the complete, all-orders in momentum, Fourier transform of the 2-body 1PM potential requires the integral
\begin{equation}
\label{2bdyPMFT}
    \int \frac{\text{d}^d \vec{q}}{(2\pi)^d} \frac{e^{-i\vec{q}\cdot \vec{x}_{12}}}{\left(\frac{(\vec{p}_1+\vec{p}_2)\cdot \vec{q}}{E_1+E_2}\right)^2-|\vec{q}|^2} = \frac{1}{4\pi r_{12}} \frac{1}{\sqrt{1-\left(\frac{\vec{p}_1+\vec{p}_2}{E_1+E_2}\right)^2}}\left[1+\frac{\left(\frac{(\vec{p}_1+\vec{p}_2)\cdot \vec{x}_{12}}{E_1+E_2}\right)^2}{\left(1-\left(\frac{\vec{p}_1+\vec{p}_2}{E_1+E_2}\right)^2\right)r_{12}^2}\right]^{-1/2}.
\end{equation}
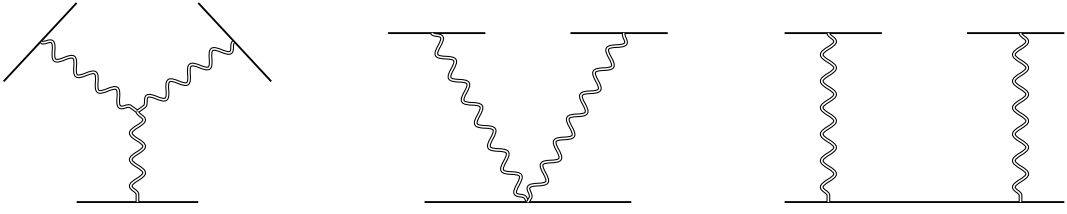
\begin{figure}
\centering
\begin{subfigure}{}
\centering
    \begin{tikzpicture}[scale=0.8]
    \draw[line width=0.7pt] (-2.2,0)--(-1,1.3);
    \draw[line width=0.7pt] (1,1.3)--(2.2,0);
    \draw[line width=0.7pt] (-1,-2)--(1,-2);
    \draw[graviton] (-1.58,0.65)--(0,-0.5);
    \draw[graviton] (1.58,0.65)--(0,-0.5);
    \draw[graviton] (0,-0.5)--(0,-2);
    \end{tikzpicture}
\end{subfigure}
\hspace{10mm}
\begin{subfigure}{}
\centering
    \begin{tikzpicture}[scale=0.8]
    \draw[line width=0.7pt] (-2.3,0.8)--(-0.7,0.8);
    \draw[line width=0.7pt] (0.7,0.8)--(2.3,0.8);
    \draw[line width=0.7pt] (-1.7,-2)--(1.7,-2);
    \draw[graviton] (-1.58,0.8)--(0,-2);
    \draw[graviton] (1.58,0.8)--(0,-2);
    \end{tikzpicture}
\end{subfigure}
\hspace{10mm}
\begin{subfigure}{}
\centering
    \begin{tikzpicture}[scale=0.8]
    \draw[line width=0.7pt] (-2.3,0.8)--(-0.7,0.8);
    \draw[line width=0.7pt] (0.7,0.8)--(2.3,0.8);
    \draw[line width=0.7pt] (-2.3,-2)--(2.3,-2);
    \draw[graviton] (-1.58,0.8)--(-1.58,-2);
    \draw[graviton] (1.58,0.8)--(1.58,-2);
    \end{tikzpicture}
\end{subfigure}
 \caption{Feynman diagrams contributing to 3-body scattering of massive scalars (solid lines) in general relativity at $\mathcal{O}\left(G^2\right)$: \textit{(left)} Y-type graph, \textit{(center)} V-type graph, \textit{(right)} U-type graph. Double wavy lines denote gravitons.}
 \label{fig:GR}
\end{figure}

For $N>2$ the problem of Fourier transforming to position space is highly non-trivial and we are unable to evaluate the PM Fourier transform integrals in closed form. We will therefore proceed by expanding in the non-relativistic limit $|\vec{p}| \ll m$ before integrating. For $N=3$ the type of scalar integrals we require take the form
\begin{equation}
    \label{YFT}
    \int \frac{\text{d}^d \vec{q}_1}{(2\pi)^d} \int \frac{\text{d}^d \vec{q}_2}{(2\pi)^d} \int \frac{\text{d}^d \vec{q}_3}{(2\pi)^d} e^{-i\left(\vec{q}_1\cdot \vec{x}_1+\vec{q}_2\cdot \vec{x}_2+\vec{q}_3\cdot\vec{x}_3\right)}\frac{(2\pi)^d \delta^{(d)}\left(\vec{q}_1+\vec{q}_2+\vec{q}_3\right)}{|\vec{q}_1|^{\alpha_1}|\vec{q}_2|^{\alpha_2}|\vec{q}_3|^{\alpha_3}}.
\end{equation}
For graphs with V-type topology, Figure \ref{fig:GR}, the integrand simplifies since for one of the propagators, $\alpha_i=0$. The integration over the corresponding $\vec{q}_i$ is then trivial, and the remaining integrals can be calculated using (\ref{2bdyFourier}). For example 
\begin{align}
    &\int \frac{\text{d}^d \vec{q}_1}{(2\pi)^d} \int \frac{\text{d}^d \vec{q}_2}{(2\pi)^d} \int \frac{\text{d}^d \vec{q}_3}{(2\pi)^d} e^{-i\left(\vec{q}_1\cdot \vec{x}_1+\vec{q}_2\cdot \vec{x}_2+\vec{q}_3\cdot\vec{x}_3\right)}\frac{(2\pi)^d \delta^{(d)}\left(\vec{q}_1+\vec{q}_2+\vec{q}_3\right)}{|\vec{q}_1|^{\alpha_1}|\vec{q}_2|^{\alpha_2}} \nonumber\\
    &= \frac{1}{2^{\alpha_1+\alpha_2} \pi^{d}} \frac{\Gamma\left(\frac{d-\alpha_1}{2}\right)}{\Gamma\left(\frac{\alpha_1}{2}\right)}\frac{\Gamma\left(\frac{d-\alpha_2}{2}\right)}{\Gamma\left(\frac{\alpha_2}{2}\right)} \frac{1}{r_{13}^{d-\alpha_1} r_{23}^{d-\alpha_2}},
\end{align}
where $r_{ij} = |\vec{x}_i-\vec{x}_j|$. For graphs with Y-type topology all $\alpha_i\neq 0$ and we need to evaluate these integrals using a different method. The idea is to rewrite the 3-momentum conserving delta function by introducing an integral over an auxiliary spatial point
\begin{equation}
    \label{deltatoint}
    (2\pi)^d \delta^{(d)}\left(\vec{q}_1+\vec{q}_2+\vec{q}_3\right) = \int \text{d}^d \vec{x}_0 \; e^{i(\vec{q}_1+\vec{q}_2+\vec{q}_3)\cdot \vec{x}_0}.
\end{equation}
The integral (\ref{YFT}) then factors into a product for which each factor can be evaluated using (\ref{2bdyFourier})
\begin{align}
    &\int \text{d}^d \vec{x}_0 \int \frac{\text{d}^d \vec{q}_1}{(2\pi)^d} \int \frac{\text{d}^d \vec{q}_2}{(2\pi)^d} \int \frac{\text{d}^d \vec{q}_3}{(2\pi)^d} \frac{e^{i\left(\vec{q}_1\cdot \vec{x}_{01}+\vec{q}_2\cdot \vec{x}_{02}+\vec{q}_3\cdot\vec{x}_{03}\right)}}{|\vec{q}_1|^{\alpha_1}|\vec{q}_2|^{\alpha_2}|\vec{q}_3|^{\alpha_3}} \nonumber\\
    &= \int \text{d}^d \vec{x}_0 \left(\int \frac{\text{d}^d \vec{q}_1}{(2\pi)^d} \frac{e^{i\vec{q}_1 \cdot \vec{x}_{01}}}{|\vec{q}_1|^{\alpha_1}}\right)\left(\int \frac{\text{d}^d \vec{q}_2}{(2\pi)^d} \frac{e^{i\vec{q}_2 \cdot \vec{x}_{02}}}{|\vec{q}_2|^{\alpha_2}}\right)\left(\int \frac{\text{d}^d \vec{q}_3}{(2\pi)^d} \frac{e^{i\vec{q}_3 \cdot \vec{x}_{03}}}{|\vec{q}_3|^{\alpha_3}}\right) \nonumber\\
    &= \frac{1}{2^{\alpha_1+\alpha_2+\alpha_3} \pi^{3d/2}} \frac{\Gamma\left(\frac{d-\alpha_1}{2}\right)\Gamma\left(\frac{d-\alpha_2}{2}\right)\Gamma\left(\frac{d-\alpha_3}{2}\right)}{\Gamma\left(\frac{\alpha_1}{2}\right)\Gamma\left(\frac{\alpha_2}{2}\right)\Gamma\left(\frac{\alpha_3}{2}\right)} I^d_3\left[\frac{d-\alpha_1}{2},\frac{d-\alpha_2}{2},\frac{d-\alpha_3}{2}\right],
\end{align}
where   
\begin{equation}
\label{I3int}
    I^d_3\left[a_1,a_2,a_3\right] = \int \text{d}^d \vec{x}_0 \frac{1}{x_{01}^{2a_1}x_{02}^{2a_2}x_{03}^{2a_3}}.
\end{equation}
The non-trivial part of this expression (\ref{I3int}) is a Euclidean Feynman triangle integral with generically non-integer power propagators. A formal expression for these integrals was obtained using Mellin-Barnes methods, the general result is expressed in terms of the Appell hypergeometric series $F_4$ \cite{Boos:1990rg}. In the present context, this series expression has a region of convergence \cite{Chu:2008xm}
\begin{equation}
    r_{12}+r_{13}< r_{23}.
\end{equation}
Interestingly, this inequality is the \textit{complement} of the region defined by the triangle inequality, and so the formal expression converges nowhere in the physical region. It should be possible to analytically continue the result to the physical region using the methods of \cite{Ananthanarayan:2020xut}. While the general result is very complicated, the integral may simplify for the specific values of $\alpha_i$ that appear in low-order PM calculations when evaluated in $d=3-2\epsilon$ up to $\mathcal{O}\left(\epsilon^0\right)$. This family of integrals is also well-known to arise in the calculation of CFT 3-point functions in momentum space \cite{Bzowski:2013sza,Bzowski:2015yxv}. In this context they are sometimes referred to as \textit{triple-K} integrals since they can be reduced to a single paramter integral over a product of three Bessel K functions.

At $\mathcal{O}\left(G^2\right)$ we need Fourier transforms (\ref{YFT}) with $\alpha_i \in 2\mathds{Z}$, and therefore the integral (\ref{I3int}) with $a_i \in \mathds{Z}+\frac{1}{2}$.\footnote{The unusual half-integer propagator powers are a consequence of the fact that $d=3$, is an odd integer. If the number of spatial dimensions was an even integer, then the corresponding Fourier transform integrals would have $a_i\in \mathds{Z}$, that is they would be ordinary triangle Feynman integrals, see for example \cite{Gutowski:2001xa}. Amusingly, it appears the $N$-body problem in GR is simpler in odd spacetime dimensions!} Inspired by \cite{Ohta:1974pq}, the authors of \cite{Loebbert:2020aos} conjectured that for $a_i \in \mathds{Z}+\frac{1}{2}$ and $a_1+a_2+a_3\leq \frac{3}{2}$
\begin{equation}
\label{halfintegerints}
    I^{3-2\epsilon}_3\left[a_1,a_2,a_3\right] = \frac{A}{2\epsilon}+B + C\log\left(r_{12}+r_{13}+r_{23}\right) +\mathcal{O}\left(\epsilon\right),
\end{equation}
where $A,B$ and $C$ are homogeneous polynomials in $r_{ij}$ of degree fixed by dimensional analysis. In \cite{Loebbert:2020aos} an approach to fixing the finitely many unknown coefficients in the Ansatz (\ref{halfintegerints}) using a Yangian differential equation was proposed. From the triple-K representation this sub-class of integrals can be evaluated directly \cite{Bzowski:2015yxv,Bzowski:2020lip}, the results agree with the Yangian approach.

Here we provide an alternative approach, with a transparent physical interpretation. We expand the required integral in the \textit{hierarchical limit}, meaning we calculate the integral (\ref{I3int}) as a series expansion around the limit
\begin{equation}
    \label{hierarchy}
    r_{12} \ll r_{13} \sim r_{23},
\end{equation}
using the method of regions \cite{Beneke:1997zp}. First we shift $\vec{x}_0$ to be centered on body 1 so that the integral takes the form 
\begin{equation}
    \int \text{d}^d \vec{x}_0 \frac{1}{x_{0}^{2a_1}|\vec{x}_{0}+\vec{x}_{12}|^{2a_2}|\vec{x}_{0}+\vec{x}_{13}|^{2a_3}}.
\end{equation}
We find there are two non-trivial regions that contribute 
\begin{align}
    &\text{Region I}: \hspace{5mm} \vec{x}_0 \sim \vec{x}_{12}, \hspace{10mm}\text{Region II}: \hspace{5mm} \vec{x}_0 \sim \vec{x}_{13}.
\end{align}
Physically, region I corresponds to the integration point being near the bodies 1 and 2 and region II near body 3. In the method of regions we are instructed to expand the integrand in the small parameter (\ref{hierarchy}) in each region, integrate term-by-term and then sum together the result. In region I we make the expansion
\begin{align}
    |\vec{x}_0+\vec{x}_{13}|^{-2a_3} &= \frac{1}{r_{13}^{2a_3}}\left(1+\frac{2\vec{x}_0\cdot \vec{x}_{13}}{r_{13}^2}+\frac{|\vec{x}_0|^2}{r_{13}^2}\right)^{-a_3} \nonumber\\
    &= \frac{1}{r_{13}^{2a_3}} - \frac{2a_3 (\vec{x}_0\cdot \vec{x}_{13})}{r_{13}^{2(a_3+1)}} + \frac{2a_3(a_3+1)(\vec{x}_0\cdot \vec{x}_{13})^2-a_3 r_{13}^2|\vec{x}_0|^2}{r_{13}^{2(a_3+2)}} +...
\end{align}
In region II we make the expansion 
\begin{align}
    |\vec{x}_0+\vec{x}_{12}|^{-2a_2} &= \frac{1}{|\vec{x}_0|^{2a_2}}\left(1+\frac{2\vec{x}_0\cdot \vec{x}_{12}}{|\vec{x}_0|^2}+\frac{r_{12}^2}{|\vec{x}_0|^2}\right)^{-a_2} \nonumber\\
    &= \frac{1}{|\vec{x}_0|^{2a_2}} - \frac{2a_2 (\vec{x}_0\cdot \vec{x}_{12})}{|\vec{x}_0|^{2(a_2+1)}} + \frac{2a_2(a_2+1)(\vec{x}_0\cdot \vec{x}_{12})^2-a_2 r_{12}^2|\vec{x}_0|^2}{|\vec{x}_0|^{2(a_2+2)}} +...
\end{align}
Each term in the integral can be evaluated using
\begin{equation}
    \int \text{d}^d x_0 \frac{1}{|\vec{x}_{0}|^{2a_1}|\vec{x}_{0}+\vec{x}|^{2a_2}} = \pi^{d/2} \frac{\Gamma\left(a_1+a_2-\frac{d}{2}\right)\Gamma\left(\frac{d}{2}-a_1\right)\Gamma\left(\frac{d}{2}-a_2\right)}{\Gamma\left(a_1\right)\Gamma\left(a_2\right)\Gamma\left(d-a_1-a_2\right)}|\vec{x}|^{d-2a_1-2a_2},
\end{equation}
after reducing tensor numerators using
\begin{equation}
    \vec{x}_{0}\cdot \vec{x} = \frac{1}{2}\left(|\vec{x}_0+\vec{x}|^2-|\vec{x}_0|^2-|\vec{x}|^2\right).
\end{equation}
By matching terms against the series expansion of (\ref{halfintegerints}) we can fix all unknown coefficients and verify the validity of the ansatz. Unlike the series obtained using Mellin-Barnes methods \cite{Boos:1990rg}, the hierarchical expansion converges in the physical region and has a straightforward physical interpretation. It may be especially useful at $\mathcal{O}\left(G^3\right)$, where the necessary integrals (with $a_1\in \mathds{Z}$ and $a_2, a_3 \in \mathds{Z}+\frac{1}{2}$) are not known in closed form.
\begin{figure}
\centering
    \begin{tikzpicture}[scale=2]
    \draw[-][line width=0.7pt] (-0.45,0.1)--(0.0,0.8);
    \draw[graviton] (0.5,0)--(-0.2,0.5);
    \draw[graviton] (0.5,0)--(-0.2,-0.5);
    \draw[-][line width=0.7pt] (-0.45,-0.1)--(0.0,-0.8);
    \draw[graviton] (0.5,0)--(1.3,0);
    \draw[graviton] (1.3,0)--(2,0.5);
    \draw[graviton] (1.3,0)--(2,-0.5);
    \draw[-][line width=0.7pt] (2.25,-0.1)--(1.8,-0.8);
    \draw[-][line width=0.7pt] (2.25,0.1)--(1.8,0.8);
    \end{tikzpicture}
 \caption{Feynman diagram contributing to 4-body scattering of massive scalars in general relativity at $\mathcal{O}\left(G^3\right)$. }
 \label{fig:4bdy}
\end{figure}
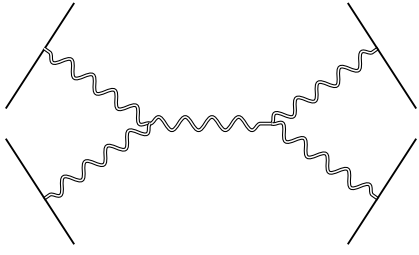

For higher-multiplicity tree diagrams, similar Fourier transform integrals are required. For $N\geq 4$ we encounter diagrams with \textit{internal} graviton lines (graviton lines not connected to any matter lines). We introduce a corresponding internal momentum, which is integrated over, and a 3-momentum conserving delta function at each cubic graviton vertex. Using the identity (\ref{deltatoint}) on each delta function we can reduce the $N$-body Fourier transform integral to an $N-2$ loop Feynman integral. For example, for the 4-body diagram shown in Figure \ref{fig:4bdy}, at leading PN order, we require a 2-loop integral
\begin{align}
    &\left(\prod_{j=1}^4 \int \frac{\text{d}^d \vec{q}_j}{(2\pi)^d} e^{-i\vec{q}_j\cdot \vec{x}_j}\right) \frac{(2\pi)^d \delta^{(d)}\left(\vec{q}_1+\vec{q}_2+\vec{q}_3+\vec{q}_4\right)}{|\vec{q}_1|^{\alpha_1}|\vec{q}_2|^{\alpha_2}|\vec{q}_1+\vec{q}_2|^{\beta}|\vec{q}_3|^{\alpha_3}|\vec{q}_4|^{\alpha_4}}\nonumber\\
    &=\left(\prod_{j=1}^4 \int \frac{\text{d}^d \vec{q}_j}{(2\pi)^d} e^{-i\vec{q}_j\cdot \vec{x}_j}\right) \int \frac{\text{d}^d \vec{k} }{(2\pi)^d} \frac{(2\pi)^d \delta^{(d)}\left(\vec{q}_1+\vec{q}_2-\vec{k}\right)(2\pi)^d \delta^{(d)}\left(\vec{q}_3+\vec{q}_4+\vec{k}\right)}{|\vec{q}_1|^{\alpha_1}|\vec{q}_2|^{\alpha_2}|\vec{k}|^{\beta}|\vec{q}_3|^{\alpha_3}|\vec{q}_4|^{\alpha_4}} \nonumber\\
    &= \frac{\Gamma\left(\frac{d-\beta}{2}\right)}{2^{\beta} \pi^{d/2}\Gamma\left(\frac{\beta}{2}\right)}\left(\prod_{j=1}^4\frac{\Gamma\left(\frac{d-\alpha_j}{2}\right)}{2^{\alpha_j} \pi^{d/2}\Gamma\left(\frac{\alpha_j}{2}\right)}\right)\nonumber\\
    &\hspace{10mm}\times\int  \frac{\text{d}^d \vec{x}_0 \text{d}^d \vec{y}_0}{|\vec{x}_{0}-\vec{x}_1|^{d-\alpha_1}|\vec{x}_{0}-\vec{x}_2|^{d-\alpha_2}|\vec{x}_{0}-\vec{y}_0|^{d-\beta}|\vec{y}_{0}-\vec{x}_3|^{d-\alpha_3}|\vec{y}_{0}-\vec{x}_4|^{d-\alpha_4}}.
\end{align}
The case $\alpha_i=\beta =2$ is needed for the 2PN 4-body potential \cite{Chu:2008xm}, and has never been evaluated in closed form. 

\section{3-Body Dynamics in General Relativity and Einstein-Maxwell}
\label{sec:3bdycalc}

 In this section we apply the general $N$-body framework described above to derive the 3-body potential $V^{(3)}_{ijk}$ in Einstein-Maxwell at ${\cal O}(G^2)$ or 2PM. The full dynamics of 3-particles at 2PM also requires the 2PM contribution to the 2-body potentials $V^{(2)}_{ij}$, which we do not include here (see e.g.~\cite{Cheung:2018wkq} for results albeit in COM frame). We present the results for Einstein-Maxwell (electrically charged scalars). The corresponding result for general relativity can be obtained by simply setting $Q_1=Q_2=Q_3=0$.

 \subsection{Full Theory Amplitudes}\label{sec:FTamplitudes}
\begin{figure}
\centering
\begin{subfigure}{}
\centering
    \begin{tikzpicture}[scale=0.8]
    \draw[line width=0.7pt] (-2.2,0)--(-1,1.3);
    \draw[line width=0.7pt] (1,1.3)--(2.2,0);
    \draw[line width=0.7pt] (-1,-2)--(1,-2);
    \draw[vector] (-1.58,0.65)--(0,-0.5);
    \draw[vector] (1.58,0.65)--(0,-0.5);
    \draw[graviton] (0,-0.5)--(0,-2);
    \end{tikzpicture}
\end{subfigure}
\hspace{10mm}
\begin{subfigure}{}
\centering
    \begin{tikzpicture}[scale=0.8]
    \draw[line width=0.7pt] (-2.3,0.8)--(-0.7,0.8);
    \draw[line width=0.7pt] (0.7,0.8)--(2.3,0.8);
    \draw[line width=0.7pt] (-1.7,-2)--(1.7,-2);
    \draw[graviton] (-1.58,0.8)--(0,-2);
    \draw[vector] (1.58,0.8)--(0,-2);
    \end{tikzpicture}
\end{subfigure}
\hspace{10mm}
\begin{subfigure}{}
\centering
    \begin{tikzpicture}[scale=0.8]
    \draw[line width=0.7pt] (-2.3,0.8)--(-0.7,0.8);
    \draw[line width=0.7pt] (0.7,0.8)--(2.3,0.8);
    \draw[line width=0.7pt] (-2.3,-2)--(2.3,-2);
    \draw[graviton] (-1.58,0.8)--(-1.58,-2);
    \draw[vector] (1.58,0.8)--(1.58,-2);
    \end{tikzpicture}
\end{subfigure}
\vspace{10mm}
\begin{subfigure}{}
\centering
    \begin{tikzpicture}[scale=0.8]
    \draw[line width=0.7pt] (-2.3,0.8)--(-0.7,0.8);
    \draw[line width=0.7pt] (0.7,0.8)--(2.3,0.8);
    \draw[line width=0.7pt] (-1.7,-2)--(1.7,-2);
    \draw[vector] (-1.58,0.8)--(0,-2);
    \draw[vector] (1.58,0.8)--(0,-2);
    \end{tikzpicture}
\end{subfigure}
\hspace{10mm}
\begin{subfigure}{}
\centering
    \begin{tikzpicture}[scale=0.8]
    \draw[line width=0.7pt] (-2.3,0.8)--(-0.7,0.8);
    \draw[line width=0.7pt] (0.7,0.8)--(2.3,0.8);
    \draw[line width=0.7pt] (-2.3,-2)--(2.3,-2);
    \draw[vector] (-1.58,0.8)--(-1.58,-2);
    \draw[vector] (1.58,0.8)--(1.58,-2);
    \end{tikzpicture}
\end{subfigure}
 \caption{Additional Feynman diagrams contributing to the 3-body scattering of charged massive scalars in Einstein-Maxwell at $\mathcal{O}\left(G^2\right)$. Single wavy lines correspond to photons, and double wavy lines to gravitons.}
 \label{fig:EM}
\end{figure}
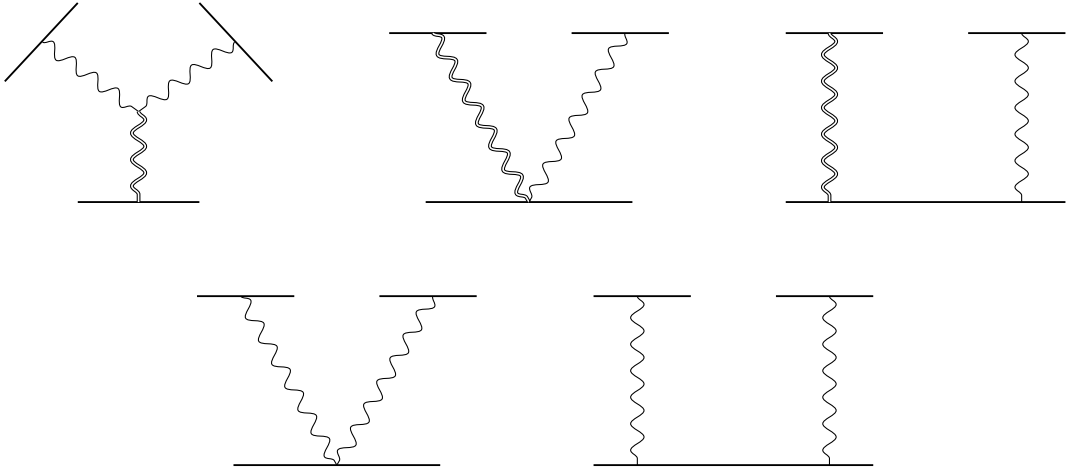

 To calculate $V_{ijk}^{(3)}$ at 2PM, we need both the  3-to-3 scattering amplitude at 2PM and the 2-to-2 scattering amplitudes at 1PM. We consider scattering of non-spinning (scalar) particles with masses $m_i$ and charges $Q_i$ for $i=1,2,3$, and derive the necessary amplitudes using standard Feynman diagram methods. The necessary Feynman rules are given in Appendix~\ref{app:FeynRules}. The (relativistically normalized) 2-body amplitude, expanded to classical order is
\begin{equation}
\label{2bdyEMamplitude}
    \mathcal{M}^{(2)}_{ij} = \frac{16\pi G \left[m_i^2 m_j^2 - 2\left(p_i\cdot p_j\right)^2+Q_i Q_j \left(
   p_i\cdot p_j\right)\right]}{q_i^2} + \mathcal{O}\left(|\vec{q}|^{-1}\right).
\end{equation}
For the 3-body amplitude, we organize the result according to the topology of the Feynman graphs that appear in Fig. \ref{fig:GR} and \ref{fig:EM}
\begin{equation}
\label{3bdyEMamplitude}
    \mathcal{M}^{(3)} = \sum_{(i,j,k)\in S_3} \left[\mathcal{M}^{(3)}_{\text{Y},ijk}+\mathcal{M}^{(3)}_{\text{V},ijk}+\mathcal{M}^{(3)}_{\text{U},ijk}\right],
\end{equation}
where
\begingroup
\allowdisplaybreaks
\begin{align}
    \mathcal{M}^{(3)}_{\text{Y},ijk} &=  -\frac{256 \pi^2 G^2}{q_i^2 q_j^2 q_k^2}\nonumber\\
    &\hspace{5mm}\times\biggr[\left(q_i\cdot
   q_j\right) \left(2 m_k^2(p_i\cdot
   p_j)^{2}+2 m_i^2(p_j\cdot p_k)^{2}-4
   \left(p_i\cdot p_j\right) \left(p_i\cdot
   p_k\right) \left(p_j\cdot
   p_k\right)-m_i^2 m_j^2 m_k^2\right)\nonumber\\
   &\hspace{10mm}+\left(p_k\cdot q_i\right) \left(p_k\cdot
   q_j\right) \left(m_i^2 m_j^2-2(p_i\cdot
   p_j)^{2}\right)-2 m_k^2
   \left(p_i\cdot p_j\right) \left(p_i\cdot
   q_j\right) \left(p_j\cdot
   q_i\right)\nonumber\\
   &\hspace{10mm}+\left(p_j\cdot q_i\right)
   \left(p_k\cdot q_j\right) \left(4
   \left(p_i\cdot p_j\right) \left(p_i\cdot
   p_k\right)-2 m_i^2
   \left(p_j\cdot p_k\right)\right)\nonumber\\
   &\hspace{10mm}+Q_iQ_j\biggr(\frac{1}{2} m_k^2 \left(p_i\cdot q_j\right)
   \left(p_j\cdot q_i\right)-\left(q_i\cdot
   q_j\right) \left(\frac{1}{2} m_k^2 \left(p_i\cdot
   p_j\right)-\left(p_i\cdot p_k\right)
   \left(p_j\cdot p_k\right)\right)\nonumber\\
   &\hspace{25mm}-2
   \left(p_j\cdot p_k\right) \left(p_i\cdot
   q_j\right) \left(p_k\cdot
   q_i\right)+\left(p_i\cdot p_j\right)
   \left(p_k\cdot q_i\right) \left(p_k\cdot
   q_j\right)\biggr)\biggr] \nonumber\\
   &\hspace{20mm}+\mathcal{O}\left(|\vec{q}|^{-3}\right),\\
   \mathcal{M}^{(3)}_{\text{V},ijk} &= \frac{64\pi^2 G^2}{q_i^2 q_k^2}\biggr[16
   \left(p_i\cdot p_j\right)
   \left(p_i\cdot p_k\right)
   \left(p_j\cdot p_k\right)-8 m_i^2
  (p_j\cdot p_k)^{2}\nonumber\\
   &\hspace{25mm}+Q_i Q_j^2 Q_k
   \left(p_i\cdot p_k\right)-8 Q_i Q_j
   \left(p_i\cdot p_k\right)
   \left(p_j\cdot p_k\right)\biggr] +\mathcal{O}\left(|\vec{q}|^{-3}\right),\\
   \mathcal{M}^{(3)}_{\text{U},ijk} &= \frac{128\pi^2 G^2}{q_i^2 q_k^2[(p_j+q_i)^2-m_j^2]}\nonumber\\
   &\hspace{5mm} \times\biggr[4 m_j^2 m_k^2 \left(p_i\cdot p_j\right) \left(p_i\cdot p_j-p_j\cdot
   q_i\right)+4 m_i^2 m_j^2 \left(p_j\cdot p_k\right) \left(2 p_k\cdot
   q_i-p_j\cdot q_k+p_j\cdot p_k\right)\nonumber\\
   &\hspace{10mm}+8 \left(p_i\cdot
   p_j\right) \left(p_j\cdot p_k\right) \left(\left(p_j\cdot p_k\right)
   \left(p_j\cdot q_i\right)+\left(p_i\cdot p_j\right) \left(-2 p_k\cdot
   q_i+p_j\cdot q_k-p_j\cdot p_k\right)\right)\nonumber\\
   &\hspace{10mm}\left.-2 m_i^2 m_j^4 m_k^2\right.\nonumber\\
   &\hspace{10mm}+Q_i Q_j \left[-2 \left(p_j\cdot q_i\right) (p_j\cdot p_k)^{2}+m_j^2 m_k^2 \left(p_j\cdot
   q_i-2 p_i\cdot p_j\right)\right.\nonumber\\
   &\hspace{25mm}\left.+4 \left(p_i\cdot p_j\right)
   \left(p_j\cdot p_k\right) \left(2 p_k\cdot q_i-p_j\cdot
   q_k+p_j\cdot p_k\right)\right]\nonumber\\
   &\hspace{10mm}+Q_j Q_k \left[m_i^2 m_j^2 \left(-2 p_k\cdot q_i+p_j\cdot q_k-2 p_j\cdot
   p_k\right)\right.\nonumber\\
   &\hspace{25mm}\left.+2 \left(p_i\cdot p_j\right) \left(\left(p_i\cdot p_j\right) \left(2
  p_k\cdot q_i-p_j\cdot q_k+2 p_j\cdot p_k\right)-2
   \left(p_j\cdot p_k\right) \left(p_j\cdot q_i\right)\right)\right]\nonumber\\
   &\hspace{10mm}+Q_i Q_j^2 Q_k \left[\left(p_j\cdot p_k\right) \left(p_j\cdot q_i\right)+\left(p_i\cdot p_j\right)
   \left(-2 \left(p_k\cdot q_i\right)+p_j\cdot q_k-2 \left(p_j\cdot
   p_k\right)\right)\right]\biggr] \nonumber\\
   &\hspace{20mm}+\mathcal{O}\left(|\vec{q}|^{-3}\right).
\end{align}
\endgroup

\subsection{Effective Potential} 
 
Using the general matching framework described in Section \ref{sec:framework} we can now use the (classical) scattering amplitudes (\ref{2bdyEMamplitude}) and (\ref{3bdyEMamplitude}) to calculate the classical, conservative, momentum space potential for 3 charged bodies in Einstein-Maxwell at 2PM. The general form of the potential is 
\begin{align}
    V\left(\{\vec{p},\vec{q}\}\right) &= \sum_{(i,j,k)\in S_3}\left[(2\pi)^6\delta^{(3)}\left(\vec{q}_i+\vec{q}_j\right)\delta^{(3)}\left(\vec{q}_k\right)\times \frac{1}{2} \times V^{(2)}_{ij}\left(\{\vec{p},\vec{q}\}\right)\right.\nonumber\\
    &\hspace{25mm}\left.+(2\pi)^3\delta^{(3)}\left(\vec{q}_i+\vec{q}_j+\vec{q}_k\right)\times V^{(3)}_{ijk}\left(\{\vec{p},\vec{q}\}\right)\right].
\end{align}

We begin with the intrinsic 2-body potential $V^{(2)}_{ij}$ at 1PM. As discussed in Section~\ref{subsec:gauge}, it is convenient to fix the ambiguity in the definition of this object by working in generalized isotropic gauge. Applying the general matching framework of Section \ref{sec:matching} with the amplitude (\ref{2bdyEMamplitude}) gives the 1PM potential in a general reference frame
\begin{equation}
    V_{ij}^{(2)}\left(\vec{p}_i,\vec{p}_j,\vec{q}_i,\vec{q}_j\right) = \frac{4 \pi  G \left[E_i E_j(\rho_{ij}-1)(2 E_i E_j
  (\rho_{ij}-1)+Q_i Q_j)-m_i^2 m_j^2\right]}{E_i E_j\left[\left(\frac{E_i \tau_{ii}+E_j\tau_{ji}}{E_i+E_j}\right)^2-|\vec{q}_i|^2\right]},
\end{equation}
where 
\begin{equation}
    \label{sigmatau}
    E_i = \sqrt{|\vec{p}_i|^2+m_i^2}, \hspace{10mm} \rho_{ij} = \frac{\vec{p}_i \cdot \vec{p}_j}{E_i E_j}, \hspace{10mm} \tau_{ij} = \frac{\vec{p}_i \cdot \vec{q}_j}{E_i}.
\end{equation}
It is straightforward to verify that in the $Q_i=0$ limit, and in the COM frame $\vec{p}_i+\vec{p}_j=0$, this reduces to the expected result \cite{Cheung:2018wkq}. This 1PM, general frame, potential can be expressed in position space using (\ref{2bdyPMFT}). The compactness of this expression can be contrasted with the result of \cite{Ledvinka:2008tk}, calculated in a different gauge.  

The calculation of the intrinsic 3-body potential at 2PM follows from the general matching framework of Section \ref{sec:matching}, with some additional subtleties arising from the spurious matter singularities present in both the full theory and iteration contributions as discussed in Section \ref{sec:matterpole}. Further details are given in Appendix \ref{app:3bdSubtraction}.

The intrinsic 3-body potential consists of four pieces
\begin{equation}
    \label{3bdypotential}
    V_{ijk}^{(3)} = V^{(3)}_{\text{Y},ijk} + V^{(3)}_{\text{V},ijk} + V^{(3)}_{\text{anti-matter},ijk} + V^{(3)}_{\text{subtraction},ijk}.
\end{equation}
The \textit{Y-graph} and \textit{V-graph} contributions correspond straightforwardly to the full-theory graphs with the associated topology and are given by
\begin{align}
\label{YVpotential}
    V^{(3)}_{\text{Y},ijk} &=  \frac{1}{8E_i E_j E_k} \frac{256 \pi^2 G^2}{q_i^2 q_j^2 q_k^2}\nonumber\\
    &\hspace{5mm}\times\biggr[\left(q_i\cdot
   q_j\right) \left(2 m_k^2(p_i\cdot
   p_j)^{2}+2 m_i^2(p_j\cdot p_k)^{2}-4
   \left(p_i\cdot p_j\right) \left(p_i\cdot
   p_k\right) \left(p_j\cdot
   p_k\right)-m_i^2 m_j^2 m_k^2\right)\nonumber\\
   &\hspace{10mm}+\left(p_k\cdot q_i\right) \left(p_k\cdot
   q_j\right) \left(m_i^2 m_j^2-2(p_i\cdot
   p_j)^{2}\right)-2 m_k^2
   \left(p_i\cdot p_j\right) \left(p_i\cdot
   q_j\right) \left(p_j\cdot
   q_i\right)\nonumber\\
   &\hspace{10mm}+\left(p_j\cdot q_i\right)
   \left(p_k\cdot q_j\right) \left(4
   \left(p_i\cdot p_j\right) \left(p_i\cdot
   p_k\right)-2 m_i^2
   \left(p_j\cdot p_k\right)\right)\nonumber\\
   &\hspace{10mm}+Q_iQ_j\biggr(\frac{1}{2} m_k^2 \left(p_i\cdot q_j\right)
   \left(p_j\cdot q_i\right)-\left(q_i\cdot
   q_j\right) \left(\frac{1}{2} m_k^2 \left(p_i\cdot
   p_j\right)-\left(p_i\cdot p_k\right)
   \left(p_j\cdot p_k\right)\right)\nonumber\\
   &\hspace{25mm}-2
   \left(p_j\cdot p_k\right) \left(p_i\cdot
   q_j\right) \left(p_k\cdot
   q_i\right)+\left(p_i\cdot p_j\right)
   \left(p_k\cdot q_i\right) \left(p_k\cdot
   q_j\right)\biggr)\biggr],\nonumber\\
    V^{(3)}_{\text{V},ijk} &=  \frac{1}{8E_i E_j E_k} \frac{64\pi^2 G^2}{q_i^2 q_k^2}\biggr[8 m_i^2
  (p_j\cdot p_k)^{2}-16
   \left(p_i\cdot p_j\right)
   \left(p_i\cdot p_k\right)
   \left(p_j\cdot p_k\right)\nonumber\\
   &\hspace{35mm}-Q_i Q_j^2 Q_k
   \left(p_i\cdot p_k\right)+8 Q_i Q_j
   \left(p_i\cdot p_k\right)
   \left(p_j\cdot p_k\right)\biggr].
\end{align}
The 4-vector momentum transfer in these expressions should be understood as a compact shorthand for
\begin{equation}    
    \label{q4vec}
    q_l^\mu \equiv \left(\frac{\vec{p}_l \cdot \vec{q}_l}{E_l}, \vec{q}_l\right).
\end{equation}
With this definition the expressions (\ref{YVpotential}) are manifestly classical or $\mathcal{O}\left(|\vec{q}|^{-4}\right)$. The \textit{anti-matter} contribution corresponds to the negative-energy part of the matter propagator in the full theory U-graph, while the \textit{subtraction} contributions correspond to the difference between the positive-energy part of the U-graph matter propagator and the corresponding iteration graph. These have a more complicated form
\begin{align}
    \label{amsub}
    V^{(3)}_{\text{anti-matter},ijk} &=  -\frac{1}{2E_j} \frac{\mathcal{N}^{(L)}_{ijk}\mathcal{N}^{(R)}_{ijk}}{q_{ijk}^{(L)} q_{ijk}^{(R)}} \nonumber\\
    V^{(3)}_{\text{subtraction},ijk}
     &= -\frac{\mathcal{N}_{ijk}^{(L)}\delta \mathcal{N}_{ijk}^{(R)} +\mathcal{N}_{ijk}^{(R)}\delta \mathcal{N}_{ijk}^{(L)}}{\hat{q}_{ijk}^{(L)} \hat{q}_{ijk}^{(R)}} + \frac{\delta q_{ijk}^{(L)} \mathcal{N}_{ijk}^{(L)} \mathcal{N}_{ijk}^{(R)}}{\hat{q}_{ijk}^{(L)} q_{ijk}^{(L)} \hat{q}_{ijk}^{(R)}} + \frac{\delta q_{ijk}^{(R)} \mathcal{N}_{ijk}^{(L)} \mathcal{N}_{ijk}^{(R)}}{\hat{q}_{ijk}^{(L)} \hat{q}_{ijk}^{(R)}q_{ijk}^{(R)}}\nonumber\\
     &\hspace{10mm}+ \frac{\Delta E_{ijk} \delta q_{ijk}^{(L)} \delta q_{ijk}^{(R)} \mathcal{N}_{ijk}^{(L)} \mathcal{N}_{ijk}^{(R)}}{\hat{q}_{ijk}^{(L)} q_{ijk}^{(L)} \hat{q}_{ijk}^{(R)}q_{ijk}^{(R)}},
\end{align}
where these expressions depend on a set of \textit{universal} or theory independent functions
\begingroup
\allowdisplaybreaks
\begin{align}
    \label{universal}
    q_{ijk}^{(L)} &= \tau_{ii}^2-|\vec{q}_i|^{2}+\frac{\tau_{ii}^3-|\vec{q}_i|^{2} \tau_{ii}}{E_i}\nonumber\\
    q_{ijk}^{(R)} &= \tau_{kk}^2-|\vec{q}_k|^2 + \frac{\tau_{kk}^3-|\vec{q}_k|^2 \tau_{kk}}{E_k}\nonumber\\
    \hat{q}_{ijk}^{(L)} &= \frac{(E_j \tau_{ji}+E_i \tau_{ii})^2}{(E_i+E_j)^2}-|\vec{q}_i|^2\nonumber\\
    \hat{q}_{ijk}^{(R)} &= \frac{(E_j \tau_{jk}+E_k \tau_{kk})^2}{(E_j+E_k)^2}-|\vec{q}_k|^2 \nonumber\\
    &\hspace{5mm}+\frac{2(E_j \tau_{jk}+E_k \tau_{kk}) \left((E_j+E_k) \left(\vec{q}_i\cdot \vec{q}_k\right)-(\tau_{ii}-\tau_{jk}+\tau_{kk})(E_j \tau_{jk}+E_k \tau_{kk})\right)}{(E_j+E_k)^3}\nonumber\\
    \delta q_{ijk}^{(L)} &=-\frac{E_j(E_j(\tau_{ji}+\tau_{ii})+2 E_i \tau_{ii})}{(E_i+E_j)^2} \nonumber\\
    &\hspace{5mm}-\frac{-E_j |\vec{q}_i|^2(E_i+E_j)+\tau_{ii}^2
   \left(2 E_i E_j+2 E_i^2+E_j^2\right)+2 E_i \tau_{ii}(E_i+E_j) \tau_{ji}+E_i E_j
   \tau_{ji}^2}{2 E_i(E_i+E_j)^2} \nonumber\\
    \delta q_{ijk}^{(R)} &= \frac{E_j(E_j(\tau_{jk}+\tau_{kk})+2 E_k \tau_{kk})}{(E_j+E_k)^2}\nonumber\\
    &\hspace{5mm}+\frac{1}{2 E_k(E_j+E_k)^3}\left[E_j E_k \left(2(E_j+E_k)
   \left(\vec{q}_i\cdot \vec{q}_k\right)+\tau_{jk}(2 \tau_{ii}
  (E_k-E_j)+(3 E_j-E_k) \tau_{jk})\right)\right.\nonumber\\
   &\hspace{35mm}\left.-E_j |\vec{q}_k|^2(E_j+E_k)^2+2
   E_k^2 \tau_{kk}(2 E_k \tau_{ii}+(3 E_j-E_k)
   \tau_{jk})\right.\nonumber\\
   &\hspace{35mm}\left.+\tau_{kk}^2 \left(3 E_j^2 E_k+4 E_j
   E_k^2+E_j^3+6 E_k^3\right)\right]\nonumber\\
    \Delta E_{ijk} &= \frac{1}{2}(-\tau_{ji}+\tau_{ii}+\tau_{jk}-\tau_{kk}) \nonumber\\
    &\hspace{5mm}+ \frac{1}{4
   E_i E_j E_k}\left[E_k \left(E_i \left(2 \left(\vec{q}_i\cdot \vec{q}_k\right)-2 \tau_{ii} \tau_{jk}+\tau_{ji}^2+\tau_{jk}^2\right)-|\vec{q}_i|^2(E_i+E_j)+E_j \tau_{ii}^2\right)\right.\nonumber\\
   &\hspace{28mm}\left.+E_i |\vec{q}_k|^2(E_j+E_k)-E_i E_j \tau_{kk}^2-2 E_i E_k \tau_{kk} \tau_{jk}\right],
\end{align}
as well the important \textit{non-universal} or theory dependent functions
\begin{align}
    \label{nonuniversal}
    \mathcal{N}_{ijk}^{(L)} &= \frac{4 \pi  G}{E_i E_j} \left[m_i^2 m_j^2-Q_i  Q_j E_i E_j(\rho_{ij}-1)-2 E_i^2 E_j^2
  (\rho_{ij}-1)^2\right]\nonumber\\
   &\hspace{5mm}+\frac{2 \pi  G} {E_i^2 E_j^2}\left[m_i^2 m_j^2(E_j \tau_{ii}-E_i \tau_{ji})+Q_i Q_j E_i 
   E_j(E_i(\rho_{ij}-1) \tau_{ji}+E_j(\tau_{ji}-\tau
  _{ii} \rho_{ij}))\right.\nonumber\\
  &\hspace{20mm}\left.-2 E_i^2 E_j^2(\rho_{ij}-1)(E_j
  (\tau_{ii}(\rho_{ij}+1)-2 \tau_{ji})-E_i(\rho_{ij}-1) \tau_{ji})\right]\nonumber\\
    \mathcal{N}_{ijk}^{(R)} &= \frac{4\pi G}{E_j E_k}\left[m_j^2 m_k^2-Q_j Q_k E_j  E_k(\rho_{jk}-1)-2 E_j^2 E_k^2
  (\rho_{jk}-1)^2\right]\nonumber\\
   &\hspace{5mm}+\frac{2 \pi  G} {E_j^2 E_k^2}\left[\frac{1}{2} m_j^2 m_k^2(E_k(-2 \tau_{ii}+\tau_{jk}-2 \tau_{kk})+E
  _j \tau_{kk})\right.\nonumber\\
  &\hspace{20mm}\left.+\frac{1}{2} Q_j Q_kE_j  E_k(E_k(\rho_{jk}(2 \tau_{ii}-\tau_{jk}+2 \tau_{kk})-2(\tau_{ki}+\tau_{kk})+\tau
  _{jk})+E_j(\tau_{jk}-\tau_{kk} \rho_{jk}))\right.\nonumber\\
  &\hspace{20mm}\left.+E_j^2 E
  _k^2(\rho_{jk}-1)(E_k(2 \tau_{ii}(\rho_{jk}+1)-4 \tau
  _{ki}-\rho_{jk} \tau_{jk}+2 \tau_{kk}(\rho_{jk}-1)+\tau_{jk})\right.\nonumber\\
  &\hspace{50mm}\left.+E_j
  (2 \tau_{jk}-\tau_{kk}(\rho_{jk}+1)))\right]\nonumber\\
    \delta \mathcal{N}_{ijk}^{(L)} &= \frac{2 \pi  G}{E_iE_j} E_j\left[4 E_iE_j(\rho_{ij}-1)+Q_i Q_j\right]\nonumber\\
    \delta \mathcal{N}_{ijk}^{(R)} &= -\frac{2 \pi  G}{E_j E_k}(E_j+2 E_k)\left[4 E_j E_k(\rho_{jk}-1)+Q_j Q_k\right].
\end{align}
\endgroup
The notation used in these expressions is defined in (\ref{sigmatau}). For the sake of notational brevity the expressions (\ref{amsub}) are written in a form that is not manifestly classical, but does manifest the cancellation of the spurious matter singularity. As discussed in detail in Appendix \ref{app:3bdSubtraction}, the symmetry properties of (\ref{universal}) and (\ref{nonuniversal}) ensure the cancellation of super-classical or $\mathcal{O}\left(|\vec{q}|^{-5}\right)$ terms after summing over $(i,j,k)\in S_3$. Quantum or $\mathcal{O}\left(|\vec{q}|^{-3}\right)$ terms in (\ref{amsub}) are likewise only included for brevity and should be discarded. 

Note that, as discussed in Section \ref{subsec:gauge}, we have fixed the gauge ambiguity in the the 3-body potential (\ref{3bdypotential}) in a convenient but ultimately arbitrary way. This ambiguity does not affect physical, gauge invariant observables calculated from the potential. It is unknown to us if there is a unique ``simplest" way to resolve this ambiguity for $N>2$ body dynamics, analogous to generalized isotropic gauge for the 2-body potential.  

In the expression (\ref{3bdypotential}) we have chosen to emphasize the explicit separation of matter and anti-matter contributions and the cancellation of super-classical contributions between diagrams. Alternatively we can derive compact forms of the potential; in general relativity ($Q_1=Q_2=Q_3=0$) for example
\begin{align}
\label{3bodypotentialalt}
\hat{V}_{ij}^{(2)} &= { 4 \pi G m_i^2 m_j^2 (2 \sigma_{ij}^2-1)  \over E_i E_j (\omega_{ij}^2 -  |\vec{q}_i|^2)  }  \,, \nonumber \\ 
\hat{V}_{ijk}^{(3)} 
&=  {32\pi^2 G^2 m_i^2 m_j^2 m_k^2 \over E_i E_j E_k q_i^2 q_j^2 q_k^2}  \left[   {q_i^2 (1-4 \sigma_{ij} \sigma_{jk} \sigma_{ik}  ) \over 2} +   \frac{(p_k \cdot q_i)^2 (2\sigma_{ij}^2-1)}{m_k^2}  - \frac{4  (p_j \cdot q_i) (p_k \cdot q_i) \sigma_{ij} \sigma_{ik} }{m_j m_k}  \right]   \nonumber \\
&\hspace{5mm} - \hat{V}_{ij}^{(2)} \hat{V}_{kj}^{(2)} \left[ {1 \over 2 E_j} - { 2 \omega_{kj}^2 \over (\omega_{kj}^2 - |\vec{q}_k|^2 ) E_{jk} } + {E_j q_i \cdot q_k (\omega_{ii} + \omega_{ij} )( \omega_{kk} + \omega_{kj}) \over 2q_i^2 q_k^2 E_{ij} E_{jk} } \right. \nonumber \\
&\hspace{27mm} \left. + \frac{E_j}{E_{jk}^2(\omega_{kj}^2 - |\vec{q}_k|^2 )}\left(2\omega_{ii} \omega_{kj}+q_i\cdot q_k \left(1-\frac{(\omega_{kk}+\omega_{kj})^2}{q_k^2}\right)\right)\right]\nonumber\\
&\hspace{5mm} + 16\pi G m_j m_k \hat{V}_{ij}^{(2)} \sigma_{jk} \left[ {1 \over E_j (\omega_{kj}^2 - |\vec{q}_k|^2 )}  + {p_k \cdot q_i \over  E_k q_k^2 } \bigg( {\omega_{ii} + \omega_{ij} \over q_i^2 E_{ij} } -{ \omega_{kk} + \omega_{kj} \over (\omega_{kj}^2 - |\vec{q}_k|^2)E_{jk}} \bigg) \right],
\end{align}
where 
\begin{equation}
    \sigma_{ij} = \frac{p_i\cdot p_j}{m_i m_j} ,\hspace{10mm} \omega_{ij} = \frac{(\vec{p}_i+\vec{p}_j)\cdot \vec{q}_i}{E_i + E_j}, \hspace{10mm} E_{ij} = E_i + E_j \,,
\end{equation}
and as before this should be understood as a function of 3-momenta with the 4-momentum transfer defined as in (\ref{q4vec}). The first line in (\ref{3bodypotentialalt}) is proportional to the sum of full theory amplitudes $\mathcal{M}^{(3)}_{\text{Y},ijk}  +   \mathcal{M}^{(3)}_{\text{V},ijk} $, while the remaining terms are the subtraction contributions. The result $\hat{V}_{ijk}^{(3)}$ in (\ref{3bodypotentialalt}) is pole-free, manifestly classical, and is physically equivalent to $V_{ijk}^{(3)}$ in (\ref{3bdypotential}) for the case of general relativity. Similar results can be derived in Einstein-Maxwell.

Finally, the potential (\ref{3bdypotential}) is expressed in momentum space. To simulate the real-time dynamics of 3-bodies we need to first Fourier transform to position space. As discussed in Section \ref{sec:NbdyFT}, the required 3-body PM integrals are unknown in closed form. Expanding order-by-order in small $\vec{p}$, the integrals needed for (\ref{3bdypotential}) can be evaluated using either differential equations methods \cite{Loebbert:2020aos}, or by the hierarchical expansion using the method of regions as described in Section \ref{sec:NbdyFT}. 

\section{Comparison with Known Results}
\label{sec:comparison}

Given the complicated form of the explicit expression (\ref{3bdypotential}) for the 3-body potential, it is useful to verify the calculation against known results and alternative methods of calculation. In this section we describe three non-overlapping points of comparison: \textit{state-of-the-art post-Newtonian results in general relativity}, \textit{probe (test mass) dynamics on multi-center black hole backgrounds} and \textit{slowly moving extremal black hole dynamics via the moduli space approximation}. In each case we find exact agreement to the calculated order. 

\subsection{Post-Newtonian Results in General Relativity}
\label{sec:PN}

The momentum space PM expression (\ref{3bdypotential}) contains $\mathcal{O}(G^2)$ 3-body interactions at all orders in velocity (momentum). To verify that our formula is correct (in some choice of coordinates) we compare it to state-of-the art PN potentials, known up to $\mathcal{O}(G^2 p^4)$ \cite{Loebbert:2020aos}. Since the potential is gauge dependent, we do not expect it to coincide exactly with expressions in the literature. To relate potentials in different gauges we can either construct an explicit change of coordinates \cite{Bern:2019crd}, or use both potentials to calculate the same gauge invariant observable. In this section we will take the latter approach and calculate gauge invariant scattering amplitudes up to classical order. 

The leading 1PN 3-body corrections were calculated long-ago by Einstein, Infeld and Hoffmann \cite{Einstein:1938yz}. These consist of 3-body interactions at $\mathcal{O}\left(G^2 p^0\right)$ and 2-body interactions at $\mathcal{O}\left(G p^2\right)$. In momentum space the 1PN potential, in the gauge used in \cite{Landau:1975pou}, is given explicitly by
\begin{align}  
\label{1PN}
    V_{\text{1PN}} &= \sum_{(i,j,k)\in S_3} \left[\delta^{(3)}\left(\vec{q}_i+\vec{q}_j\right)\delta^{(3)}\left(\vec{q}_k\right)\times\frac{4\pi G m_i m_j}{|\vec{q}_i|^2}\left( 1+\frac{3|\vec{p}_i|^2}{m_i^2}+\frac{\vec{p}_i\cdot \vec{p}_j}{m_i m_j}-\frac{(\vec{p}_i\cdot \vec{q}_i)(\vec{p}_j\cdot \vec{q}_i)}{m_i m_j|\vec{q}_i|^2}\right)\right.\nonumber\\
    &\hspace{20mm}\left.+\delta^{(3)}\left(\vec{q}_i+\vec{q}_j+\vec{q}_k\right)\times\frac{16\pi^2 G^2 m_i m_j m_k}{|\vec{q}_i|^2|\vec{q}_j|^2}\right].
\end{align}
There is also a 2-body $\mathcal{O}\left(G^2 p^0\right)$ contribution at 1PN order, but since this contributes to the 3-body scattering amplitude beginning at $\mathcal{O}\left(G^3\right)$ we can ignore it for this calculation. Clearly the 2-body part of this potential is not calculated in the generalized isotropic gauge defined in Section \ref{subsec:gauge}. Since a general gauge transformation will mix the 2- and 3-body parts of the potential, we do not expect the 3-body parts of (\ref{1PN}) to agree exactly with (\ref{3bdypotential}) either. To show that these expressions are physically equivalent we use (\ref{1PN}) together with the Born series (\ref{Born}) to calculate a classical scattering amplitude, and compare this with the expected full-theory result (\ref{3bdypotential}). Since our 2PM potential produces a scattering amplitude in agreement with (\ref{3bdyEMamplitude}) by construction, we conclude that the Einstein-Infeld-Hoffmann potential is related to (\ref{3bdypotential}) by a coordinate transformation. 

The 2PN 3-body corrections were calculated in a series of papers \cite{Ohta:1974pq,SCHAFER1987336}. We will compare with the explicit form given in \cite{Loebbert:2020aos}; transforming the 3-body potential at $\mathcal{O}\left(G^2 p^2\right)$ and 2-body potential at $\mathcal{O}\left(G p^4\right)$ given in that paper to momentum space we find\footnote{This expression is produced after correcting a small typo in the published version of \cite{Loebbert:2020aos}. In the final line of equation (52) the numerator should read $9(\mathbf{n}_{ij}\cdot \mathbf{v}_{ij})^2-9\mathbf{v}_{ij}^2\color{red}-\color{black}2(\mathbf{n}_{ij}\cdot \mathbf{v}_{ik})^2\color{red}+\color{black}2\mathbf{v}_{ik}^2$. We are very grateful to Florian Loebbert, Jan Plefka, Canxin Shi and Tianheng Wang for communicating with us and confirming this.}  
\begingroup
\allowdisplaybreaks
\begin{align}  
\label{2PN}
    &V_{\text{2PN}} =\sum_{(i,j,k)\in S_3} \Biggr[\delta^{(3)}\left(\vec{q}_i+\vec{q}_j\right)\delta^{(3)}\left(\vec{q}_k\right) \times \frac{4\pi G m_i m_j}{|\vec{q}_i|^2}\nonumber\\
    &\hspace{18mm}\times\biggr(-\frac{(\vec{p}_i\cdot \vec{p}_j)^2}{m_i^2 m_j^2}-\frac{|\vec{p}_i|^2 |\vec{p}_j|^2}{8 m_i^2 m_j^2}+\frac{5 |\vec{p}_i|^2}{16
   m_i^4}+\frac{5 |\vec{p}_j|^2}{16 m_j^4}-\frac{|\vec{p}_i|^2(\vec{p}_i\cdot \vec{q}_i)(\vec{p}_j\cdot \vec{q}_i)}{2 m_i^3 m_j |\vec{q}_i|^2}\nonumber\\
   &\hspace{25mm}+\frac{2
  (\vec{p}_i\cdot \vec{p}_j)(\vec{p}_i\cdot \vec{q}_i) \left(\vec{p}_j\cdot \vec{q}_i\right)}{m_i^2 m_j^2 |\vec{q}_i|^2}-\frac{|\vec{p}_j|^2(\vec{p}_i\cdot \vec{q}_i) \left(\vec{p}_j\cdot \vec{q}_i\right)}{2 m_i m_j^3 |\vec{q}_i|^2}-\frac{(\vec{p}_i\cdot \vec{q}_i)^{2} (\vec{p}_j\cdot \vec{q}_i)^{2}}{2 m_i^2 m_j^2 |\vec{q}_i|^4}\biggr) \nonumber\\
  &\hspace{15mm}+\delta^{(3)}\left(\vec{q}_i+\vec{q}_j+\vec{q}_k\right)\times \frac{16\pi^2 G^2 m_i m_j m_k}{|\vec{q}_i|^2|\vec{q}_j|^2}\nonumber\\
  &\hspace{20mm}\times \biggr(\frac{4 \vec{p}_i\cdot \vec{p}_j}{m_i m_j}-\frac{8 \vec{p}_i\cdot \vec{p}_k}{m_i m_k}+\frac{3
   |\vec{p}_i|^2}{2 m_i^2}+\frac{9 |\vec{p}_k|^2}{4 m_k^2}+\frac{(\vec{p}_i\cdot\vec{q}_i)^2}{m_i^2 |\vec{q}_i|^2}-\frac{3 (\vec{p}_i\cdot\vec{q}_i)
   (\vec{p}_k\cdot\vec{q}_j)}{m_i m_k |\vec{q}_i|^2}\nonumber\\
   &\hspace{25mm}+\frac{4 (\vec{p}_i\cdot\vec{q}_i)
   (\vec{p}_i\cdot\vec{q}_j)}{m_i^2 |\vec{q}_i|^2}-\frac{4 (\vec{p}_i\cdot\vec{q}_i)
   (\vec{p}_j\cdot\vec{q}_i)}{m_i m_j |\vec{q}_i|^2}+\frac{3 (\vec{p}_i\cdot\vec{q}_i)
   (\vec{p}_k\cdot\vec{q}_i)}{m_i m_k |\vec{q}_i|^2}\nonumber\\
   &\hspace{25mm}+\frac{(\vec{q}_i\cdot \vec{q}_j) (\vec{p}_i\cdot \vec{q}_i) (\vec{p}_j\cdot \vec{q}_j)}{2
   m_i m_j |\vec{q}_i|^2 |\vec{q}_j|^2}-\frac{(\vec{q}_i\cdot \vec{q}_j) (\vec{p}_i\cdot \vec{q}_i)^2}{m_i^2
   |\vec{q}_i|^4}\biggr)\nonumber\\
   &\hspace{15mm}+\delta^{(3)}\left(\vec{q}_i+\vec{q}_j+\vec{q}_k\right)\times \frac{32\pi^2 G^2 m_i m_j m_k}{|\vec{q}_i|^2|\vec{q}_j|^2 |\vec{q}_k|^2}\nonumber\\
   &\hspace{20mm}\times \biggr(\frac{ (\vec{p}_i\cdot\vec{q}_j) (\vec{p}_i\cdot\vec{q}_k)}{m_i^2}-\frac{2 \left(\vec{p}_j\cdot
   \vec{q}_j\right) (\vec{p}_i\cdot\vec{q}_k)}{m_i m_j}-\frac{8 (\vec{p}_i\cdot\vec{q}_i)
   (\vec{p}_i\cdot\vec{q}_j)}{m_i^2}\nonumber\\
   &\hspace{25mm}+\frac{4 (\vec{p}_j\cdot\vec{q}_i) \left(\vec{p}_i\cdot
   \vec{q}_j\right)}{m_i m_j}+\frac{5 (\vec{p}_i\cdot\vec{q}_i) (\vec{p}_j\cdot\vec{q}_j)}{m_i m_j}\biggr)\Biggr].
\end{align}
\endgroup
Again, we are also ignoring a 2-body $\mathcal{O}\left(G^2 p^2\right)$ contribution that does not mix with the 3-body interaction at this PN order. Calculating a gauge invariant scattering amplitude using this expression we find complete agreement with our 2PM expression (\ref{3bdypotential}). 

Finally, the current state-of-the art is the explicit 2-body $\mathcal{O}\left(G p^6\right)$ and 3-body $\mathcal{O}\left(G^2 p^4\right)$ contributions to the 3PN potential calculated in \cite{Loebbert:2020aos}. The expression for the position space Lagrangian is quite lengthy (around 5000 terms), but with some assistance from a computer algebra system can be converted to a momentum space Hamiltonian and subsequently a scattering amplitude. Once again we find perfect agreement with our 2PM result (\ref{3bdypotential}).

\subsection{Probe Scattering on a Majumdar-Papapetrou Background}
\label{sec:probe}

A complementary perturbative approach to gravitational $N$-body dynamics is to consider an expansion in a small \textit{mass ratio}, but keeping terms to all orders in $G$ and $v$. 

Let's begin by reviewing the setup in the 2-body case. We designate one of the bodies (the background) with mass $m_1$ and the other (the probe) with mass $m_2$ and $m_2/m_1 \ll 1$. To leading order in this small dimensionless parameter, and in the absence of spin and finite-size effects, the dynamics are captured by a probe particle of mass $m_2$ in geodesic motion on a background Schwarzschild metric of mass $m_1$. Explicitly, it is convenient to work with the isotropic gauge form of the metric, in $(+,-,-,-)$ signature, with a COM at $\vec{x}_1$
\begin{equation}
    ds^2 = g_{tt}\left(\vec{x}\right)dt^2 + g_{rr}\left(\vec{x}\right)d\vec{x}\cdot d\vec{x},
\end{equation}
where
\begin{equation}
    g_{tt}\left(\vec{x}\right) = \frac{1-\frac{G m_1}{2|\vec{x}-\vec{x}_1|}}{1+\frac{G m_1}{2|\vec{x}-\vec{x}_1|}}, \hspace{10mm} g_{rr}\left(\vec{x}\right) = -1-\frac{G m_1}{2|\vec{x}-\vec{x}_1|}.
\end{equation}
The motion of a point particle of mass $m_2$ on this background is given by extremizing the worldline action
\begin{equation}
    S_{\text{probe}} = -m_2 \int \text{d}s_2,
\end{equation}
where $s_2$ is the proper time measured by the probe. Fixing the affine parameter on the worldline to coincide with the time measured by an asymptotic observer, this defines a probe Hamiltonian 
\begin{equation}
    H^{\text{2-body}}_{\text{probe}} = \left[\left(\frac{1-\frac{G m_1}{2r_{12}}}{1+\frac{Gm_1}{2r_{12}}}\right)\left(m_2^2+\frac{|\vec{p}_2|^2}{1+\frac{Gm_1}{2r_{12}}}\right)\right]^{1/2}.
\end{equation}
This Hamiltonian describes a relativistic particle of mass $m_2$ interacting with a fixed potential. Order-by-order in $G$ we calculate a classical scattering amplitude using the Born series (\ref{Born}), and this should agree with the two-body amplitude calculated using the full PM Hamiltonian, to leading order in the small mass ratio. Up to 2PM (1-loop) for 2-bodies, the probe limit completely fixes the Hamiltonian for \textit{arbitrary} mass ratio \cite{Cheung:2020gbf}. 

For the 3-body case there are more possibilities. The simplest is to again consider body 1 as a background, and treat bodies 2 and 3 as probes. This means we are considering a mass hierarchy $m_1 \gg m_2 \sim m_3$. The calculation is identical to the above and leads to a Hamiltonian 
\begin{align}
    &H^{\text{3-body}}_{\text{probe}}= \left[\left(\frac{1-\frac{G m_1}{2r_{12}}}{1+\frac{Gm_1}{2r_{12}}}\right)\left(m_2^2+\frac{|\vec{p}_2|^2}{1+\frac{Gm_1}{2r_{12}}}\right)\right]^{1/2} + \left[\left(\frac{1-\frac{G m_1}{2r_{13}}}{1+\frac{Gm_1}{2r_{13}}}\right)\left(m_3^2+\frac{|\vec{p}_3|^2}{1+\frac{Gm_1}{2r_{13}}}\right)\right]^{1/2}.
\end{align}
While we can certainly use this to calculate a 3-body classical scattering amplitude and compare with the 2PM result (\ref{3bdypotential}), such a comparison is not interesting. There are no terms that depend on the coordinates of all 3 bodies, \textit{i.e.} there are no genuine 3-body interactions. This is not surprising, treating two of the bodies as probes means that they do not interact with each other. It would be very interesting to calculate a physical observable, \textit{e.g.} a classical impulse, to next-to-leading order in this mass ratio using the self-force formalism \cite{Barack:2018yvs}. We will not pursue this here.

For a non-trivial probe calculation we need to consider a different hierarchy of the form $m_1 \sim m_2 \gg m_3$. This means we treat body 3 as a probe on a \textit{multi-center} black hole background of bodies 1 and 2. Unfortunately, in pure GR such solutions are not known analytically.\footnote{There do exist formal solutions of the vacuum Einstein equations describing 2 or more collinear Schwarzschild black holes in static equillibrium \cite{1964NCim...33..331I}. In these solutions the gravitational attraction is balanced by the (negative) tension of a string-like \textit{strut} between the black holes at which the metric has a naked conical singularity \cite{Costa:2000kf}. Due to the pathological nature of this solution, it is not clear if a probe scattering on this background can be related to some limit of the general 3-body potential.} In Einstein-Maxwell theory however, such solutions are possible and are given by the well-known \textit{Majumdar-Papapetrou (MP)} solution \cite{Majumdar:1947eu,Papaetrou:1947ib}
\begin{align}
    \label{MPsol}
    ds^2 = f(\vec{x})^{-2} dt^2 - f(\vec{x})^2 d\vec{x}\cdot d\vec{x},\hspace{10mm} A = \left(1-f(\vec{x})^{-1}\right)dt,
\end{align}
where
\begin{equation}
    f(\vec{x}) = 1+\frac{G m_1}{|\vec{x}-\vec{x}_1|}+\frac{G m_2}{|\vec{x}-\vec{x}_2|}.
\end{equation}
This solution describes two (or more generally $N$) extremal, $Q_i=m_i$, black holes in exact static equilibrium at positions $\vec{x}_1$ and $\vec{x}_2$. We also allow the probe to have a charge $Q_3$ which is assumed to be small compared to $m_1$ and $m_2$ in Planck units. This modifies the point particle effective action as
\begin{equation}
    S_{\text{charged probe}} = -m_3 \int \text{d}s_3 + Q_3\int A.
\end{equation}
The result is a Hamiltonian for a charged probe on an MP background
\begin{align}
    \label{probe}
    H_{\text{MP}} &= \left(1+\frac{Gm_1}{r_{13}}+\frac{Gm_2}{r_{23}}\right)^{-2}\left[|\vec{p}_3|^2 +m_3^2\left(1+\frac{Gm_1}{r_{13}}+\frac{Gm_2}{r_{23}}\right)^2\right]^{1/2}\nonumber\\
    &\hspace{7mm}+Q_3\left[1- \left(1+\frac{Gm_1}{r_{13}}+\frac{Gm_2}{r_{23}}\right)^{-1}\right].
\end{align}
Expanding to 2PM order and Fourier transforming to momentum space, we find the potential
\begin{align}
    V_{\text{MP}} &=  \delta^{(3)}\left(\vec{q}_1+\vec{q}_3\right)\delta^{(3)}\left(\vec{q}_2\right)\left[\frac{4\pi G  m_1 \left(Q_3 E_3 -2|\vec{p}_3|^2-m_3^2\right)}{E_3|\vec{q}_1|^2 }\right]\nonumber\\
    &\hspace{3mm}+ \delta^{(3)}\left(\vec{q}_2+\vec{q}_3\right)\delta^{(3)}\left(\vec{q}_1\right)\left[\frac{4\pi G  m_2 \left(Q_3 E_3 -2|\vec{p}_3|^2-m_3^2\right)}{E_3|\vec{q}_2|^2 }\right]\nonumber\\
    &\hspace{3mm}+\delta^{(3)}\left(\vec{q}_1+\vec{q}_2+\vec{q}_3\right)\left[\frac{16 \pi ^2 G^2 m_1 m_2 \left(6 E_3^4-2 E_3^3 Q_3-3 E_3^2 m_3^2-m_3^4\right)}{E_3^3 |\vec{q}_1|^2 |\vec{q}_2|^2}\right].
\end{align}
From this potential we calculate a scattering amplitude using the Born series (\ref{Born}). Beginning with the 2PM Hamiltonian (\ref{3bdypotential}), if we take the $Q_i\rightarrow m_i$ limit for $i=1,2$, expand to leading order in the hierarchy $m_1\sim m_2 \gg m_3 \sim Q_3$, and then calculate a scattering amplitude, we find exact agreement with the probe calculation at $\mathcal{O}\left(G^2 p_3^\infty\right)$. 

Interestingly, if we calculate the 2PM amplitude (\ref{3bdyEMamplitude}) for generic $m_i$ and $Q_i$, and \textit{then} take the limits above we do not find agreement. We will discuss this non-commuting order-of-limits in the next subsection.

\subsection{Extremal Black Hole Scattering in the Moduli Space Approximation}
\label{sec:MSA}

Long ago, Manton \cite{Manton:1981mp} proposed an interesting approach for describing the scattering of slowly-moving BPS solitons. Consider a field theory model with \textit{static} $N$-soliton solutions $\phi\left(x^\mu,\{X\}\right)$, where $\{X\}$ denotes a set of collective coordinates or moduli including the spatial positions of the solitons $\{\vec{x}_i\}$. Manton proposed that up to $\mathcal{O}\left(v^2\right)$, but non-perturbatively in all other parameters, the dynamics of the $N$-soliton system is captured by an effective $0+1$-dimensional non-linear sigma model on the moduli space with target space metric given by the natural moduli space metric calculated from the overlap of field theory zero modes~\cite{Manton:1981mp}. This effective model can then be used to calculate physical observables for slowly-moving dynamical solitons, an approach sometimes called the \textit{moduli space approximation (MSA)}. In the original context the solitons considered were BPS magnetic monopole solutions of a Yang-Mills-Higgs model \cite{Bogomolny:1975de,Prasad:1975kr}. By an indirect argument, Atiyah and Hitchin obtained the 2-monopole moduli space metric as an exact expression in $g_{\text{YM}}$ and used it to analyze monopole-monopole scattering at small-velocity but arbitrary impact parameter \cite{Atiyah1985LowES}. This approach has since been generalized to a variety of physical systems \cite{Ruback:1987sg, Gutowski:1998bn}, including, importantly for this paper, extremal black holes in an Einstein-Maxwell-Dilaton model \cite{Ferrell:1987gf,Shiraishi:1992nq}. 

Following Manton's proposal, Ferrell and Eardley were able to calculate the exact moduli space metric for $N$ extremal Reissner-Nordstr\"om black holes in Einstein-Maxwell theory by perturbing around the Majumdar-Papapetrou solution (\ref{MPsol}) \cite{Ferrell:1987gf}. The resulting effective action describes the interactions of $N$ extremal black holes at $\mathcal{O}\left(G^\infty v^2\right)$\footnote{Given that this action is exact in $G$ it is surprising that it contains no terms at higher-order than $G^3$. This is a special feature of extremal black hole solutions that can be uplifted to intersecting BPS brane solutions of $11d$ supergravity \cite{Gibbons:1997iy}. Generic extremal black hole solutions of the Einstein-Maxwell-Dilaton model have an MSA effective action with terms at all orders in $G$ \cite{Shiraishi:1992nq}.}, and is given by
\begin{align}  
\label{MSA}
    S_{\text{MSA}} &= \int \text{d}t \;\Biggr[\sum_{i=1}^N \frac{1}{2}m_i |\vec{v}_i|^2 +\frac{3}{8\pi} \int \text{d}^3\vec{x}_0 \left(1+\sum_{i=1}^N \frac{Gm_i}{x_{0i}}\right)^2\sum_{j,k=1}^N \frac{G m_j m_k}{x_{0j}x_{0k}}\nonumber\\
    &\hspace{39mm}\times \left[\frac{1}{2}\left(\vec{x}_{0j}\cdot \vec{x}_{0k}\right)|\vec{v}_j-\vec{v}_k|^2-\left(\vec{x}_{0j} \times \vec{x}_{0k}\right)\cdot \left(\vec{v}_j\times \vec{v}_k\right)\right]\Biggr],
\end{align}
where $x_{0i}=|\vec{x}_0-\vec{x}_i|$. From the effective action we calculate the momentum space potential up to $\mathcal{O}\left(G^2 p^2\right)$,
\begin{align}
    \label{MSAH}
    &V_{\text{MSA}} = \sum_{(i,j,k)\in S_3}\Biggr[\delta^{(3)}\left(\vec{q}_i+\vec{q}_j\right)\delta^{(3)}\left(\vec{q}_k\right)\left[-\frac{3\pi G m_i m_j}{|\vec{q}_i|^2}\left(\frac{\vec{p}_i}{m_i}-\frac{\vec{p}_j}{m_j}\right)^2\right] \nonumber\\
    &\hspace{20mm}+\delta^{(3)}\left(\vec{q}_i+\vec{q}_j+\vec{q}_k\right)\Biggr[\frac{72\pi^2 G^2 m_i m_j }{|\vec{q}_i|^2 |\vec{q}_j|^2}\left(\frac{m_k \left(\vec{p}_i\cdot \vec{p}_j\right)}{m_i
   m_j}-\frac{2 \left(\vec{p}_i\cdot
   \vec{p}_k\right)}{m_i}+\frac{|\vec{p}_k|^2}{m_k}\right)\nonumber\\
    &\hspace{55mm}+\frac{48 \pi^2 G^2 m_i m_j m_k}{|\vec{q}_i|^2|\vec{q}_j|^2|\vec{q}_k|^2}\left(\frac{\left(\vec{p}_i\cdot\vec{q}_j\right) \left(\vec{p}_j\cdot \vec{q}_i\right)}{m_i m_j}-\frac{\left(\vec{p}_i\cdot \vec{q}_i\right)
   \left(\vec{p}_j\cdot\vec{q}_j\right)}{m_i m_j}\right.\nonumber\\
   &\left.\hspace{55mm}+\frac{|\vec{p}_j|^2
   \left(\vec{q}_j\cdot \vec{q}_k\right)}{m_j^2}-\frac{\left(\vec{p}_j\cdot
   \vec{p}_k\right) \left(\vec{q}_j\cdot \vec{q}_k\right)}{m_j m_k}\right)\Biggr]\Biggr]+\mathcal{O}\left(G^3 p^2\right).
\end{align}
Using this potential we calculate a classical scattering amplitude for three extremal black holes up to $\mathcal{O}\left(G^2 p^2\right)$. We find that this agrees exactly with the scattering amplitude calculated using the 2PM Einstein-Maxwell Hamiltonian (\ref{3bdypotential}) truncated to this order, in the extremal limit $Q_i=m_i$. Since (\ref{MSA}) contains interactions to all orders in $G$, we can use it to calculate amplitudes and compare with $N$-body Hamiltonians obtained from scattering amplitudes at all loop orders. This may be a valuable cross-check for calculations at $\mathcal{O}\left(G^3\right)$ and beyond. 

As mentioned in Section \ref{sec:probe} the procedures of taking the large mass limit and calculating an amplitude from a Hamiltonian do not commute. To compare directly with the probe Hamiltonian (\ref{probe}) we need to take the large mass limit of the 2PM Hamiltonian \textit{first}, and then calculate a scattering amplitude treating only the probe as a dynamical degree of freedom. It is possible however to match the calculation taken in the opposite order, where the amplitude is calculated from the general 2PM Hamiltonian (\ref{3bdypotential}) first, and then the large mass limit is taken. In this case we need to treat the background black holes as dynamical but slowly moving, with interactions given by the MSA potential (\ref{MSA}). The correct effective Hamiltonian is found to be a hybrid of the probe potential (\ref{probe}) and the MSA potential (\ref{MSAH}), which at 2PM and in momentum space is given by
\begin{align}
\label{hybrid}
    H_{\text{hybrid}} &= H_{\text{probe}}+ \frac{|\vec{p}_1|^2}{2m_1}+ \frac{|\vec{p}_2|^2}{2m_2}+ \delta^{(3)}\left(\vec{q}_1+\vec{q}_2\right)\delta^{(3)}\left(\vec{q}_3\right)\left[-\frac{6\pi G m_1 m_2}{|\vec{q}_1|^2}\left(\frac{\vec{p}_1}{m_1}-\frac{\vec{p}_2}{m_2}\right)^2\right].
\end{align}
The mis-match between the order-of-limits arises because of the iteration graphs shown in Figure \ref{fig:EFTamps}. The matter propagator, at classical order, contains an $\mathcal{O}\left({|\vec{p}_{1,2}|\hspace{0.05mm}}^{-2}\right)$ kinematic singularity that cancels against the $\mathcal{O}\left({|\vec{p}_{1,2}|\hspace{0.05mm}}^{2}\right)$ MSA two-body interaction (\ref{hybrid}), giving a non-zero contribution in the $|\vec{p}_{1,2}|\rightarrow 0$ limit. It would be interesting to understand if a similar order-of-limits problem exists in physical observables such as the linear impulse on the probe.

\section{Discussion}
\label{sec:discussion}

The main results of this paper are the development of a general approach to calculating classical $N$-body potentials using scattering amplitudes, and the explicit calculation of the intrinsic 3-body potential (\ref{3bdypotential}) in Einstein-Maxwell at 2PM in momentum space. We have presented a robust set of checks against known results in various limiting cases and find complete agreement. We believe the amplitudes based methods developed in this paper will be a powerful tool for understanding general relativistic $N$-body dynamics. There are many important future directions that should be pursued to further develop this program.

The most obvious direction is to use these methods to calculate conservative potentials at higher powers in $G$. At the next order beyond the calculation presented in this paper, $\mathcal{O}\left(G^3\right)$ or 3PM, there are contributions to the general $N$-body potential from 2-body scattering at 2-loops, 3-body scattering at 1-loop and 4-body scattering at tree-level. The 3-body pieces will involve a more complicated iteration subtraction calculation involving EFT 1-loop diagrams with triangle topology as well as non-1PI diagrams. It will be important for such future calculations to develop a systematic approach to cancelling non-physical non-analytic features order-by-order in PM perturbation theory. The 4-body pieces at this order are tree-level and should be calculable using a generalization of the procedure described in this paper. In the PN expansion, the static $\mathcal{O}\left(G^3 v^0\right)$ 3-body potential is known \cite{SCHAFER1987336} as well as the 4-body potential in an unevaluated form \cite{Chu:2008xm}. 

Possibly the most severe bottleneck for obtaining analytic \textit{position space} results for $N$-body potentials (in the PN expansion) is the evaluation of the class of Euclidean Feynman integrals described in Section \ref{sec:NbdyFT}. Without analytic expressions, numerical simulations of real-time $N$-body dynamics may require computationally expensive numerical Fourier transform integrals at each time-step (though this is almost certainly still less computationally expensive than a full numerical relativity simulation \cite{Campanelli:2007ea, Lousto:2007rj, Galaviz:2010mx}). For 3-bodies at 3PM, obtaining analytic results would require at least the calculation of (\ref{I3int}) for the case $a_1\in \mathds{Z}$ and $a_2, a_3 \in \mathds{Z}+\frac{1}{2}$.\footnote{Preliminary investigations for explicit values of the variables $r_{12}$, $r_{13}$ and $r_{23}$ suggest the space of functions in this case involves elliptic functions. We are grateful to Michael Ruf for discussions on this point.} We are optimistic that with powerful modern Feynman integration technology, especially methods based on differential equations \cite{Kotikov:1990kg,Bern:1992em,Remiddi:1997ny,Gehrmann:1999as,Henn:2014qga}, it will be possible to push the state-of-the art for analytic $N$-body potentials. Alternatively, in Section \ref{sec:NbdyFT} we have argued that it should be possible to make progress by specializing to the hierarchical limit of 3-body dynamics ($r_{12}\ll r_{13} \sim r_{23}$) and demonstrated that the method of regions is a powerful tool in this context for evaluating the integrals (\ref{I3int}) order-by-order in this hierarchy. 

The position space potential is most relevant for the dynamics of bound systems for which the PN expansion is appropriate. For unbound systems (scattering dynamics) the bodies involved may generally have relativistic momenta, and so the PM (velocity resummed) expansion is appropriate. In this context the calculation of an analytic position space potential may be very difficult; even at 2PM the ``Y-graph" Fourier transform integrals are not obviously related to Feynman integrals. Fortunately the construction of a potential in this context may be unnecessary. By applying the formalism of KMOC \cite{Kosower:2018adc} to $N$-body scattering amplitudes it should be possible to calculate gauge invariant physical conservative observables such as linear and angular impulses, as well as dissipative observables such as gravitational waveforms, directly from scattering amplitudes. Not only would this strategy by-pass the difficulty of solving equations of motion for a PM Hamiltonian, it also avoids the gauge ambiguities described in Section \ref{subsec:gauge} by working exclusively with physical quantities. It is tantalizing to imagine that these unbound $N$-body observables may be related to bound observables by analytic continuation along the lines of the boundary-to-bound dictionary introduced in \cite{Kalin:2019rwq}, if perhaps only in certain special symmetrical configurations. 

Finally it would be interesting to apply the methods described in this paper to the calculation of spin-dependent $N$-body potentials following the recent successful use of amplitudes based methods for the spinning 2-body problem \cite{Vaidya:2014kza, Bern:2020buy}. To our knowledge, even the leading-order in PN spin-dependent (intrinsic) 3-body potential has never been calculated. We leave this and the other interesting open problems described above to future work.

\vspace{3mm}
\noindent \textbf{Acknowledgements}

\vspace{3mm}
We would like to thank Zvi Bern, Florian Loebbert, Smadar Naoz, Jan Plefka, Michael Ruf, Canxin Shi and Tianheng Wang for useful discussions and comments on the draft. CRTJ and MS are supported by the Department of Energy under Award Number DE-SC0009937. MS is also supported by the Alfred P. Sloan Foundation. We are grateful to the Mani L. Bhaumik Institute for Theoretical Physics. CRTJ acknowledges the support of the Munich Institute for Astro-, Particle and BioPhysics (MIAPbP) which is funded by the Deutsche Forschungsgemeinschaft (DFG, German Research Foundation) under Germany´s Excellence Strategy – EXC-2094 – 390783311.

\appendix

\section{Feynman Rules for Einstein-Maxwell}
\label{app:FeynRules}

At $\mathcal{O}\left(G^2\right)$ the necessary 3-body scattering amplitudes can be calculated using Feynman diagrams. For reference the explicit Feynman rules we need for Einstein-Maxwell amplitude $i \mathcal{M}_N$ are given below in \textit{Feynman} gauge for the photon (single wavy line) and \textit{de Donder} gauge for the graviton (double wavy line). Where appropriate, the arrow on the dashed scalar line indicates both the flow of electric charge and the direction of momentum flow. Note that the electric charge $Q_i$ is given in Planck units, where $Q_i=m_i$ corresponds to extremality.
\begingroup
\allowdisplaybreaks
\begin{center}
    \begin{tikzpicture}
    \begin{scope}[xshift=0cm, yshift=0cm]
    \draw[scalar] (0,0)--(2,0);
    \node at (5,0) {$=$}; 
    \node at (9,0.3) {
    \begin{minipage}{3cm}
    \begin{align*}
        \frac{i}{p^2-m^2}
    \end{align*}
    \end{minipage}
    };
    \end{scope}
    \begin{scope}[xshift=0cm, yshift=-2cm]
    \draw[vector] (0,0)--(2,0);
    \node at (-0.1,0.3) {\footnotesize $\mu$};
    \node at (2.1,0.3) {\footnotesize $\nu$};
    \node at (5,0) {$=$}; 
    \node at (9,0.3) {
    \begin{minipage}{3cm}
    \begin{align*}
        \frac{-i\eta_{\mu\nu}}{p^2}
    \end{align*}
    \end{minipage}
    };
    \end{scope}
    \begin{scope}[xshift=0cm, yshift=-4cm]
    \draw[graviton] (0,0)--(2,0);
    \node at (-0.1,0.3) {\footnotesize $\mu,\nu$};
    \node at (2.1,0.3) {\footnotesize $\rho,\sigma$};
    \node at (5,0) {$=$}; 
    \node at (9,0.3) {
    \begin{minipage}{3cm}
    \begin{align*}
        \frac{i}{2}\left[\frac{\eta_{\mu\rho}\eta_{\nu\sigma}+\eta_{\mu\sigma}\eta_{\nu\rho}-\eta_{\mu\nu}\eta_{\rho\sigma}}{p^2}\right]
    \end{align*}
    \end{minipage}
    };
    \end{scope}
    \begin{scope}[xshift=0cm, yshift=-7cm]
    \draw[scalar] (0.133,-0.5)--(1,0);
    \draw[scalar] (1,0)--(1.866,-0.5);
    \draw[vector] (1,0)--(1,1);
    \node at (1,1.2) {\footnotesize $\mu$};
    \node at (0.4,-0.1) {\footnotesize $p_1$};
    \node at (1.7,-0.1) {\footnotesize $p_2$};
    \node at (5,0) {$=$}; 
    \node at (9,0.3) {
    \begin{minipage}{3cm}
    \begin{align*}
        -i \sqrt{4\pi G} Q_i \left(p_1+p_2\right)^\mu
    \end{align*}
    \end{minipage}
    };
    \end{scope}
    \begin{scope}[xshift=0cm, yshift=-10cm]
    \draw[scalar] (0.2,-0.8)--(1,0);
    \draw[scalar] (1,0)--(1.8,-0.8);
    \draw[vector] (1,0)--(0.2,0.8);
    \draw[vector] (1,0)--(1.8,0.8);
    \node at (0.1,0.9) {\footnotesize $\mu$};
    \node at (1.9,0.9) {\footnotesize $\nu$};
    \node at (5,0) {$=$}; 
    \node at (9,0.3) {
    \begin{minipage}{3cm}
    \begin{align*}
        8i \pi G Q_i^2 \eta_{\mu\nu}
    \end{align*}
    \end{minipage}
    };
    \end{scope}
    \begin{scope}[xshift=0cm, yshift=-13cm]
    \draw[scalar] (0.133,-0.5)--(1,0);
    \draw[scalar] (1,0)--(1.866,-0.5);
    \draw[graviton] (1,0)--(1,1);
    \node at (1,1.2) {\footnotesize $\mu, \nu$};
    \node at (0.4,-0.1) {\footnotesize $p_1$};
    \node at (1.7,-0.1) {\footnotesize $p_2$};
    \node at (5,0) {$=$}; 
    \node at (11,0.3) {
    \begin{minipage}{3cm}
    \begin{align*}
        i\sqrt{8\pi G}\left[\eta^{\mu\nu}\left(p_1\cdot p_2 - m_i^2\right)-p_1^\mu p_2^\nu-p_1^\nu p_2^\mu\right]
    \end{align*}
    \end{minipage}
    };
    \end{scope}
    \begin{scope}[xshift=0cm, yshift=-16cm]
    \draw[scalar] (0.2,-0.8)--(1,0);
    \draw[scalar] (1,0)--(1.8,-0.8);
    \draw[graviton] (1,0)--(0.2,0.8);
    \draw[graviton] (1,0)--(1.8,0.8);
    \node at (0.1,0.9) {\footnotesize $\mu,\nu$};
    \node at (1.9,0.9) {\footnotesize $\rho,\sigma$};
    \node at (0.4,-0.35) {\footnotesize $p_1$};
    \node at (1.7,-0.35) {\footnotesize $p_2$};
    \node at (5,0) {$=$}; 
    \node at (11,0.3) {
    \begin{minipage}{3cm}
    \begin{align*}
        &8 i \pi  G \left[\left(p_1\cdot p_2-m^2\right)
   \left(\eta^{\mu \nu } \eta^{\rho \sigma }-\eta^{\mu \sigma } \eta^{\nu \rho }-\eta^{\mu \rho } \eta^{\nu \sigma
   }\right)\right.\nonumber\\
   &\hspace{10mm}\left.-\eta^{\mu \nu }
   \left(p_1^{\sigma } p_2^{\rho
   }+p_1^{\rho } p_2^{\sigma }\right)+\eta^{\mu
   \rho } \left(p_1^{\sigma } p_2^{\nu
   }+p_1^{\nu } p_2^{\sigma }\right)\right.\nonumber\\
   &\hspace{10mm}\left.+\eta^{\mu \sigma }
   \left(p_1^{\rho } p_2^{\nu }+p_1^{\nu }
   p_2^{\rho }\right)+\eta^{\nu \rho } \left(p_1^{\sigma }
   p_2^{\mu }+p_1^{\mu } p_2^{\sigma
   }\right)\right.\nonumber\\
   &\hspace{10mm}\left.+\eta^{\nu \sigma } \left(p_1^{\rho } p_2^{\mu
   }+p_1^{\mu } p_2^{\rho }\right)-\eta^{\rho \sigma }
   \left(p_1^{\nu } p_2^{\mu }+p_1^{\mu }
   p_2^{\nu }\right)\right]
    \end{align*}
    \end{minipage}
    };
    \end{scope}
     \begin{scope}[xshift=0cm, yshift=-19cm]
    \draw[scalar] (0.2,-0.8)--(1,0);
    \draw[scalar] (1,0)--(1.8,-0.8);
    \draw[graviton] (1,0)--(0.2,0.8);
    \draw[vector] (1,0)--(1.8,0.8);
    \node at (0.1,0.9) {\footnotesize $\mu,\nu$};
    \node at (1.9,0.9) {\footnotesize $\rho$};
    \node at (0.4,-0.35) {\footnotesize $p_1$};
    \node at (1.7,-0.35) {\footnotesize $p_2$};
    \node at (5,0) {$=$}; 
    \node at (11,0.3) {
    \begin{minipage}{3cm}
    \begin{align*}
        -4 i \sqrt{2} \pi  G Q_i \left[\eta^{\mu \nu }
   \left(p_1+p_2\right)^{\rho }-\eta^{\mu \rho }
   \left(p_1+p_2\right)^{\nu }-\eta^{\nu \rho }
   \left(p_1+p_2\right)^{\mu }\right]
    \end{align*}
    \end{minipage}
    };
    \end{scope}
    \end{tikzpicture}
    \begin{tikzpicture}
    \begin{scope}[xshift=0cm, yshift=-22cm]
    \draw[vector] (0.133,-0.5)--(1,0);
    \draw[vector] (1,0)--(1.866,-0.5);
    \draw[graviton] (1,0)--(1,1);
    \node at (1,1.2) {\footnotesize $\mu, \nu$};
    \node at (0,-0.6) {\footnotesize $\rho$};
    \node at (2,-0.6) {\footnotesize $\sigma$};
    \node at (0.3,0.1) {\footnotesize $p_1$};
    \node at (1.7,0.1) {\footnotesize $p_2$};
    \draw[->] (0.7,0.1)--(0.3,-0.2);
    \draw[->] (1.3,0.1)--(1.7,-0.2);
    \node at (5,0) {$=$}; 
    \node at (11,0.3) {
    \begin{minipage}{3cm}
    \begin{align*}
        & i \sqrt{8 \pi G} \left[ p_1^{\sigma }p_2^{\mu
   } \eta^{\nu \rho }-p_1^{\sigma }p_2^{\rho } \eta^{\mu \nu
   }+ p_1^{\rho }p_2^{\mu } \eta^{\nu \sigma
   }-p_1^{\rho }p_2^{\sigma } \eta^{\mu \nu }\right.\nonumber\\
   &\hspace{12mm}\left.+
   p_1^{\nu }p_2^{\sigma } \eta^{\mu \rho }+p_1^{\nu }p_2^{\rho }
   \eta^{\mu \sigma }-p_1^{\nu }p_2^{\mu } \eta^{\rho \sigma
   }\right.\nonumber\\
   &\hspace{12mm}\left.-\left(p_1\cdot p_2\right) \left(\eta^{\mu
   \sigma } \eta^{\nu \rho }+\eta^{\mu \rho } \eta^{\nu \sigma }-\eta^{\mu \nu }
   \eta^{\rho \sigma }\right)\right]
    \end{align*}
    \end{minipage}
    };
    \end{scope}
    \begin{scope}[xshift=0cm, yshift=-25cm]
    \draw[graviton] (0.133,-0.5)--(1,0);
    \draw[graviton] (1,0)--(1.866,-0.5);
    \draw[graviton] (1,0)--(1,1);
    \node at (1,1.2) {\footnotesize $\mu_1, \nu_1$};
    \node at (0,-0.7) {\footnotesize $\mu_2, \nu_2$};
    \node at (2,-0.7) {\footnotesize $\mu_3, \nu_3$};
    \node at (0.3,0.1) {\footnotesize $p_2$};
    \node at (1.7,0.1) {\footnotesize $p_3$};
    \node at (1.5,0.7) {\footnotesize $p_1$};
    \draw[->] (1.2,0.4)--(1.2,0.9);
    \draw[->] (0.7,0.1)--(0.3,-0.2);
    \draw[->] (1.3,0.1)--(1.7,-0.2);
    \node at (5,0) {$=$}; 
    \node at (10,0.3) {
    \begin{minipage}{3cm}
    \begin{align*}
        i\sqrt{2\pi G} \;V_{hhh}^{\mu_1 \nu_1 \mu_2 \nu_2 \mu_3 \nu_3}\left(p_1,p_2,p_3\right)
    \end{align*}
    \end{minipage}
    };
    \end{scope}
    \end{tikzpicture}
\end{center}
\endgroup
The cubic graviton vertex function can be compactly expressed by introducing a set of auxilliary 4-vectors $z_i^\mu$ 
\begin{equation}
    V_{hhh}^{\mu_1 \nu_1 \mu_2 \nu_2 \mu_3 \nu_3}\left(p_1,p_2,p_3\right) = \frac{1}{2^3}\frac{\partial^2}{\partial z_{1\mu_1}\partial z_{1\nu_1}}\frac{\partial^2}{\partial z_{2\mu_2}\partial z_{2\nu_2}}\frac{\partial^2}{\partial z_{3\mu_3}\partial z_{3\nu_3}}V_{hhh}\left(p_1,p_2,p_3,z_1,z_2,z_3\right),
\end{equation}
where
\begin{align}
    &V_{hhh}\left(p_1,p_2,p_3,z_1,z_2,z_3\right) \nonumber\\
    &= \sum_{(i,j,k)\in S_3} \left[2 \left(p_i\cdot z_k\right) \left(p_j\cdot
   z_k\right) (z_i\cdot z_j)^2-4 \left(z_i\cdot z_k\right)
   \left(z_j\cdot z_k\right) \left(p_i\cdot
   z_j\right) \left(p_k\cdot z_i\right)\right.\nonumber\\
   &\hspace{20mm}\left.+\frac{1}{6}
   \left(p_i\cdot p_j+p_i\cdot
   p_k+p_j\cdot p_k\right) \left(z_i^2
   \left(z_j^2 z_k^2-6 (z_j\cdot
   z_k)^2\right)+8 \left(z_i\cdot z_j\right)
   \left(z_i\cdot z_k\right) \left(z_j\cdot
   z_k\right)\right) \right].
\end{align}
As in the rest of the paper this notation denotes the sum over the $3!$ distinct permutations of $\{1,2,3\}$. Note that since this is an off-shell vertex function the vectors $z_i^\mu$ are completely general and do not satisfy any additional identities.

\section{Calculation of 3-Body Subtraction Contribution}
\label{app:3bdSubtraction}

In this appendix we provide further details of the calculation of the subtraction contribution (\ref{amsub}) to the 3-body potential in Einstein-Maxwell. \\
\\
For the calculation of the subtraction contribution (\ref{wouldbesubtraction}), it is a little simpler (but completely optional) to keep the $\mathcal{O}\left(|\vec{q}|^{-3}|\right)$ and higher terms until the end of the calculation. For 2-body dynamics at $\mathcal{O}\left(G\right)$ the \textit{quantum} potential in generalized isotropic gauge is
\begin{equation}
    \label{quantumpotential}
    V_{ij}^{(2)}\left(\vec{p}_i,\vec{p}_j,\vec{q}_i,\vec{q}_j\right) = \frac{1}{4\sqrt{E_i\left(\vec{p}_i\right) E_j\left(\vec{p}_j\right) E_i\left(\vec{p}_i-\vec{q}_i\right)E_j\left(\vec{p}_j-\vec{q}_j\right)}} \times \frac{\tilde{N}_{ij}^{}\left(\vec{p}_i,\vec{p}_j,\vec{q}_i\right)}{\left(\frac{(\vec{p}_i+\vec{p}_j)\cdot \vec{q}_i}{E_i +E_j}\right)^2-|\vec{q}_i|^2},
\end{equation}
where
\begin{align}
    \tilde{N}_{ij}^{}\left(\vec{p}_i,\vec{p}_j,\vec{q}_i\right) &= 4 \pi  G \left[-\frac{4 E_i E_j(\rho_{ij}-1) \left((E_i+E_j)^2 |\vec{q}_i|^2-(E_i \tau_{ii}+E_j \tau_{ji})^2+Q_i Q_j(E_i+E_j)^2\right)}{(E_i+E_j)^2}\right.\nonumber\\
    &\hspace{15mm}\left.+Q_i Q_j \left(\frac{(E_i
   \tau_{ii}+E_j \tau_{ji})^2}{(E_i+E_j)^2}-|\vec{q}_i|^2\right)-8
   E_i^2 E_j^2(\rho_{ij}-1)^2+4 m_i^2 m_j^2\right].
\end{align}
For the calculation of the U-graph contributions we define the following off-shell numerator function
\begin{align}
    N_{ij}^{}\left(p_i,p_j,q_i\right) &= 4 \pi  G \left[-8(p_i\cdot p_j)^{2}+4 m_i^2
   \left(-p_j\cdot q_i-p_j^{}+2 m_j^2\right)+4
   \left(m_j^2-p_j^{2}\right) \left(p_i\cdot
   q_i\right)\right.\nonumber\\
   &\hspace{16mm}+\left.4 p_i^{2} \left(p_j\cdot
   q_i+p_j^{2}-m_j^2\right)+Q_i Q_j \left(-2
   \left(p_j\cdot q_i\right)+2 \left(p_i\cdot
   q_i\right)-q_i^{2}\right)\right.\nonumber\\
   &\hspace{16mm}\left.+4 \left(p_i\cdot
   p_j\right) \left(2 \left(p_j\cdot
   q_i\right)-2 \left(p_i\cdot
   q_i\right)+q_i^{2}+Q_i Q_j\right)\right],
\end{align}
this is simply the numerator of the off-shell 2-body Feynman diagram that gives the quantum potential (\ref{quantumpotential}). In terms of these functions, the subtraction contribution (\ref{wouldbesubtraction}) is calculated by continuing 
\begin{align}
    \label{subdiff1}
    V^{(3)}_{\text{subtraction},ijk} &\overset{!}{=}\frac{1}{8\sqrt{E_i\left(\vec{p}_i\right)E_j\left(\vec{p}_j\right)E_k\left(\vec{p}_k\right)E_i\left(\vec{p}_i-\vec{q}_i\right)E_j\left(\vec{p}_j-\vec{q}_j\right)E_k\left(\vec{p}_k-\vec{q}_k\right)}}\nonumber\\
    &\hspace{10mm}\times \frac{1}{2E_j\left(\vec{p}_j+\vec{q}_i\right)} \times  \frac{1}{E_i\left(\vec{p}_i\right)+E_j\left(\vec{p}_j\right)-E_i\left(\vec{p}_i-\vec{q}_i\right)-E_j\left(\vec{p}_j+\vec{q}_i\right)}\nonumber\\
    &\hspace{10mm}\times\left[\frac{N^{}_{ij}\left(p_i,p_j,q_i\right)N^{}_{kj}\left(p_k,p_j+q_i,q_k\right)}{\left[\left(E_i(\vec{p}_i)-E_i(\vec{p}_i-\vec{q}_i)\right)^2-|\vec{q}_i|^2\right] \left[\left(E_k(\vec{p}_k)-E_k(\vec{p}_k-\vec{q}_k)\right)^2-|\vec{q}_k|^2\right]} \right.\nonumber\\
    &\hspace{20mm}\left.- \frac{\tilde{N}^{}_{ij}\left(\vec{p}_i,\vec{p}_j,\vec{q}_i\right) \tilde{N}^{}_{kj}\left(\vec{p}_k,\vec{p}_j+\vec{q}_i,\vec{q}_k\right)}{\left[\left(\frac{(\vec{p}_i+\vec{p}_j)\cdot \vec{q}_i}{E_j(\vec{p}_j) +E_k(\vec{p}_k)}\right)^2-|\vec{q}_i|^2\right] \left[\left(\frac{((\vec{p}_j+\vec{q}_i)+\vec{p}_k)\cdot \vec{q}_k}{E_j(\vec{p}_j+\vec{q}_i) +E_k(\vec{p}_k)}\right)^2-|\vec{q}_k|^2\right]}\right],
\end{align}
away from the constraint surface of energy conservation. Pictorially, this is the contribution from difference of full-theory and EFT graphs of the form depicted in Figure \ref{fig:matterpole} with $p_L = p_i + p_j$, $p_L' = p_i-q_i$ and $p_R = p_k$. As discussed in Section \ref{sec:matterpole} the challenge involved is to manipulate this expression into a form in which the spurious matter pole cancels \textit{and} that the super-classical $\mathcal{O}\left(|\vec{q}|^{-5}\right)$ part of this expression cancel in the complete subtraction potential (after summing over $(i,j,k)\in S_3$). To engineer such a continuation we will fix the 3-body energy conservation constraint (\ref{consen}) differently for different parts of the expression. We will use the notation $\vert_{\vec{q}_j,\vec{p}_j\cdot \vec{q}_k} $ to denote the procedure of first eliminating $\vec{q}_j$ using 3-momentum conservation, followed by eliminating $\vec{p}_j\cdot \vec{q}_k$ by solving the energy conservation constraint (\ref{consen}). We make the following definitions
\begingroup
\allowdisplaybreaks
\begin{align}
    \mathcal{N}_{ijk}^{(L)} &= \frac{N^{}_{ij}\left(p_i,p_j,q_i\right) }{4\sqrt{E_i\left(\vec{p}_i\right)E_j\left(\vec{p}_j\right)E_i\left(\vec{p}_i-\vec{q}_i\right)E_j\left(\vec{p}_j+\vec{q}_i\right)}}\biggr\vert_{\vec{q}_j,\vec{p}_j\cdot \vec{q}_k} \nonumber\\
    \hat{\mathcal{N}}_{ijk}^{(L)} &= \frac{\tilde{N}^{}_{ij}\left(\vec{p}_i,\vec{p}_j,\vec{q}_i\right)}{4\sqrt{E_i\left(\vec{p}_i\right)E_j\left(\vec{p}_j\right)E_i\left(\vec{p}_i-\vec{q}_i\right)E_j\left(\vec{p}_j+\vec{q}_i\right)}} \biggr\vert_{\vec{q}_j,\vec{p}_j\cdot \vec{q}_k} \nonumber\\
    q_{ijk}^{(L)} &= \left[\left(E_i(\vec{p}_i)-E_i(\vec{p}_i-\vec{q}_i)\right)^2-|\vec{q}_i|^2\right] \biggr\vert_{\vec{q}_j,\vec{p}_j\cdot \vec{q}_k} \nonumber\\
    \hat{q}_{ijk}^{(L)} &= \left[\left(\frac{(\vec{p}_i+\vec{p}_j)\cdot \vec{q}_i}{E_i(\vec{p}_i) +E_j(\vec{p}_j)}\right)^2-|\vec{q}_i|^2\right]\biggr\vert_{\vec{q}_j,\vec{p}_j\cdot \vec{q}_k} \nonumber\\
    \Delta E_{ijk}^{(L)} &= \left[E_i\left(\vec{p}_i\right)+E_j\left(\vec{p}_j\right)-E_i\left(\vec{p}_i-\vec{q}_i\right)-E_j\left(\vec{p}_j+\vec{q}_i\right)\right]\biggr\vert_{\vec{q}_j,\vec{p}_j\cdot \vec{q}_k} \nonumber\\
    \mathcal{N}_{ijk}^{(R)} &= \frac{N^{}_{kj}\left(p_k,p_j+q_i,q_k\right)}{4\sqrt{E_k\left(\vec{p}_k\right)E_k\left(\vec{p}_k-\vec{q}_k\right)E_j\left(\vec{p}_j+\vec{q}_i\right)E_j\left(\vec{p}_j-\vec{q}_j\right)}}\biggr\vert_{\vec{q}_j,\vec{p}_j\cdot \vec{q}_i} \nonumber\\
    \hat{\mathcal{N}}_{ijk}^{(R)} &= \frac{\tilde{N}^{}_{kj}\left(\vec{p}_k,\vec{p}_j+\vec{q}_i,\vec{q}_k\right)}{4\sqrt{E_k\left(\vec{p}_k\right)E_k\left(\vec{p}_k-\vec{q}_k\right)E_j\left(\vec{p}_j+\vec{q}_i\right)E_j\left(\vec{p}_j-\vec{q}_j\right)}}\biggr\vert_{\vec{q}_j,\vec{p}_j\cdot \vec{q}_i}\nonumber\\
    q_{ijk}^{(R)} &= \left[\left(E_k(\vec{p}_k)-E_k(\vec{p}_k-\vec{q}_k)\right)^2-|\vec{q}_k|^2\right] \biggr\vert_{\vec{q}_j,\vec{p}_j\cdot \vec{q}_i}\nonumber\\
    \hat{q}_{ijk}^{(R)} &= \left[\left(\frac{((\vec{p}_j+\vec{q}_i)+\vec{p}_k)\cdot \vec{q}_k}{E_j(\vec{p}_j+\vec{q}_i) +E_k(\vec{p}_k)}\right)^2-|\vec{q}_k|^2\right]\biggr\vert_{\vec{q}_j,\vec{p}_j\cdot \vec{q}_i}\nonumber\\
    \Delta E_{ijk}^{(R)} &= \left[E_i\left(\vec{p}_i\right)+E_j\left(\vec{p}_j\right)-E_i\left(\vec{p}_i-\vec{q}_i\right)-E_j\left(\vec{p}_j+\vec{q}_i\right)\right]\biggr\vert_{\vec{q}_j,\vec{p}_j\cdot \vec{q}_i}.
\end{align}
\endgroup
With these definitions, (\ref{subdiff1}) is equivalent to
\begin{equation}
    \label{subdiff2}
    V^{(3)}_{\text{subtraction},ijk} \overset{!}{=}\frac{1}{\Delta E_{ijk}^{(L)}}\left[\frac{\mathcal{N}_{ijk}^{(L)} \mathcal{N}_{ijk}^{(R)}}{q_{ijk}^{(L)}q_{ijk}^{(R)}}-\frac{\hat{\mathcal{N}}_{ijk}^{(L)} \hat{\mathcal{N}}_{ijk}^{(R)}}{\hat{q}_{ijk}^{(L)}\hat{q}_{ijk}^{(R)}}\right].
\end{equation}
As discussed in Section \ref{sec:matterpole}, consistency requires the hatted and un-hatted quantities differ by an expression proportional to the matter pole. This inspires us to define 
\begin{align}
    \hat{\mathcal{N}}_{ijk}^{(L)} &= \mathcal{N}_{ijk}^{(L)} + \Delta E_{ijk}^{(L)} \delta \mathcal{N}_{ijk}^{(L)} \nonumber\\
    \hat{\mathcal{N}}_{ijk}^{(R)} &= \mathcal{N}_{ijk}^{(R)} + \Delta E_{ijk}^{(R)} \delta \mathcal{N}_{ijk}^{(R)} \nonumber\\
    \hat{q}_{ijk}^{(L)} &= q_{ijk}^{(L)} + \Delta E_{ijk}^{(L)} \delta q_{ijk}^{(L)}\nonumber\\
    \hat{q}_{ijk}^{(R)} &= q_{ijk}^{(R)} + \Delta E_{ijk}^{(R)} \delta q_{ijk}^{(R)}.
\end{align}
We also note the following
\begin{equation}
    \Delta E_{ijk}^{(L)} \overset{!}{=} \Delta E_{ijk}^{(R)} \overset{!}{=} \Delta E_{ijk} \equiv \frac{1}{2}\left[\Delta E_{ijk}^{(L)}  + \Delta E_{ijk}^{(R)} \right].
\end{equation}
With some elementary algebra we derive the the expression 
\begin{align}
    \label{subtractionquantum}
    V^{(3)}_{\text{subtraction},ijk} &\overset{!}{=} -\frac{\delta\mathcal{N}_{ijk}^{(L)}\delta \mathcal{N}_{ijk}^{(R)}\Delta E_{ijk}}{\hat{q}_{ijk}^{(L)} \hat{q}_{ijk}^{(R)}}-\frac{\mathcal{N}_{ijk}^{(L)}\delta \mathcal{N}_{ijk}^{(R)} +\mathcal{N}_{ijk}^{(R)}\delta \mathcal{N}_{ijk}^{(L)}}{\hat{q}_{ijk}^{(L)} \hat{q}_{ijk}^{(R)}} + \frac{\delta q_{ijk}^{(L)} \mathcal{N}_{ijk}^{(L)} \mathcal{N}_{ijk}^{(R)}}{q_{ijk}^{(L)} \hat{q}_{ijk}^{(L)} \hat{q}_{ijk}^{(R)}} \nonumber\\
    &\hspace{10mm}+ \frac{\delta q_{ijk}^{(R)} \mathcal{N}_{ijk}^{(L)} \mathcal{N}_{ijk}^{(R)}}{\hat{q}_{ijk}^{(L)} q_{ijk}^{(R)}\hat{q}_{ijk}^{(R)}}+ \frac{\Delta E_{ijk} \delta q_{ijk}^{(L)} \delta q_{ijk}^{(R)} \mathcal{N}_{ijk}^{(L)} \mathcal{N}_{ijk}^{(R)}}{q_{ijk}^{(L)} \hat{q}_{ijk}^{(L)} q_{ijk}^{(R)}\hat{q}_{ijk}^{(R)} },
\end{align}
for which the matter pole has cancelled, at the price of introducing some (harmless) doubled graviton poles. Explicit expressions for all of these functions for Einstein-Maxwell theory, expanded in $q$ to the order relevant for the calculation of the classical potential, are given in (\ref{universal}) and (\ref{nonuniversal}).

What remains is to demonstrate the cancellation of the super-classical contribution. To see this we examine the leading $q$-scaling of each individual object. In general we expect (and confirm from the explicit expressions (\ref{universal}) and (\ref{nonuniversal}))
\begin{equation}
    \label{qscaling}
    \mathcal{N}^{(L/R)} \sim q^0, \hspace{5mm} \delta \mathcal{N}^{(L/R)} \sim q^0, \hspace{5mm} q^{(L/R)} \sim q^2, \hspace{5mm} \delta q^{(L/R)} \sim q^1, \hspace{5mm} \Delta E \sim q^1.
\end{equation}
By inspection, the first term in (\ref{subtractionquantum}) is purely quantum and so can be dropped giving (\ref{amsub}), the final three terms contain super-classical contributions. These cancel when we sum this expression with the crossed $(i\leftrightarrow k)$ graph, by virtue of the following symmetry properties
\begin{align}
    \label{relabelsym}
    \mathcal{N}^{(L/R)}_{ijk} &= \mathcal{N}^{(R/L)}_{kji} + \mathcal{O}\left(q^1\right) \nonumber\\
    q^{(L/R)}_{ijk} &= q^{(R/L)}_{kji} + \mathcal{O}\left(q^3\right) \nonumber\\
    \hat{q}^{(L/R)}_{ijk} &= \hat{q}^{(R/L)}_{kji} + \mathcal{O}\left(q^3\right) \nonumber\\
    \delta q^{(L/R)}_{ijk} &= -\delta q^{(R/L)}_{kji} + \mathcal{O}\left(q^2\right) \nonumber\\
    \Delta E_{ijk} &= -\Delta E_{kji} + \mathcal{O}\left(q^2\right).
\end{align}
From this we can see that the final three terms above are \textit{odd} under $i\leftrightarrow k$ at leading order in $q$, and so the super-classical contributions cancel.

\bibliographystyle{JHEP}
\bibliography{cite.bib}

\providecommand{\href}[2]{#2}\begingroup\raggedright\begin{thebibliography}{100}

\bibitem{Poincare}
H.~Poincar\'e, \emph{Sur le probl\`eme des trois corps et les \'equations de la
  dynamique}, {\emph{Acta Mathematica} {\bfseries 13} (1890) 1}.

\bibitem{PhysRevLett.110.114301}
M.~\ifmmode~\check{S}\else \v{S}\fi{}uvakov and V.~Dmitra\ifmmode
  \check{s}\else \v{s}\fi{}inovi\ifmmode~\acute{c}\else \'{c}\fi{}, \emph{Three
  classes of newtonian three-body planar periodic orbits},
  \href{https://doi.org/10.1103/PhysRevLett.110.114301}{\emph{Phys. Rev. Lett.}
  {\bfseries 110} (2013) 114301}.

\bibitem{Campanelli:2007ea}
M.~Campanelli, C.O.~Lousto and Y.~Zlochower, \emph{{Close encounters of three
  black holes}}, \href{https://doi.org/10.1103/PhysRevD.77.101501}{\emph{Phys.
  Rev. D} {\bfseries 77} (2008) 101501}
  [\href{https://arxiv.org/abs/0710.0879}{{\ttfamily 0710.0879}}].

\bibitem{Lousto:2007rj}
C.O.~Lousto and Y.~Zlochower, \emph{{Foundations of multiple black hole
  evolutions}}, \href{https://doi.org/10.1103/PhysRevD.77.024034}{\emph{Phys.
  Rev. D} {\bfseries 77} (2008) 024034}
  [\href{https://arxiv.org/abs/0711.1165}{{\ttfamily 0711.1165}}].

\bibitem{Galaviz:2010mx}
P.~Galaviz, B.~Bruegmann and Z.~Cao, \emph{{Numerical evolution of multiple
  black holes with accurate initial data}},
  \href{https://doi.org/10.1103/PhysRevD.82.024005}{\emph{Phys. Rev. D}
  {\bfseries 82} (2010) 024005}
  [\href{https://arxiv.org/abs/1004.1353}{{\ttfamily 1004.1353}}].

\bibitem{Pretorius:2005gq}
F.~Pretorius, \emph{{Evolution of binary black hole spacetimes}},
  \href{https://doi.org/10.1103/PhysRevLett.95.121101}{\emph{Phys. Rev. Lett.}
  {\bfseries 95} (2005) 121101}
  [\href{https://arxiv.org/abs/gr-qc/0507014}{{\ttfamily gr-qc/0507014}}].

\bibitem{Campanelli:2005dd}
M.~Campanelli, C.O.~Lousto, P.~Marronetti and Y.~Zlochower, \emph{{Accurate
  evolutions of orbiting black-hole binaries without excision}},
  \href{https://doi.org/10.1103/PhysRevLett.96.111101}{\emph{Phys. Rev. Lett.}
  {\bfseries 96} (2006) 111101}
  [\href{https://arxiv.org/abs/gr-qc/0511048}{{\ttfamily gr-qc/0511048}}].

\bibitem{Baker:2005vv}
J.G.~Baker, J.~Centrella, D.-I.~Choi, M.~Koppitz and J.~van Meter,
  \emph{{Gravitational wave extraction from an inspiraling configuration of
  merging black holes}},
  \href{https://doi.org/10.1103/PhysRevLett.96.111102}{\emph{Phys. Rev. Lett.}
  {\bfseries 96} (2006) 111102}
  [\href{https://arxiv.org/abs/gr-qc/0511103}{{\ttfamily gr-qc/0511103}}].

\bibitem{Buonanno:1998gg}
A.~Buonanno and T.~Damour, \emph{{Effective one-body approach to general
  relativistic two-body dynamics}},
  \href{https://doi.org/10.1103/PhysRevD.59.084006}{\emph{Phys. Rev. D}
  {\bfseries 59} (1999) 084006}
  [\href{https://arxiv.org/abs/gr-qc/9811091}{{\ttfamily gr-qc/9811091}}].

\bibitem{Buonanno:2000ef}
A.~Buonanno and T.~Damour, \emph{{Transition from inspiral to plunge in binary
  black hole coalescences}},
  \href{https://doi.org/10.1103/PhysRevD.62.064015}{\emph{Phys. Rev. D}
  {\bfseries 62} (2000) 064015}
  [\href{https://arxiv.org/abs/gr-qc/0001013}{{\ttfamily gr-qc/0001013}}].

\bibitem{Mino:1996nk}
Y.~Mino, M.~Sasaki and T.~Tanaka, \emph{{Gravitational radiation reaction to a
  particle motion}},
  \href{https://doi.org/10.1103/PhysRevD.55.3457}{\emph{Phys. Rev. D}
  {\bfseries 55} (1997) 3457}
  [\href{https://arxiv.org/abs/gr-qc/9606018}{{\ttfamily gr-qc/9606018}}].

\bibitem{Quinn:1996am}
T.C.~Quinn and R.M.~Wald, \emph{{An Axiomatic approach to electromagnetic and
  gravitational radiation reaction of particles in curved space-time}},
  \href{https://doi.org/10.1103/PhysRevD.56.3381}{\emph{Phys. Rev. D}
  {\bfseries 56} (1997) 3381}
  [\href{https://arxiv.org/abs/gr-qc/9610053}{{\ttfamily gr-qc/9610053}}].

\bibitem{Poisson:2011nh}
E.~Poisson, A.~Pound and I.~Vega, \emph{{The Motion of point particles in
  curved spacetime}}, \href{https://doi.org/10.12942/lrr-2011-7}{\emph{Living
  Rev. Rel.} {\bfseries 14} (2011) 7}
  [\href{https://arxiv.org/abs/1102.0529}{{\ttfamily 1102.0529}}].

\bibitem{Barack:2018yvs}
L.~Barack and A.~Pound, \emph{{Self-force and radiation reaction in general
  relativity}}, \href{https://doi.org/10.1088/1361-6633/aae552}{\emph{Rept.
  Prog. Phys.} {\bfseries 82} (2019) 016904}
  [\href{https://arxiv.org/abs/1805.10385}{{\ttfamily 1805.10385}}].

\bibitem{Einstein:1938yz}
A.~Einstein, L.~Infeld and B.~Hoffmann, \emph{{The Gravitational equations and
  the problem of motion}}, \href{https://doi.org/10.2307/1968714}{\emph{Annals
  Math.} {\bfseries 39} (1938) 65}.

\bibitem{Einstein:1940mt}
A.~Einstein and L.~Infeld, \emph{{The Gravitational equations and the problem
  of motion. 2.}}, \href{https://doi.org/10.2307/1969015}{\emph{Annals Math.}
  {\bfseries 41} (1940) 455}.

\bibitem{Ohta:1973je}
T.~Ohta, H.~Okamura, T.~Kimura and K.~Hiida, \emph{{Physically acceptable
  solution of einstein's equation for many-body system}},
  \href{https://doi.org/10.1143/PTP.50.492}{\emph{Prog. Theor. Phys.}
  {\bfseries 50} (1973) 492}.

\bibitem{Jaranowski:1997ky}
P.~Jaranowski and G.~Schaefer, \emph{{Third postNewtonian higher order ADM
  Hamilton dynamics for two-body point mass systems}},
  \href{https://doi.org/10.1103/PhysRevD.57.7274}{\emph{Phys. Rev. D}
  {\bfseries 57} (1998) 7274}
  [\href{https://arxiv.org/abs/gr-qc/9712075}{{\ttfamily gr-qc/9712075}}].

\bibitem{Damour:1999cr}
T.~Damour, P.~Jaranowski and G.~Schaefer, \emph{{Dynamical invariants for
  general relativistic two-body systems at the third postNewtonian
  approximation}},
  \href{https://doi.org/10.1103/PhysRevD.62.044024}{\emph{Phys. Rev. D}
  {\bfseries 62} (2000) 044024}
  [\href{https://arxiv.org/abs/gr-qc/9912092}{{\ttfamily gr-qc/9912092}}].

\bibitem{Blanchet:2000nv}
L.~Blanchet and G.~Faye, \emph{{Equations of motion of point particle binaries
  at the third postNewtonian order}},
  \href{https://doi.org/10.1016/S0375-9601(00)00360-1}{\emph{Phys. Lett. A}
  {\bfseries 271} (2000) 58}
  [\href{https://arxiv.org/abs/gr-qc/0004009}{{\ttfamily gr-qc/0004009}}].

\bibitem{Damour:2001bu}
T.~Damour, P.~Jaranowski and G.~Schaefer, \emph{{Dimensional regularization of
  the gravitational interaction of point masses}},
  \href{https://doi.org/10.1016/S0370-2693(01)00642-6}{\emph{Phys. Lett. B}
  {\bfseries 513} (2001) 147}
  [\href{https://arxiv.org/abs/gr-qc/0105038}{{\ttfamily gr-qc/0105038}}].

\bibitem{Jaranowski:2015lha}
P.~Jaranowski and G.~Sch\"afer, \emph{{Derivation of local-in-time fourth
  post-Newtonian ADM Hamiltonian for spinless compact binaries}},
  \href{https://doi.org/10.1103/PhysRevD.92.124043}{\emph{Phys. Rev. D}
  {\bfseries 92} (2015) 124043}
  [\href{https://arxiv.org/abs/1508.01016}{{\ttfamily 1508.01016}}].

\bibitem{Bertotti:1956pxu}
B.~Bertotti, \emph{{On gravitational motion}},
  \href{https://doi.org/10.1007/bf02746175}{\emph{Nuovo Cim.} {\bfseries 4}
  (1956) 898}.

\bibitem{Kerr:1959zlt}
R.P.~Kerr, \emph{{The Lorentz-covariant approximation method in general
  relativity I}}, \href{https://doi.org/10.1007/bf02732767}{\emph{Nuovo Cim.}
  {\bfseries 13} (1959) 469}.

\bibitem{Bertotti:1960wuq}
B.~Bertotti and J.~Plebanski, \emph{{Theory of gravitational perturbations in
  the fast motion approximation}},
  \href{https://doi.org/10.1016/0003-4916(60)90132-9}{\emph{Annals Phys.}
  {\bfseries 11} (1960) 169}.

\bibitem{Portilla:1979xx}
M.~Portilla, \emph{{MOMENTUM AND ANGULAR MOMENTUM OF TWO GRAVITATING
  PARTICLES}}, \href{https://doi.org/10.1088/0305-4470/12/7/025}{\emph{J. Phys.
  A} {\bfseries 12} (1979) 1075}.

\bibitem{Westpfahl:1979gu}
K.~Westpfahl and M.~Goller, \emph{{GRAVITATIONAL SCATTERING OF TWO RELATIVISTIC
  PARTICLES IN POSTLINEAR APPROXIMATION}},
  \href{https://doi.org/10.1007/BF02817047}{\emph{Lett. Nuovo Cim.} {\bfseries
  26} (1979) 573}.

\bibitem{Portilla:1980uz}
M.~Portilla, \emph{{SCATTERING OF TWO GRAVITATING PARTICLES: CLASSICAL
  APPROACH}}, \href{https://doi.org/10.1088/0305-4470/13/12/017}{\emph{J. Phys.
  A} {\bfseries 13} (1980) 3677}.

\bibitem{Bel:1981be}
L.~Bel, T.~Damour, N.~Deruelle, J.~Ibanez and J.~Martin,
  \emph{{Poincar\'e-invariant gravitational field and equations of motion of
  two pointlike objects: The postlinear approximation of general relativity}},
  \href{https://doi.org/10.1007/BF00756073}{\emph{Gen. Rel. Grav.} {\bfseries
  13} (1981) 963}.

\bibitem{Westpfahl:1985tsl}
K.~Westpfahl, \emph{{High-Speed Scattering of Charged and Uncharged Particles
  in General Relativity}},
  \href{https://doi.org/10.1002/prop.2190330802}{\emph{Fortsch. Phys.}
  {\bfseries 33} (1985) 417}.

\bibitem{Damour:2016gwp}
T.~Damour, \emph{{Gravitational scattering, post-Minkowskian approximation and
  Effective One-Body theory}},
  \href{https://doi.org/10.1103/PhysRevD.94.104015}{\emph{Phys. Rev. D}
  {\bfseries 94} (2016) 104015}
  [\href{https://arxiv.org/abs/1609.00354}{{\ttfamily 1609.00354}}].

\bibitem{Goldberger:2004jt}
W.D.~Goldberger and I.Z.~Rothstein, \emph{{An Effective field theory of gravity
  for extended objects}},
  \href{https://doi.org/10.1103/PhysRevD.73.104029}{\emph{Phys. Rev. D}
  {\bfseries 73} (2006) 104029}
  [\href{https://arxiv.org/abs/hep-th/0409156}{{\ttfamily hep-th/0409156}}].

\bibitem{Gilmore:2008gq}
J.B.~Gilmore and A.~Ross, \emph{{Effective field theory calculation of second
  post-Newtonian binary dynamics}},
  \href{https://doi.org/10.1103/PhysRevD.78.124021}{\emph{Phys. Rev. D}
  {\bfseries 78} (2008) 124021}
  [\href{https://arxiv.org/abs/0810.1328}{{\ttfamily 0810.1328}}].

\bibitem{Foffa:2011ub}
S.~Foffa and R.~Sturani, \emph{{Effective field theory calculation of
  conservative binary dynamics at third post-Newtonian order}},
  \href{https://doi.org/10.1103/PhysRevD.84.044031}{\emph{Phys. Rev. D}
  {\bfseries 84} (2011) 044031}
  [\href{https://arxiv.org/abs/1104.1122}{{\ttfamily 1104.1122}}].

\bibitem{Foffa:2016rgu}
S.~Foffa, P.~Mastrolia, R.~Sturani and C.~Sturm, \emph{{Effective field theory
  approach to the gravitational two-body dynamics, at fourth post-Newtonian
  order and quintic in the Newton constant}},
  \href{https://doi.org/10.1103/PhysRevD.95.104009}{\emph{Phys. Rev. D}
  {\bfseries 95} (2017) 104009}
  [\href{https://arxiv.org/abs/1612.00482}{{\ttfamily 1612.00482}}].

\bibitem{Porto:2017dgs}
R.A.~Porto and I.Z.~Rothstein, \emph{{Apparent ambiguities in the
  post-Newtonian expansion for binary systems}},
  \href{https://doi.org/10.1103/PhysRevD.96.024062}{\emph{Phys. Rev. D}
  {\bfseries 96} (2017) 024062}
  [\href{https://arxiv.org/abs/1703.06433}{{\ttfamily 1703.06433}}].

\bibitem{Foffa:2019hrb}
S.~Foffa, P.~Mastrolia, R.~Sturani, C.~Sturm and W.J.~Torres~Bobadilla,
  \emph{{Static two-body potential at fifth post-Newtonian order}},
  \href{https://doi.org/10.1103/PhysRevLett.122.241605}{\emph{Phys. Rev. Lett.}
  {\bfseries 122} (2019) 241605}
  [\href{https://arxiv.org/abs/1902.10571}{{\ttfamily 1902.10571}}].

\bibitem{Blumlein:2019zku}
J.~Bl\"umlein, A.~Maier and P.~Marquard, \emph{{Five-Loop Static Contribution
  to the Gravitational Interaction Potential of Two Point Masses}},
  \href{https://doi.org/10.1016/j.physletb.2019.135100}{\emph{Phys. Lett. B}
  {\bfseries 800} (2020) 135100}
  [\href{https://arxiv.org/abs/1902.11180}{{\ttfamily 1902.11180}}].

\bibitem{Foffa:2019yfl}
S.~Foffa, R.A.~Porto, I.~Rothstein and R.~Sturani, \emph{{Conservative dynamics
  of binary systems to fourth Post-Newtonian order in the EFT approach II:
  Renormalized Lagrangian}},
  \href{https://doi.org/10.1103/PhysRevD.100.024048}{\emph{Phys. Rev. D}
  {\bfseries 100} (2019) 024048}
  [\href{https://arxiv.org/abs/1903.05118}{{\ttfamily 1903.05118}}].

\bibitem{Blumlein:2020pog}
J.~Bl\"umlein, A.~Maier, P.~Marquard and G.~Sch\"afer, \emph{{Fourth
  post-Newtonian Hamiltonian dynamics of two-body systems from an effective
  field theory approach}},
  \href{https://doi.org/10.1016/j.nuclphysb.2020.115041}{\emph{Nucl. Phys. B}
  {\bfseries 955} (2020) 115041}
  [\href{https://arxiv.org/abs/2003.01692}{{\ttfamily 2003.01692}}].

\bibitem{Blumlein:2020pyo}
J.~Bl\"umlein, A.~Maier, P.~Marquard and G.~Sch\"afer, \emph{{The fifth-order
  post-Newtonian Hamiltonian dynamics of two-body systems from an effective
  field theory approach: potential contributions}},
  \href{https://doi.org/10.1016/j.nuclphysb.2021.115352}{\emph{Nucl. Phys. B}
  {\bfseries 965} (2021) 115352}
  [\href{https://arxiv.org/abs/2010.13672}{{\ttfamily 2010.13672}}].

\bibitem{Bjerrum-Bohr:2018xdl}
N.E.J.~Bjerrum-Bohr, P.H.~Damgaard, G.~Festuccia, L.~Plant\'e and P.~Vanhove,
  \emph{{General Relativity from Scattering Amplitudes}},
  \href{https://doi.org/10.1103/PhysRevLett.121.171601}{\emph{Phys. Rev. Lett.}
  {\bfseries 121} (2018) 171601}
  [\href{https://arxiv.org/abs/1806.04920}{{\ttfamily 1806.04920}}].

\bibitem{Cheung:2018wkq}
C.~Cheung, I.Z.~Rothstein and M.P.~Solon, \emph{{From Scattering Amplitudes to
  Classical Potentials in the Post-Minkowskian Expansion}},
  \href{https://doi.org/10.1103/PhysRevLett.121.251101}{\emph{Phys. Rev. Lett.}
  {\bfseries 121} (2018) 251101}
  [\href{https://arxiv.org/abs/1808.02489}{{\ttfamily 1808.02489}}].

\bibitem{Kosower:2018adc}
D.A.~Kosower, B.~Maybee and D.~O'Connell, \emph{{Amplitudes, Observables, and
  Classical Scattering}},
  \href{https://doi.org/10.1007/JHEP02(2019)137}{\emph{JHEP} {\bfseries 02}
  (2019) 137} [\href{https://arxiv.org/abs/1811.10950}{{\ttfamily
  1811.10950}}].

\bibitem{Iwasaki:1971vb}
Y.~Iwasaki, \emph{{Quantum theory of gravitation vs. classical theory. -
  fourth-order potential}},
  \href{https://doi.org/10.1143/PTP.46.1587}{\emph{Prog. Theor. Phys.}
  {\bfseries 46} (1971) 1587}.

\bibitem{Iwasaki:1971iy}
Y.~Iwasaki, \emph{{Fourth-order gravitational potential based on quantum field
  theory}}, \href{https://doi.org/10.1007/BF02770190}{\emph{Lett. Nuovo Cim.}
  {\bfseries 1S2} (1971) 783}.

\bibitem{Okamura:1973my}
H.~Okamura, T.~Ohta, T.~Kimura and K.~Hiida, \emph{{Perturbation calculation of
  gravitational potentials}},
  \href{https://doi.org/10.1143/PTP.50.2066}{\emph{Prog. Theor. Phys.}
  {\bfseries 50} (1973) 2066}.

\bibitem{Amati:1990xe}
D.~Amati, M.~Ciafaloni and G.~Veneziano, \emph{{Higher Order Gravitational
  Deflection and Soft Bremsstrahlung in Planckian Energy Superstring
  Collisions}}, \href{https://doi.org/10.1016/0550-3213(90)90375-N}{\emph{Nucl.
  Phys. B} {\bfseries 347} (1990) 550}.

\bibitem{Donoghue:1993eb}
J.F.~Donoghue, \emph{{Leading quantum correction to the Newtonian potential}},
  \href{https://doi.org/10.1103/PhysRevLett.72.2996}{\emph{Phys. Rev. Lett.}
  {\bfseries 72} (1994) 2996}
  [\href{https://arxiv.org/abs/gr-qc/9310024}{{\ttfamily gr-qc/9310024}}].

\bibitem{Donoghue:1994dn}
J.F.~Donoghue, \emph{{General relativity as an effective field theory: The
  leading quantum corrections}},
  \href{https://doi.org/10.1103/PhysRevD.50.3874}{\emph{Phys. Rev. D}
  {\bfseries 50} (1994) 3874}
  [\href{https://arxiv.org/abs/gr-qc/9405057}{{\ttfamily gr-qc/9405057}}].

\bibitem{Bjerrum-Bohr:2002gqz}
N.E.J.~Bjerrum-Bohr, J.F.~Donoghue and B.R.~Holstein, \emph{{Quantum
  gravitational corrections to the nonrelativistic scattering potential of two
  masses}}, \href{https://doi.org/10.1103/PhysRevD.71.069903}{\emph{Phys. Rev.
  D} {\bfseries 67} (2003) 084033}
  [\href{https://arxiv.org/abs/hep-th/0211072}{{\ttfamily hep-th/0211072}}].

\bibitem{Bjerrum-Bohr:2013bxa}
N.E.J.~Bjerrum-Bohr, J.F.~Donoghue and P.~Vanhove, \emph{{On-shell Techniques
  and Universal Results in Quantum Gravity}},
  \href{https://doi.org/10.1007/JHEP02(2014)111}{\emph{JHEP} {\bfseries 02}
  (2014) 111} [\href{https://arxiv.org/abs/1309.0804}{{\ttfamily 1309.0804}}].

\bibitem{Kawai:1985xq}
H.~Kawai, D.C.~Lewellen and S.H.H.~Tye, \emph{{A Relation Between Tree
  Amplitudes of Closed and Open Strings}},
  \href{https://doi.org/10.1016/0550-3213(86)90362-7}{\emph{Nucl. Phys. B}
  {\bfseries 269} (1986) 1}.

\bibitem{Bern:1998sv}
Z.~Bern, L.J.~Dixon, M.~Perelstein and J.S.~Rozowsky, \emph{{Multileg one loop
  gravity amplitudes from gauge theory}},
  \href{https://doi.org/10.1016/S0550-3213(99)00029-2}{\emph{Nucl. Phys. B}
  {\bfseries 546} (1999) 423}
  [\href{https://arxiv.org/abs/hep-th/9811140}{{\ttfamily hep-th/9811140}}].

\bibitem{Bern:2008qj}
Z.~Bern, J.J.M.~Carrasco and H.~Johansson, \emph{{New Relations for
  Gauge-Theory Amplitudes}},
  \href{https://doi.org/10.1103/PhysRevD.78.085011}{\emph{Phys. Rev. D}
  {\bfseries 78} (2008) 085011}
  [\href{https://arxiv.org/abs/0805.3993}{{\ttfamily 0805.3993}}].

\bibitem{Bern:2010ue}
Z.~Bern, J.J.M.~Carrasco and H.~Johansson, \emph{{Perturbative Quantum Gravity
  as a Double Copy of Gauge Theory}},
  \href{https://doi.org/10.1103/PhysRevLett.105.061602}{\emph{Phys. Rev. Lett.}
  {\bfseries 105} (2010) 061602}
  [\href{https://arxiv.org/abs/1004.0476}{{\ttfamily 1004.0476}}].

\bibitem{Bern:2019prr}
Z.~Bern, J.J.~Carrasco, M.~Chiodaroli, H.~Johansson and R.~Roiban, \emph{{The
  Duality Between Color and Kinematics and its Applications}},
  \href{https://arxiv.org/abs/1909.01358}{{\ttfamily 1909.01358}}.

\bibitem{Bern:1994zx}
Z.~Bern, L.J.~Dixon, D.C.~Dunbar and D.A.~Kosower, \emph{{One loop n point
  gauge theory amplitudes, unitarity and collinear limits}},
  \href{https://doi.org/10.1016/0550-3213(94)90179-1}{\emph{Nucl. Phys. B}
  {\bfseries 425} (1994) 217}
  [\href{https://arxiv.org/abs/hep-ph/9403226}{{\ttfamily hep-ph/9403226}}].

\bibitem{Bern:1994cg}
Z.~Bern, L.J.~Dixon, D.C.~Dunbar and D.A.~Kosower, \emph{{Fusing gauge theory
  tree amplitudes into loop amplitudes}},
  \href{https://doi.org/10.1016/0550-3213(94)00488-Z}{\emph{Nucl. Phys. B}
  {\bfseries 435} (1995) 59}
  [\href{https://arxiv.org/abs/hep-ph/9409265}{{\ttfamily hep-ph/9409265}}].

\bibitem{Bern:1997sc}
Z.~Bern, L.J.~Dixon and D.A.~Kosower, \emph{{One loop amplitudes for e+ e- to
  four partons}},
  \href{https://doi.org/10.1016/S0550-3213(97)00703-7}{\emph{Nucl. Phys. B}
  {\bfseries 513} (1998) 3}
  [\href{https://arxiv.org/abs/hep-ph/9708239}{{\ttfamily hep-ph/9708239}}].

\bibitem{Britto:2004nc}
R.~Britto, F.~Cachazo and B.~Feng, \emph{{Generalized unitarity and one-loop
  amplitudes in N=4 super-Yang-Mills}},
  \href{https://doi.org/10.1016/j.nuclphysb.2005.07.014}{\emph{Nucl. Phys. B}
  {\bfseries 725} (2005) 275}
  [\href{https://arxiv.org/abs/hep-th/0412103}{{\ttfamily hep-th/0412103}}].

\bibitem{Bern:2007ct}
Z.~Bern, J.J.M.~Carrasco, H.~Johansson and D.A.~Kosower, \emph{{Maximally
  supersymmetric planar Yang-Mills amplitudes at five loops}},
  \href{https://doi.org/10.1103/PhysRevD.76.125020}{\emph{Phys. Rev. D}
  {\bfseries 76} (2007) 125020}
  [\href{https://arxiv.org/abs/0705.1864}{{\ttfamily 0705.1864}}].

\bibitem{Neill:2013wsa}
D.~Neill and I.Z.~Rothstein, \emph{{Classical Space-Times from the S Matrix}},
  \href{https://doi.org/10.1016/j.nuclphysb.2013.09.007}{\emph{Nucl. Phys. B}
  {\bfseries 877} (2013) 177}
  [\href{https://arxiv.org/abs/1304.7263}{{\ttfamily 1304.7263}}].

\bibitem{Chetyrkin:1981qh}
K.G.~Chetyrkin and F.V.~Tkachov, \emph{{Integration by Parts: The Algorithm to
  Calculate beta Functions in 4 Loops}},
  \href{https://doi.org/10.1016/0550-3213(81)90199-1}{\emph{Nucl. Phys. B}
  {\bfseries 192} (1981) 159}.

\bibitem{Laporta:2000dsw}
S.~Laporta, \emph{{High precision calculation of multiloop Feynman integrals by
  difference equations}},
  \href{https://doi.org/10.1142/S0217751X00002159}{\emph{Int. J. Mod. Phys. A}
  {\bfseries 15} (2000) 5087}
  [\href{https://arxiv.org/abs/hep-ph/0102033}{{\ttfamily hep-ph/0102033}}].

\bibitem{Smirnov:2008iw}
A.V.~Smirnov, \emph{{Algorithm FIRE -- Feynman Integral REduction}},
  \href{https://doi.org/10.1088/1126-6708/2008/10/107}{\emph{JHEP} {\bfseries
  10} (2008) 107} [\href{https://arxiv.org/abs/0807.3243}{{\ttfamily
  0807.3243}}].

\bibitem{Kotikov:1990kg}
A.V.~Kotikov, \emph{{Differential equations method: New technique for massive
  Feynman diagrams calculation}},
  \href{https://doi.org/10.1016/0370-2693(91)90413-K}{\emph{Phys. Lett.}
  {\bfseries B254} (1991) 158}.

\bibitem{Bern:1993kr}
Z.~Bern, L.J.~Dixon and D.A.~Kosower, \emph{{Dimensionally regulated pentagon
  integrals}}, \href{https://doi.org/10.1016/0550-3213(94)90398-0}{\emph{Nucl.
  Phys.} {\bfseries B412} (1994) 751}
  [\href{https://arxiv.org/abs/hep-ph/9306240}{{\ttfamily hep-ph/9306240}}].

\bibitem{Remiddi:1997ny}
E.~Remiddi, \emph{{Differential equations for Feynman graph amplitudes}},
  {\emph{Nuovo Cim.} {\bfseries A110} (1997) 1435}
  [\href{https://arxiv.org/abs/hep-th/9711188}{{\ttfamily hep-th/9711188}}].

\bibitem{Gehrmann:1999as}
T.~Gehrmann and E.~Remiddi, \emph{{Differential equations for two loop four
  point functions}},
  \href{https://doi.org/10.1016/S0550-3213(00)00223-6}{\emph{Nucl. Phys.}
  {\bfseries B580} (2000) 485}
  [\href{https://arxiv.org/abs/hep-ph/9912329}{{\ttfamily hep-ph/9912329}}].

\bibitem{Henn:2013pwa}
J.M.~Henn, \emph{{Multiloop integrals in dimensional regularization made
  simple}}, \href{https://doi.org/10.1103/PhysRevLett.110.251601}{\emph{Phys.
  Rev. Lett.} {\bfseries 110} (2013) 251601}
  [\href{https://arxiv.org/abs/1304.1806}{{\ttfamily 1304.1806}}].

\bibitem{Henn:2013nsa}
J.M.~Henn, A.V.~Smirnov and V.A.~Smirnov, \emph{{Evaluating single-scale and/or
  non-planar diagrams by differential equations}},
  \href{https://doi.org/10.1007/JHEP03(2014)088}{\emph{JHEP} {\bfseries 03}
  (2014) 088} [\href{https://arxiv.org/abs/1312.2588}{{\ttfamily 1312.2588}}].

\bibitem{Parra-Martinez:2020dzs}
J.~Parra-Martinez, M.S.~Ruf and M.~Zeng, \emph{{Extremal black hole scattering
  at $\mathcal{O}(G^3)$: graviton dominance, eikonal exponentiation, and
  differential equations}},
  \href{https://doi.org/10.1007/JHEP11(2020)023}{\emph{JHEP} {\bfseries 11}
  (2020) 023} [\href{https://arxiv.org/abs/2005.04236}{{\ttfamily
  2005.04236}}].

\bibitem{Antonelli:2019ytb}
A.~Antonelli, A.~Buonanno, J.~Steinhoff, M.~van~de Meent and J.~Vines,
  \emph{{Energetics of two-body Hamiltonians in post-Minkowskian gravity}},
  \href{https://doi.org/10.1103/PhysRevD.99.104004}{\emph{Phys. Rev. D}
  {\bfseries 99} (2019) 104004}
  [\href{https://arxiv.org/abs/1901.07102}{{\ttfamily 1901.07102}}].

\bibitem{Khalil:2022ylj}
M.~Khalil, A.~Buonanno, J.~Steinhoff and J.~Vines, \emph{{Energetics and
  scattering of gravitational two-body systems at fourth post-Minkowskian
  order}}, \href{https://doi.org/10.1103/PhysRevD.106.024042}{\emph{Phys. Rev.
  D} {\bfseries 106} (2022) 024042}
  [\href{https://arxiv.org/abs/2204.05047}{{\ttfamily 2204.05047}}].

\bibitem{Buonanno:2022pgc}
A.~Buonanno, M.~Khalil, D.~O'Connell, R.~Roiban, M.P.~Solon and M.~Zeng,
  \emph{{Snowmass White Paper: Gravitational Waves and Scattering Amplitudes}},
   in \emph{{2022 Snowmass Summer Study}}, 4, 2022
  [\href{https://arxiv.org/abs/2204.05194}{{\ttfamily 2204.05194}}].

\bibitem{Adamo:2022dcm}
T.~Adamo, J.J.M.~Carrasco, M.~Carrillo-Gonz\'alez, M.~Chiodaroli, H.~Elvang,
  H.~Johansson et~al., \emph{{Snowmass White Paper: the Double Copy and its
  Applications}},  in \emph{{2022 Snowmass Summer Study}}, 4, 2022
  [\href{https://arxiv.org/abs/2204.06547}{{\ttfamily 2204.06547}}].

\bibitem{Bjerrum-Bohr:2022blt}
N.E.J.~Bjerrum-Bohr, P.H.~Damgaard, L.~Plante and P.~Vanhove, \emph{{The SAGEX
  Review on Scattering Amplitudes, Chapter 13: Post-Minkowskian expansion from
  Scattering Amplitudes}},  \href{https://arxiv.org/abs/2203.13024}{{\ttfamily
  2203.13024}}.

\bibitem{Kosower:2022yvp}
D.A.~Kosower, R.~Monteiro and D.~O'Connell, \emph{{The SAGEX Review on
  Scattering Amplitudes, Chapter 14: Classical Gravity from Scattering
  Amplitudes}},  \href{https://arxiv.org/abs/2203.13025}{{\ttfamily
  2203.13025}}.

\bibitem{3PM}
Z.~Bern, C.~Cheung, R.~Roiban, C.-H.~Shen, M.P.~Solon and M.~Zeng,
  \emph{{Scattering amplitudes and the conservative Hamiltonian for binary
  systems at third post-Minkowskian order}},
  \href{https://doi.org/10.1103/PhysRevLett.122.201603}{\emph{Phys. Rev. Lett.}
  {\bfseries 122} (2019) 201603}
  [\href{https://arxiv.org/abs/1901.04424}{{\ttfamily 1901.04424}}].

\bibitem{3PMLong}
Z.~Bern, C.~Cheung, R.~Roiban, C.-H.~Shen, M.P.~Solon and M.~Zeng, \emph{{Black
  hole binary dynamics from the double copy and effective theory}},
  \href{https://arxiv.org/abs/1908.01493}{{\ttfamily 1908.01493}}.

\bibitem{Bern:2021dqo}
Z.~Bern, J.~Parra-Martinez, R.~Roiban, M.S.~Ruf, C.-H.~Shen, M.P.~Solon et~al.,
  \emph{{Scattering Amplitudes and Conservative Binary Dynamics at ${\cal
  O}(G^4)$}}, \href{https://doi.org/10.1103/PhysRevLett.126.171601}{\emph{Phys.
  Rev. Lett.} {\bfseries 126} (2021) 171601}
  [\href{https://arxiv.org/abs/2101.07254}{{\ttfamily 2101.07254}}].

\bibitem{Bern:2021yeh}
Z.~Bern, J.~Parra-Martinez, R.~Roiban, M.S.~Ruf, C.-H.~Shen, M.P.~Solon et~al.,
  \emph{{Scattering Amplitudes, the Tail Effect, and Conservative Binary
  Dynamics at $O(G^4)$}},  \href{https://arxiv.org/abs/2112.10750}{{\ttfamily
  2112.10750}}.

\bibitem{Arkani-Hamed:2017jhn}
N.~Arkani-Hamed, T.-C.~Huang and Y.-t.~Huang, \emph{{Scattering amplitudes for
  all masses and spins}},
  \href{https://doi.org/10.1007/JHEP11(2021)070}{\emph{JHEP} {\bfseries 11}
  (2021) 070} [\href{https://arxiv.org/abs/1709.04891}{{\ttfamily
  1709.04891}}].

\bibitem{Bern:2020buy}
Z.~Bern, A.~Luna, R.~Roiban, C.-H.~Shen and M.~Zeng, \emph{{Spinning black hole
  binary dynamics, scattering amplitudes, and effective field theory}},
  \href{https://doi.org/10.1103/PhysRevD.104.065014}{\emph{Phys. Rev. D}
  {\bfseries 104} (2021) 065014}
  [\href{https://arxiv.org/abs/2005.03071}{{\ttfamily 2005.03071}}].

\bibitem{Chiodaroli:2021eug}
M.~Chiodaroli, H.~Johansson and P.~Pichini, \emph{{Compton black-hole
  scattering for s \ensuremath{\leq} 5/2}},
  \href{https://doi.org/10.1007/JHEP02(2022)156}{\emph{JHEP} {\bfseries 02}
  (2022) 156} [\href{https://arxiv.org/abs/2107.14779}{{\ttfamily
  2107.14779}}].

\bibitem{Aoude:2020onz}
R.~Aoude, K.~Haddad and A.~Helset, \emph{{On-shell heavy particle effective
  theories}}, \href{https://doi.org/10.1007/JHEP05(2020)051}{\emph{JHEP}
  {\bfseries 05} (2020) 051}
  [\href{https://arxiv.org/abs/2001.09164}{{\ttfamily 2001.09164}}].

\bibitem{Jakobsen:2021zvh}
G.U.~Jakobsen, G.~Mogull, J.~Plefka and J.~Steinhoff, \emph{{SUSY in the sky
  with gravitons}}, \href{https://doi.org/10.1007/JHEP01(2022)027}{\emph{JHEP}
  {\bfseries 01} (2022) 027}
  [\href{https://arxiv.org/abs/2109.04465}{{\ttfamily 2109.04465}}].

\bibitem{Maybee:2019jus}
B.~Maybee, D.~O'Connell and J.~Vines, \emph{{Observables and amplitudes for
  spinning particles and black holes}},
  \href{https://doi.org/10.1007/JHEP12(2019)156}{\emph{JHEP} {\bfseries 12}
  (2019) 156} [\href{https://arxiv.org/abs/1906.09260}{{\ttfamily
  1906.09260}}].

\bibitem{Guevara:2019fsj}
A.~Guevara, A.~Ochirov and J.~Vines, \emph{{Black-hole scattering with general
  spin directions from minimal-coupling amplitudes}},
  \href{https://doi.org/10.1103/PhysRevD.100.104024}{\emph{Phys. Rev. D}
  {\bfseries 100} (2019) 104024}
  [\href{https://arxiv.org/abs/1906.10071}{{\ttfamily 1906.10071}}].

\bibitem{Guevara:2018wpp}
A.~Guevara, A.~Ochirov and J.~Vines, \emph{{Scattering of Spinning Black Holes
  from Exponentiated Soft Factors}},
  \href{https://doi.org/10.1007/JHEP09(2019)056}{\emph{JHEP} {\bfseries 09}
  (2019) 056} [\href{https://arxiv.org/abs/1812.06895}{{\ttfamily
  1812.06895}}].

\bibitem{Kosmopoulos:2021zoq}
D.~Kosmopoulos and A.~Luna, \emph{{Quadratic-in-spin Hamiltonian at $
  \mathcal{O} $(G$^{2}$) from scattering amplitudes}},
  \href{https://doi.org/10.1007/JHEP07(2021)037}{\emph{JHEP} {\bfseries 07}
  (2021) 037} [\href{https://arxiv.org/abs/2102.10137}{{\ttfamily
  2102.10137}}].

\bibitem{Chung:2019duq}
M.-Z.~Chung, Y.-T.~Huang and J.-W.~Kim, \emph{{Classical potential for general
  spinning bodies}}, \href{https://doi.org/10.1007/JHEP09(2020)074}{\emph{JHEP}
  {\bfseries 09} (2020) 074}
  [\href{https://arxiv.org/abs/1908.08463}{{\ttfamily 1908.08463}}].

\bibitem{Chung:2020rrz}
M.-Z.~Chung, Y.-t.~Huang, J.-W.~Kim and S.~Lee, \emph{{Complete Hamiltonian for
  spinning binary systems at first post-Minkowskian order}},
  \href{https://doi.org/10.1007/JHEP05(2020)105}{\emph{JHEP} {\bfseries 05}
  (2020) 105} [\href{https://arxiv.org/abs/2003.06600}{{\ttfamily
  2003.06600}}].

\bibitem{Chen:2021qkk}
W.-M.~Chen, M.-Z.~Chung, Y.-t.~Huang and J.-W.~Kim, \emph{{The 2PM Hamiltonian
  for binary Kerr to quartic in spin}},
  \href{https://arxiv.org/abs/2111.13639}{{\ttfamily 2111.13639}}.

\bibitem{Jakobsen:2022fcj}
G.U.~Jakobsen and G.~Mogull, \emph{{Conservative and Radiative Dynamics of
  Spinning Bodies at Third Post-Minkowskian Order Using Worldline Quantum Field
  Theory}}, \href{https://doi.org/10.1103/PhysRevLett.128.141102}{\emph{Phys.
  Rev. Lett.} {\bfseries 128} (2022) 141102}
  [\href{https://arxiv.org/abs/2201.07778}{{\ttfamily 2201.07778}}].

\bibitem{Bern:2022kto}
Z.~Bern, D.~Kosmopoulos, A.~Luna, R.~Roiban and F.~Teng, \emph{{Binary Dynamics
  Through the Fifth Power of Spin at $\mathcal{O}(G^2)$}},
  \href{https://arxiv.org/abs/2203.06202}{{\ttfamily 2203.06202}}.

\bibitem{Aoude:2022trd}
R.~Aoude, K.~Haddad and A.~Helset, \emph{{Searching for Kerr in the 2PM
  amplitude}}, \href{https://doi.org/10.1007/JHEP07(2022)072}{\emph{JHEP}
  {\bfseries 07} (2022) 072}
  [\href{https://arxiv.org/abs/2203.06197}{{\ttfamily 2203.06197}}].

\bibitem{Aoude:2022thd}
R.~Aoude, K.~Haddad and A.~Helset, \emph{{Classical gravitational
  spinning-spinless scattering at $\mathcal{O}(G^{2} S^{\infty})$}},
  \href{https://arxiv.org/abs/2205.02809}{{\ttfamily 2205.02809}}.

\bibitem{FebresCordero:2022jts}
F.~Febres~Cordero, M.~Kraus, G.~Lin, M.S.~Ruf and M.~Zeng, \emph{{Conservative
  Binary Dynamics with a Spinning Black Hole at $\mathcal{O}(G^3)$ from
  Scattering Amplitudes}},  \href{https://arxiv.org/abs/2205.07357}{{\ttfamily
  2205.07357}}.

\bibitem{Cheung:2020sdj}
C.~Cheung and M.P.~Solon, \emph{{Tidal Effects in the Post-Minkowskian
  Expansion}},
  \href{https://doi.org/10.1103/PhysRevLett.125.191601}{\emph{Phys. Rev. Lett.}
  {\bfseries 125} (2020) 191601}
  [\href{https://arxiv.org/abs/2006.06665}{{\ttfamily 2006.06665}}].

\bibitem{Haddad:2020que}
K.~Haddad and A.~Helset, \emph{{Tidal effects in quantum field theory}},
  \href{https://doi.org/10.1007/JHEP12(2020)024}{\emph{JHEP} {\bfseries 12}
  (2020) 024} [\href{https://arxiv.org/abs/2008.04920}{{\ttfamily
  2008.04920}}].

\bibitem{Aoude:2020ygw}
R.~Aoude, K.~Haddad and A.~Helset, \emph{{Tidal effects for spinning
  particles}}, \href{https://doi.org/10.1007/JHEP03(2021)097}{\emph{JHEP}
  {\bfseries 03} (2021) 097}
  [\href{https://arxiv.org/abs/2012.05256}{{\ttfamily 2012.05256}}].

\bibitem{Bern:2020uwk}
Z.~Bern, J.~Parra-Martinez, R.~Roiban, E.~Sawyer and C.-H.~Shen, \emph{{Leading
  Nonlinear Tidal Effects and Scattering Amplitudes}},
  \href{https://doi.org/10.1007/JHEP05(2021)188}{\emph{JHEP} {\bfseries 05}
  (2021) 188} [\href{https://arxiv.org/abs/2010.08559}{{\ttfamily
  2010.08559}}].

\bibitem{Cheung:2020gbf}
C.~Cheung, N.~Shah and M.P.~Solon, \emph{{Mining the Geodesic Equation for
  Scattering Data}},
  \href{https://doi.org/10.1103/PhysRevD.103.024030}{\emph{Phys. Rev. D}
  {\bfseries 103} (2021) 024030}
  [\href{https://arxiv.org/abs/2010.08568}{{\ttfamily 2010.08568}}].

\bibitem{AccettulliHuber:2020dal}
M.~Accettulli~Huber, A.~Brandhuber, S.~De~Angelis and G.~Travaglini,
  \emph{{From amplitudes to gravitational radiation with cubic interactions and
  tidal effects}},
  \href{https://doi.org/10.1103/PhysRevD.103.045015}{\emph{Phys. Rev. D}
  {\bfseries 103} (2021) 045015}
  [\href{https://arxiv.org/abs/2012.06548}{{\ttfamily 2012.06548}}].

\bibitem{DiVecchia:2020ymx}
P.~Di~Vecchia, C.~Heissenberg, R.~Russo and G.~Veneziano, \emph{{Universality
  of ultra-relativistic gravitational scattering}},
  \href{https://doi.org/10.1016/j.physletb.2020.135924}{\emph{Phys. Lett. B}
  {\bfseries 811} (2020) 135924}
  [\href{https://arxiv.org/abs/2008.12743}{{\ttfamily 2008.12743}}].

\bibitem{DiVecchia:2021bdo}
P.~Di~Vecchia, C.~Heissenberg, R.~Russo and G.~Veneziano, \emph{{The eikonal
  approach to gravitational scattering and radiation at $ \mathcal{O}
  $(G$^{3}$)}}, \href{https://doi.org/10.1007/JHEP07(2021)169}{\emph{JHEP}
  {\bfseries 07} (2021) 169}
  [\href{https://arxiv.org/abs/2104.03256}{{\ttfamily 2104.03256}}].

\bibitem{Bjerrum-Bohr:2021din}
N.E.J.~Bjerrum-Bohr, P.H.~Damgaard, L.~Plant\'e and P.~Vanhove, \emph{{The
  amplitude for classical gravitational scattering at third Post-Minkowskian
  order}}, \href{https://doi.org/10.1007/JHEP08(2021)172}{\emph{JHEP}
  {\bfseries 08} (2021) 172}
  [\href{https://arxiv.org/abs/2105.05218}{{\ttfamily 2105.05218}}].

\bibitem{Damgaard:2021ipf}
P.H.~Damgaard, L.~Plante and P.~Vanhove, \emph{{On an exponential
  representation of the gravitational S-matrix}},
  \href{https://doi.org/10.1007/JHEP11(2021)213}{\emph{JHEP} {\bfseries 11}
  (2021) 213} [\href{https://arxiv.org/abs/2107.12891}{{\ttfamily
  2107.12891}}].

\bibitem{Brandhuber:2021eyq}
A.~Brandhuber, G.~Chen, G.~Travaglini and C.~Wen, \emph{{Classical
  gravitational scattering from a gauge-invariant double copy}},
  \href{https://doi.org/10.1007/JHEP10(2021)118}{\emph{JHEP} {\bfseries 10}
  (2021) 118} [\href{https://arxiv.org/abs/2108.04216}{{\ttfamily
  2108.04216}}].

\bibitem{Herrmann:2021lqe}
E.~Herrmann, J.~Parra-Martinez, M.S.~Ruf and M.~Zeng, \emph{{Gravitational
  Bremsstrahlung from Reverse Unitarity}},
  \href{https://doi.org/10.1103/PhysRevLett.126.201602}{\emph{Phys. Rev. Lett.}
  {\bfseries 126} (2021) 201602}
  [\href{https://arxiv.org/abs/2101.07255}{{\ttfamily 2101.07255}}].

\bibitem{Herrmann:2021tct}
E.~Herrmann, J.~Parra-Martinez, M.S.~Ruf and M.~Zeng, \emph{{Radiative
  classical gravitational observables at $ \mathcal{O} $(G$^{3}$) from
  scattering amplitudes}},
  \href{https://doi.org/10.1007/JHEP10(2021)148}{\emph{JHEP} {\bfseries 10}
  (2021) 148} [\href{https://arxiv.org/abs/2104.03957}{{\ttfamily
  2104.03957}}].

\bibitem{DiVecchia:2021ndb}
P.~Di~Vecchia, C.~Heissenberg, R.~Russo and G.~Veneziano, \emph{{Radiation
  Reaction from Soft Theorems}},
  \href{https://doi.org/10.1016/j.physletb.2021.136379}{\emph{Phys. Lett. B}
  {\bfseries 818} (2021) 136379}
  [\href{https://arxiv.org/abs/2101.05772}{{\ttfamily 2101.05772}}].

\bibitem{Heissenberg:2021tzo}
C.~Heissenberg, \emph{{Infrared divergences and the eikonal exponentiation}},
  \href{https://doi.org/10.1103/PhysRevD.104.046016}{\emph{Phys. Rev. D}
  {\bfseries 104} (2021) 046016}
  [\href{https://arxiv.org/abs/2105.04594}{{\ttfamily 2105.04594}}].

\bibitem{Alessio:2022kwv}
F.~Alessio and P.~Di~Vecchia, \emph{{Radiation reaction for spinning black-hole
  scattering}},
  \href{https://doi.org/10.1016/j.physletb.2022.137258}{\emph{Phys. Lett. B}
  {\bfseries 832} (2022) 137258}
  [\href{https://arxiv.org/abs/2203.13272}{{\ttfamily 2203.13272}}].

\bibitem{Naoz:2012bx}
S.~Naoz, B.~Kocsis, A.~Loeb and N.~Yunes, \emph{{Resonant Post-Newtonian
  Eccentricity Excitation in Hierarchical Three-body Systems}},
  \href{https://doi.org/10.1088/0004-637X/773/2/187}{\emph{Astrophys. J.}
  {\bfseries 773} (2013) 187}
  [\href{https://arxiv.org/abs/1206.4316}{{\ttfamily 1206.4316}}].

\bibitem{Lim:2020cvm}
H.~Lim and C.L.~Rodriguez, \emph{{Relativistic three-body effects in
  hierarchical triples}},
  \href{https://doi.org/10.1103/PhysRevD.102.064033}{\emph{Phys. Rev. D}
  {\bfseries 102} (2020) 064033}
  [\href{https://arxiv.org/abs/2001.03654}{{\ttfamily 2001.03654}}].

\bibitem{Martinez:2020lzt}
M.A.S.~Martinez et~al., \emph{{Black Hole Mergers from Hierarchical Triples in
  Dense Star Clusters}},
  \href{https://doi.org/10.3847/1538-4357/abba25}{\emph{Astrophys. J.}
  {\bfseries 903} (2020) 67}
  [\href{https://arxiv.org/abs/2009.08468}{{\ttfamily 2009.08468}}].

\bibitem{Fragione:2020gly}
G.~Fragione, M.A.S.~Martinez, K.~Kremer, S.~Chatterjee, C.L.~Rodriguez, C.S.~Ye
  et~al., \emph{{Demographics of triple systems in dense star clusters}},
  \href{https://doi.org/10.3847/1538-4357/aba89b}{\emph{Astrophys. J.}
  {\bfseries 900} (2020) 16}
  [\href{https://arxiv.org/abs/2007.11605}{{\ttfamily 2007.11605}}].

\bibitem{Ohta:1974pq}
T.~Ohta, H.~Okamura, K.~Hiida and T.~Kimura, \emph{{Higher order gravitational
  potential for many-body system}},
  \href{https://doi.org/10.1143/PTP.51.1220}{\emph{Prog. Theor. Phys.}
  {\bfseries 51} (1974) 1220}.

\bibitem{SCHAFER1987336}
G.~Sch{\"a}fer, \emph{Three-body hamiltonian in general relativity},
  \href{https://doi.org/https://doi.org/10.1016/0375-9601(87)90389-6}{\emph{Physics
  Letters A} {\bfseries 123} (1987) 336}.

\bibitem{Ledvinka:2008tk}
T.~Ledvinka, G.~Schaefer and J.~Bicak, \emph{{Relativistic Closed-Form
  Hamiltonian for Many-Body Gravitating Systems in the Post-Minkowskian
  Approximation}},
  \href{https://doi.org/10.1103/PhysRevLett.100.251101}{\emph{Phys. Rev. Lett.}
  {\bfseries 100} (2008) 251101}
  [\href{https://arxiv.org/abs/0807.0214}{{\ttfamily 0807.0214}}].

\bibitem{Loebbert:2020aos}
F.~Loebbert, J.~Plefka, C.~Shi and T.~Wang, \emph{{Three-body effective
  potential in general relativity at second post-Minkowskian order and
  resulting post-Newtonian contributions}},
  \href{https://doi.org/10.1103/PhysRevD.103.064010}{\emph{Phys. Rev. D}
  {\bfseries 103} (2021) 064010}
  [\href{https://arxiv.org/abs/2012.14224}{{\ttfamily 2012.14224}}].

\bibitem{Ferrell:1987gf}
R.C.~Ferrell and D.M.~Eardley, \emph{{Slow motion scattering and coalescence of
  maximally charged black holes}},
  \href{https://doi.org/10.1103/PhysRevLett.59.1617}{\emph{Phys. Rev. Lett.}
  {\bfseries 59} (1987) 1617}.

\bibitem{Lippmann:1950zz}
B.A.~Lippmann and J.~Schwinger, \emph{{Variational Principles for Scattering
  Processes. I}}, \href{https://doi.org/10.1103/PhysRev.79.469}{\emph{Phys.
  Rev.} {\bfseries 79} (1950) 469}.

\bibitem{Cristofoli:2019neg}
A.~Cristofoli, N.E.J.~Bjerrum-Bohr, P.H.~Damgaard and P.~Vanhove,
  \emph{{Post-Minkowskian Hamiltonians in general relativity}},
  \href{https://doi.org/10.1103/PhysRevD.100.084040}{\emph{Phys. Rev. D}
  {\bfseries 100} (2019) 084040}
  [\href{https://arxiv.org/abs/1906.01579}{{\ttfamily 1906.01579}}].

\bibitem{Bern:2019crd}
Z.~Bern, C.~Cheung, R.~Roiban, C.-H.~Shen, M.P.~Solon and M.~Zeng, \emph{{Black
  Hole Binary Dynamics from the Double Copy and Effective Theory}},
  \href{https://doi.org/10.1007/JHEP10(2019)206}{\emph{JHEP} {\bfseries 10}
  (2019) 206} [\href{https://arxiv.org/abs/1908.01493}{{\ttfamily
  1908.01493}}].

\bibitem{Boos:1990rg}
E.E.~Boos and A.I.~Davydychev, \emph{{A Method of evaluating massive Feynman
  integrals}}, \href{https://doi.org/10.1007/BF01016805}{\emph{Theor. Math.
  Phys.} {\bfseries 89} (1991) 1052}.

\bibitem{Chu:2008xm}
Y.-Z.~Chu, \emph{{The n-body problem in General Relativity up to the second
  post-Newtonian order from perturbative field theory}},
  \href{https://doi.org/10.1103/PhysRevD.79.044031}{\emph{Phys. Rev. D}
  {\bfseries 79} (2009) 044031}
  [\href{https://arxiv.org/abs/0812.0012}{{\ttfamily 0812.0012}}].

\bibitem{Ananthanarayan:2020xut}
B.~Ananthanarayan, S.~Friot, S.~Ghosh and A.~Hurier, \emph{{New analytic
  continuations for the Appell $F_4$ series from quadratic transformations of
  the Gauss $_{2}F_1$ function}},
  \href{https://arxiv.org/abs/2005.07170}{{\ttfamily 2005.07170}}.

\bibitem{Bzowski:2013sza}
A.~Bzowski, P.~McFadden and K.~Skenderis, \emph{{Implications of conformal
  invariance in momentum space}},
  \href{https://doi.org/10.1007/JHEP03(2014)111}{\emph{JHEP} {\bfseries 03}
  (2014) 111} [\href{https://arxiv.org/abs/1304.7760}{{\ttfamily 1304.7760}}].

\bibitem{Bzowski:2015yxv}
A.~Bzowski, P.~McFadden and K.~Skenderis, \emph{{Evaluation of conformal
  integrals}}, \href{https://doi.org/10.1007/JHEP02(2016)068}{\emph{JHEP}
  {\bfseries 02} (2016) 068}
  [\href{https://arxiv.org/abs/1511.02357}{{\ttfamily 1511.02357}}].

\bibitem{Gutowski:2001xa}
J.~Gutowski and G.~Papadopoulos, \emph{{Three body interactions, angular
  momentum and black hole moduli spaces}},
  \href{https://doi.org/10.1088/0264-9381/19/3/305}{\emph{Class. Quant. Grav.}
  {\bfseries 19} (2002) 493}
  [\href{https://arxiv.org/abs/hep-th/0107252}{{\ttfamily hep-th/0107252}}].

\bibitem{Bzowski:2020lip}
A.~Bzowski, \emph{{TripleK: A Mathematica package for evaluating triple-K
  integrals and conformal correlation functions}},
  \href{https://doi.org/10.1016/j.cpc.2020.107538}{\emph{Comput. Phys. Commun.}
  {\bfseries 258} (2021) 107538}
  [\href{https://arxiv.org/abs/2005.10841}{{\ttfamily 2005.10841}}].

\bibitem{Beneke:1997zp}
M.~Beneke and V.A.~Smirnov, \emph{{Asymptotic expansion of Feynman integrals
  near threshold}},
  \href{https://doi.org/10.1016/S0550-3213(98)00138-2}{\emph{Nucl. Phys. B}
  {\bfseries 522} (1998) 321}
  [\href{https://arxiv.org/abs/hep-ph/9711391}{{\ttfamily hep-ph/9711391}}].

\bibitem{Landau:1975pou}
L.D.~Landau and E.M.~Lifschits, \emph{{The Classical Theory of Fields}},
  vol.~Volume 2 of \emph{Course of Theoretical Physics}, Pergamon Press, Oxford
  (1975).

\bibitem{1964NCim...33..331I}
W.~{Israel} and K.A.~{Khan}, \emph{{Collinear particles and bondi dipoles in
  general relativity}}, \href{https://doi.org/10.1007/BF02750196}{\emph{Il
  Nuovo Cimento} {\bfseries 33} (1964) 331}.

\bibitem{Costa:2000kf}
M.S.~Costa and M.J.~Perry, \emph{{Interacting black holes}},
  \href{https://doi.org/10.1016/S0550-3213(00)00577-0}{\emph{Nucl. Phys. B}
  {\bfseries 591} (2000) 469}
  [\href{https://arxiv.org/abs/hep-th/0008106}{{\ttfamily hep-th/0008106}}].

\bibitem{Majumdar:1947eu}
S.D.~Majumdar, \emph{{A class of exact solutions of Einstein's field
  equations}}, \href{https://doi.org/10.1103/PhysRev.72.390}{\emph{Phys. Rev.}
  {\bfseries 72} (1947) 390}.

\bibitem{Papaetrou:1947ib}
A.~Papaetrou, \emph{{A Static solution of the equations of the gravitational
  field for an arbitrary charge distribution}}, {\emph{Proc. Roy. Irish Acad.
  A} {\bfseries 51} (1947) 191}.

\bibitem{Manton:1981mp}
N.S.~Manton, \emph{{A Remark on the Scattering of BPS Monopoles}},
  \href{https://doi.org/10.1016/0370-2693(82)90950-9}{\emph{Phys. Lett. B}
  {\bfseries 110} (1982) 54}.

\bibitem{Bogomolny:1975de}
E.B.~Bogomolny, \emph{{Stability of Classical Solutions}}, {\emph{Sov. J. Nucl.
  Phys.} {\bfseries 24} (1976) 449}.

\bibitem{Prasad:1975kr}
M.K.~Prasad and C.M.~Sommerfield, \emph{{An Exact Classical Solution for the 't
  Hooft Monopole and the Julia-Zee Dyon}},
  \href{https://doi.org/10.1103/PhysRevLett.35.760}{\emph{Phys. Rev. Lett.}
  {\bfseries 35} (1975) 760}.

\bibitem{Atiyah1985LowES}
M.F.~Atiyah and N.J.~Hitchin, \emph{Low energy scattering of non-abelian
  monopoles}, {\emph{Physics Letters A} {\bfseries 107} (1985) 21}.

\bibitem{Ruback:1987sg}
P.J.~Ruback, \emph{{$\sigma$ Model Solitons and Their Moduli Space Metrics}},
  \href{https://doi.org/10.1007/BF01224905}{\emph{Commun. Math. Phys.}
  {\bfseries 116} (1988) 645}.

\bibitem{Gutowski:1998bn}
J.~Gutowski and G.~Papadopoulos, \emph{{The Moduli spaces of world volume brane
  solitons}}, \href{https://doi.org/10.1016/S0370-2693(98)00608-X}{\emph{Phys.
  Lett. B} {\bfseries 432} (1998) 97}
  [\href{https://arxiv.org/abs/hep-th/9802186}{{\ttfamily hep-th/9802186}}].

\bibitem{Shiraishi:1992nq}
K.~Shiraishi, \emph{{Moduli space metric for maximally charged dilaton black
  holes}}, \href{https://doi.org/10.1016/0550-3213(93)90648-9}{\emph{Nucl.
  Phys. B} {\bfseries 402} (1993) 399}
  [\href{https://arxiv.org/abs/1407.5377}{{\ttfamily 1407.5377}}].

\bibitem{Gibbons:1997iy}
G.W.~Gibbons, G.~Papadopoulos and K.S.~Stelle, \emph{{HKT and OKT geometries on
  soliton black hole moduli spaces}},
  \href{https://doi.org/10.1016/S0550-3213(97)00599-3}{\emph{Nucl. Phys. B}
  {\bfseries 508} (1997) 623}
  [\href{https://arxiv.org/abs/hep-th/9706207}{{\ttfamily hep-th/9706207}}].

\bibitem{Bern:1992em}
Z.~Bern, L.J.~Dixon and D.A.~Kosower, \emph{{Dimensionally regulated one loop
  integrals}}, \href{https://doi.org/10.1016/0370-2693(93)90400-C}{\emph{Phys.
  Lett. B} {\bfseries 302} (1993) 299}
  [\href{https://arxiv.org/abs/hep-ph/9212308}{{\ttfamily hep-ph/9212308}}].

\bibitem{Henn:2014qga}
J.M.~Henn, \emph{{Lectures on differential equations for Feynman integrals}},
  \href{https://doi.org/10.1088/1751-8113/48/15/153001}{\emph{J. Phys. A}
  {\bfseries 48} (2015) 153001}
  [\href{https://arxiv.org/abs/1412.2296}{{\ttfamily 1412.2296}}].

\bibitem{Kalin:2019rwq}
G.~K\"alin and R.A.~Porto, \emph{{From Boundary Data to Bound States}},
  \href{https://doi.org/10.1007/JHEP01(2020)072}{\emph{JHEP} {\bfseries 01}
  (2020) 072} [\href{https://arxiv.org/abs/1910.03008}{{\ttfamily
  1910.03008}}].

\bibitem{Vaidya:2014kza}
V.~Vaidya, \emph{{Gravitational spin Hamiltonians from the S matrix}},
  \href{https://doi.org/10.1103/PhysRevD.91.024017}{\emph{Phys. Rev. D}
  {\bfseries 91} (2015) 024017}
  [\href{https://arxiv.org/abs/1410.5348}{{\ttfamily 1410.5348}}].

\end{thebibliography}\endgroup
\end{document}